\documentclass[12pt]{iopart}
\usepackage{graphicx,comment}
\usepackage{dcolumn}
\usepackage{bm}
\usepackage{hyperref}

\usepackage{float}
\usepackage{caption}
\usepackage{subcaption}
\usepackage{xcolor,marginnote,upgreek}
\usepackage{mathtools,amsfonts}
\usepackage[percent]{overpic}
\usepackage{booktabs}

\newcommand{\A}{\mathcal{A}}
\newcommand{\E}{\mathcal{E}}
\newcommand{\V}{\mathcal{V}}
\newcommand{\p}{\mathbf{p}}
\newcommand{\q}{\mathbf{q}}
\newcommand{\up}{\uparrow}
\newcommand{\down}{\downarrow}

\DeclareMathOperator{\sech}{sech}

\begin{document}

\title{Reading qubits with sequential weak measurements: limits of information extraction}
\author{Cesar Lema$^1$, Aleix Bou-Comas$^{2,3}$, Atithi Acharya$^4$, Vadim Oganesyan$^{3,5,\ast}$, Anirvan M. Sengupta$^{4,6}$}
\address{$^1$ Department of Applied Physics, Stanford University, Palo Alto, CA 94305, USA}
\address{$^2$ The Graduate Center, CUNY, New York, New York 10016, USA}
\address{$^3$ Institute of Fundamental Physics IFF-CSIC, Calle Serrano 113b, Madrid 28006, Spain}
\address{$^4$ Department of Physics and Astronomy, Rutgers University, Piscataway, NJ 08854, USA}
\address{$^5$ Department of Physics and Astronomy, College of Staten Island, CUNY, Staten Island, New York 10314, USA }
\address{$^6$ Center for Computational Mathematics and Center for Computational Quantum Physics, Flatiron Institute, New York, NY 10010 USA}
\ead{vadim.oganesyan@csi.cuny.edu}
\noindent$^\ast$ Corresponding author.

\begin{abstract}

We study the information physics of quantum trajectories based on weak measurements in order to address the optimal achievable performance in qubit configuration readout for two realistic models of single qubit readout: (i) Model I  is informationally complete, but without intrinsic dynamics; (ii) Model II is informationally incomplete weak measurements with intrinsic dynamics.
We
use mutual information to characterize how much
information about the initial state is encoded in the measurement record. 
Using a fixed discrete time-step formulation, we compute the mutual information while varying the measurement strength, duration of measurement record, and the relative strength of intrinsic dynamics in our measurement schemes.
We observe and exploit the emergence of continuum scaling and the Stochastic Master Equation in the weak measurement limit.
We develop a perturbative analytic expansion in
the measurement efficiency parameter to calculate
mutual information, which captures qualitative and quantitative features of the numerical data.

Both models exhibit clear bounds on information extraction as limiting values of the scaling function.
Our analysis obtains these bounds and also flags optimal conditions on measurement strength and/or duration required to saturate them, as determined by intrinsic precessional dynamics (in Model II).
Our results should be useful both for quantum device operation and optimization and also, possibly, for improving the performance of recent machine learning approaches for qubit and multiqubit configuration readout in current Noisy Intermediate-Scale Quantum (NISQ) experiment regimes.

\end{abstract}

\maketitle

\section{\label{sec:intro} Introduction}

Quantum hardware and experiments will allow us to probe fundamental questions about the quantum world and build technologies to solve previously intractable problems. 
High fidelity readout --- inferring the initial state of a quantum system from measurements --- is a crucial subroutine required for accessing the vast information contained in these quantum systems.
In practical schemes, such as dispersive qubit readout of superconducting circuits \cite{Oliver_2019}, optical readout of solid-state spin qubits \cite{TVQrosenthal2024single}, or homodyne measurement in quantum optics \cite{leonhardt1997measuring},
the quantum measurements are weak generalized measurements where the measured state evolves through small stochastic perturbations while an observer gains small amounts of information about the state.
Furthermore, in the current Noisy Intermediate-Scale Quantum (NISQ) hardware regime, inherent dynamics and noise can lead to nontrivial evolution and loss of accessible information about the initial state.
For readout, these effects ultimately lead to less than perfect inference of the initial state, even when going beyond filter-based methods \cite{gambetta2007protocols} and using sophisticated machine learning frameworks like reservoir computing \cite{angelatos2021reservoir}.

In this paper, we take a step back and ask a more fundamental question: given a scheme of sequential weak measurements, how much intrinsic information does the measurement record have about the initial state? We want to pose this question under the ideal circumstance of knowing everything about the quantum dynamics, the measurement process, and the possibilities for the initial states. 
This way we separate out the issues of characterizing a system from the fundamental information physics and limitations of the measurement setting. For concreteness, we work throughout with the binary qubit state space $\{|\!\up\rangle,|\!\down\rangle\}$, for which the Holevo bound caps the information any measurement record can convey about the initial state at one bit; how closely a given measurement scheme approaches this ceiling is the central quantitative question we address.

We study this question by two different methods, focusing on two realistic models of qubit readout.
We first estimate the mutual information, a classical information theoretic measure \cite{ash2012information, Cover2006}, between the qubit and the measurement record. These estimates have error bounds with a probabilistic guarantee, and allows to study the dependence of mutual information on the number of sequential measurements.
We also use mutual information to upper bound the accuracy \cite{meyen2016relation} of initial state recovery for any prediction scheme. Then, by posing the initial state readout as a supervised learning problem, we consider the accuracy of the Bayes optimal classifier, the optimal prediction scheme, and how it changes with the number of sequential measurements. 

We apply these methods to several systems continuously monitored, building on the work of Oreshkov and Brun \cite{oreshkov2005weak}. One of the systems is a qubit measured sequentially with an informationally complete weak measurement scheme. The second system has the qubit repeatedly measured with an informationally incomplete weak measurement scheme but in presence of nontrivial unitary evolution. This system has been studied by Tang and Li \cite{tang2020phase}, who claim a possible phase transition in measurement records with certain initialization schemes.

We find that, for many generic systems with measurement, there is a time-scale beyond which the information about the initial state becomes irrelevant for measurement statistics. Thus, longer measurements marginally yield very little extra information about the initial state. This phenomenon has important consequences for recent physics-agnostic statistical (machine learning) methods for initial state recovery. For a fixed number of trajectories, if we provide very long measurement records, the method starts overfitting, trying to extract signals from late measurements which are effectively independent of the initial state. Physics-aware approaches can avoid this pitfall.

\subsection{Prior Work}

The single qubit configuration readout, inferring the initial state of a qubit from continuous-in-time noisy measurements, has a long history \cite{gambetta2007protocols,schuster2007circuit, Oliver_2019}. As multiqubit readout methods gain prominence \cite{jeffrey2014fast}, new problems related to such inference are being investigated. For example, \cite{angelatos2021reservoir} show that, within superconducting quantum circuit implementations, while the most widely employed readout setup is that of Quantum Non-Demolition (QND) dispersive measurement for a single qubit, the multi-qubit system has non-trivial cross-talk leading to non-QND effects. These effects ultimately lead to less than perfect inference of the initial state, even when the authors go beyond filter-based methods and use sophisticated machine learning frameworks like reservoir computing \cite{angelatos2021reservoir}.

Previous works have also shown theoretical limits on the performance of qubit readout. However these were shown for specific prediction and recovery schemes, such as continuous measurements with excited qubit decay using filter predictors \cite{gambetta2007protocols}. Here we give a general upper bound for any prediction/estimation scheme. Furthermore, the system considered in \cite{gambetta2007protocols} had intrinsic dissipation. In contrast, we consider systems where the intrinsic dynamics is not dissipative (the best case scenario for information recovery) and still indicate the possibility of information loss.

\subsection{Organization of the paper}

In Section~\ref{sec:setup}, we setup discrete time quantum trajectories, define the two models of weak quantum measurements we focus on in this work: 1) information complete and 2) informationally incomplete with unitary dynamics, and present the Stochastic Master Equations describing their continuous time dynamics in the weak measurement limit. Lastly, we introduce our key measure of information about initial states in measurement records --- mutual information.
In Section~\ref{sec:SME_simulation}, we numerically verify the continuous time dynamics of our measurement models and show the emergence of mutual information plateaus in the weak measurement limit. In Section~\ref{sec:SME_theory}, we study three setting where we can compute mutual information that give us insight into information loss. Here, we also present a perturbative approach to estimate the mutual information plateau value in general settings. Finally in Section~\ref{sec:state_recovery}, we apply our theory to initial state recovery and show useful dynamical timescales for learning in ``physics informed learning".

\section{\label{sec:setup} Definitions and models}

\subsection{\label{subsec:setup} Quantum trajectories}

Quantum trajectories describe how the state of  a quantum system evolves under sequential measurements, conditioned on a specific measurement record.
The quantum system starts in an initial state $\rho_0 \in \mathcal{D}(\mathcal{H})$, i.e. an element of the set of all density operators on the Hilbert space $\mathcal{H}$ of dimension $N$. 
A (generalized) measurement scheme consists of a set of measurement Kraus operators $\{ K_a \}_{a\in O}$ or, equivalently, a POVM system $\{ K^\dagger_aK_a\}_{a\in O}$, where $O$ is the set of outcomes, with $K_a \in \mathbb{C}^{N \times N}$ such that they form a resolution of identity $\sum_a K^\dag_a K_a = I$. A sequence of these measurements on the initial state results in the measurement record $a_{1:T} \coloneqq (a_1,..., a_T)$.

We know that the measurement probability,
\begin{align}
    \nonumber
    \Pr\left[ a_{1:T} | \rho_{0} \right] 
    &=
    \Tr\left[  
    \left( K_{a_T} \cdots K_{a_1} \right)
    \rho_0
    \left( K_{a_T} \cdots K_{a_1} \right)^\dag
    \right]
    \\
    \nonumber
    &=
    \Tr\left[  
    K_{a_1}^\dag \cdots K_{a_T}^\dag 
    K_{a_T} \cdots K_{a_1}
    \rho_0
    \right]
    \\
    &=
    \Tr\left[  
    M_{a_{1:T}}
    \rho_0
    \right]
\end{align}
where $M_{a_{1:T}} \coloneqq \left( K_{a_T} \cdots K_{a_1} \right)^\dag \left( K_{a_T} \cdots K_{a_1} \right)$ are positive semidefinite. These operators also form a resolution of identity $\sum_{ a_{1:T}} M_{a_{1:T}} = I$ since so do the Kraus operators at each step. We could thus think of $\{ M_{a_{1:T}} \}$ as POVM's themselves. This is what we would have in the `measurement only' scenario.

We will also consider having concurrent unitary evolution under a non-trivial Hamiltonian $H$. Consider the measurements are taken in intervals of time $\tau$ and let the unitary operator for evolution over the time interval $\Delta t$ be
\begin{equation}
U:=\exp\left(-iH\Delta t\right),    
\end{equation}
setting $\hbar=1$, then the previous measurement probability becomes
\begin{align}
    \nonumber
    &\Pr\left[ a_{1:T} | \rho_{0} \right] 
    \\
    \nonumber
    &=
    \Tr\left[  
    \left( K_{a_T} U\cdots K_{a_1} U\right)
    \rho_0
    \left( K_{a_T}U \cdots K_{a_1} U\right)^\dag
    \right]
    \\
    \nonumber
    &=
    \Tr\left[  
    U^\dag K_{a_1}^\dag \cdots U^\dag K_{a_T}^\dag 
    K_{a_T}U \cdots K_{a_1}U
    \rho_0
    \right]
    \\
    \label{Eq: Evolution}
    &=
    \Tr\left[  
    \tilde M_{a_{1:T}}
    \rho_0
    \right]
\end{align}
where now 
$\tilde M_{a_{1:T}} \coloneqq \left( K_{a_T}U \cdots K_{a_1} U\right)^\dag \left( K_{a_T}U \cdots K_{a_1}U \right)$. 
This is essentially like defining a new measurement scheme with the set of Kraus operators 
$\tilde K_a=K_aU$ since 
\begin{equation}
    \sum\limits_a \tilde K^\dag_a \tilde K_a 
    =
    U^\dag(\sum\limits_a K^\dag_a K_a)U 
    =
    U^\dag U
    =
    I.
\end{equation} 
As a result 
$ \sum\limits_{a_{1:T} } \tilde M_{a_{1:T}} = I$
and the set $\{ \tilde M_{a_{1:T}} \}$ can also be thought of as POVM's themselves. 

Under sequential measurements, the state evolves stochastically and conditioned on the measurement outcomes as
\begin{gather}
    \rho_t 
    =
    \frac{ 
    ( \tilde{K}_{a_t} \cdots \tilde{K}_{a_1} )
    \rho_0
    ( \tilde{K}_{a_t} \cdots \tilde{K}_{a_1} )^\dag
    }
    {
    \Tr\left[  
    ( \tilde{K}_{a_t} \cdots \tilde{K}_{a_1} )
    \rho_0
    ( \tilde{K}_{a_t} \cdots \tilde{K}_{a_1} )^\dag
    \right]
    }.
\end{gather}

With this setup in place, we now define the models of sequential weak quantum measurements we focus on in this work.

\subsection{Weak measurement models}

We consider sequential weak measurements schemes built upon universal weak measurements of observables \cite{oreshkov2005weak}. Universal weak measurements are a generalized measurement that decompose the projective measurement of any observable $\mathcal{O}$ into a sequence of weak measurements. They have a diffusive behavior such that each measurement output only gives a little information about the measured state while only slightly perturbing it. 

In this work, we will restrict ourselves to qubit systems and operators $\mathcal{O}$ with eigenvalues in the set $ \{1, -1\}$. Then, in a universal weak measurement scheme, each measurement step is a generalized measurement described by the set of Kraus operators $\{K^{\mathcal{O}}_{y}(x) \}$, or equivalently
the POVM $\{ K^{\mathcal{O} \dag}_{y}(x) K^{\mathcal{O}}_{y}(x) \}$, with
\begin{equation} \label{eq:Kraus-O}
    \small
    K^{\mathcal{O}}_{y} (x)
    =
    \sqrt{
    \frac{e^{y x}}{e^{x}+e^{-x}}
    }
    P^{\mathcal{O}}_{+}
    +
    \sqrt{
    \frac{e^{-yx}}{e^{x}+e^{-x}}
    }
    P^{\mathcal{O}}_{-}
\end{equation}
for measurement output $y \in \{1, -1\}$, fixed measurement strength $x \in \mathbb{R}$, and projectors $P^{\mathcal{O}}_{\pm}$ of the observable $\mathcal{O}$ to the eigenspaces corresponding to eigenvalues $\pm 1$, respectively. The extreme limits in $|x|$ are the projective limit with $
|x| \to \infty$ and the weak measurement limit with $|x| \ll 1 $.

\subsubsection{Model I: Informationally complete measurements}

First, we introduce an informationally complete weak measurement model aiming to learn everything about an initial qubit state without any unitary dynamics. In this case, we weakly measure all Pauli operators $\sigma \in \{ X, Y, Z \}$.
This measurement model is described by the set of six Kraus operators $K^{\sigma}_{y} (x)$, which are extensions of the Kraus operators in Eq.~\ref{eq:Kraus-O}, with
\begin{equation}
    K^{\sigma}_{y} (x)
    \coloneqq
    \frac{1}{\sqrt{3}}
    \left[
    \sqrt{
    \frac{e^{y x}}{e^{x}+e^{-x}}
    }
    P^{\sigma}_{+}
    +
    \sqrt{
    \frac{e^{-yx}}{e^{x}+e^{-x}}
    }
    P^{\sigma}_{-}
    \right]
\end{equation}
where $\sigma \in \{ X, Y, Z \}$, the measurement output $y\in \{ 1, -1\}$, fixed measurement strength $x \in \mathbb{R}$, and projectors $P^{\sigma}_{\pm}$ onto the eigenspaces of $\sigma$.

Then, the qubit undergoes sequential measurements which provide information about the qubit state along all three Pauli $\sigma$-axis in the Bloch sphere. These measurements do not particularly preserve any component of the initial state.

\subsubsection{Model II: Informationally incomplete measurements with unitary evolution}

Our second model, based on the work of \cite{tang2020phase}, corresponds to the generic case of measuring an observable with non-commuting Hamiltonian evolution. We consider an informationally incomplete weak measurement, equivalent to measuring a single observable and the universal weak measurement in Eq.~\ref{eq:Kraus-O}, while the system evolves under a non-trivial Hamiltonian. Specifically, 
we weakly measure the spin observable, or Pauli-$Z$ operator, while the qubit undergoes a unitary evolution that rotates the state along a different direction of the measurement. The rotation is implemented by applying a transverse magnetic field to a spin through the Hamiltonian
\begin{equation}
    H = \omega \frac{X}{2},
\end{equation}
setting $\hbar=1$. This measurement model is described by the set of Kraus operators
\begin{align}
    K^{Z}_y(x,\phi)
    \coloneqq
    \left[
    \sqrt{
    \frac{e^{y x}}{e^{x}+e^{-x}}
    }
    \hat{P}^{Z}_{+}
    +
    \sqrt{
    \frac{e^{-yx}}{e^{x}+e^{-x}}
    }
    \hat{P}^{Z}_{-}
    \right]
    e^{-i
    \frac{\phi}{2} X}
\end{align}
where we define the precession angle, the angle of rotation about the $x$-axis on the Bloch sphere, over time $\Delta t$ to be $\phi := \omega\Delta t$, the measurement output $y\in \{ 1, -1\}$, and fixed measurement strength $x \in \mathbb{R}$.

In Eqs.~(6)--(9), $x$ is a \emph{dimensionless} per-step measurement strength controlling the information gained per click, while the physical timescale enters through $\tau$, a characteristic measurement time with units of time. As we will see in the next subsection, the discrete-time models become continuous-time stochastic master equations when the per-step strength is fine and frequent: $\Delta t/\tau = x^2/12$ for Model~I and $\Delta t/\tau = x^2/4$ for Model~II. Equivalently, the weak-measurement continuum limit is taken by sending $x\to 0$ and $\Delta t\to 0$ while holding the measurement rate $\Gamma\sim x^2/\tau$ (and, in Model~II, the dimensionless ratio $\phi/x^2$) fixed.

\subsection{\label{subsec:sme-eta-one}Continuous time limits of models I \& II}

In this subsection, we present the measurement dynamics of our measurement models in the weak measurement limit, namely $|x| \ll 1$. These dynamics are governed by Stochastic Master Equations (SME). The general form in the diffusive limit \cite{barchielli2009quantum,ROUCHON2022252} and the explicit derivation for our models are  given in Appendix ~\ref{sec:WML}. Here, we present the main results.

For model I, we introduce the dimensionless timestep as $\tfrac{\Delta t}{\tau}:=\tfrac{x^2}{12}$ and the SME is
\begin{align} \label{eq:model_1_sme_main_text}
    \nonumber
    d\rho_t
    =&\frac{1}{\tau}
    \sum_{i}
    \left( 
    \sigma_i \rho_t \sigma_i
    -
    \rho_t
    \right)
    \, dt
    \\
    +
    &\frac{1}{\sqrt{\tau}} 
    \sum_{i}
    \left(
    \left[ \sigma_i \rho_t + \rho_t \sigma_i  \right]
    -
    \frac{2}{\tau}
    \Tr\left[ \sigma_i \rho_t \right] \rho_t
    \right)
    \, dW_{ti},
    \\
    dy_{ti} 
    &=
    \frac{2} {\tau}
    \text{Tr}\left[ \sigma_i \rho_t \right]
    \, dt
    +
    \frac{1}{\sqrt{\tau}}
    \, dW_{t i}.
\end{align}
For model II, we introduce the precession angle as $\phi \coloneqq \omega \Delta t$ and the dimensionless time-step $\tfrac{\Delta t}{\tau} := \tfrac{x^2}{4}$ so 
\begin{align}
\tfrac{\phi}{x^2}:=a=\tfrac{\omega\tau}{4}
\label{eq:adefn}
\end{align}
has to be kept fixed as we take the continuous time limit. With this parameterization the SME is,
\begin{align}
    \nonumber
    d \rho_t
    =&
    -i\frac{\omega}{2}
    [X,\rho_t] 
    \, dt
    +
    \frac{1}{\tau}
    \left(Z \rho_t Z  - \rho_t \right) 
    \, dt
    \\
    &+ \frac{1}{\sqrt{\tau}}
    \left( 
    \left[ Z\rho_t + \rho_t Z \right]
    - 2 \text{Tr}\left[ Z \rho_t \right] \rho_t
    \right)
    \, dW_t ,
    \\
    dy_t
    &
    = \frac{2} {\tau}
    \text{Tr}\left[ Z \rho_t \right]
    \, dt 
    + \frac{1}{\sqrt{\tau}}
    \, dW_t.
\end{align}
Model II SME has been studied in detail by \cite{lin2023asymptotic}.
These are stochastic differential equations for dynamics of the state $\rho_t$ and the continuous measurement output $y_t$, where $dt$ is deterministic infinitesimal time and $dW_t$ is the stochastic Wiener differential.

In the SME literature, it is conventional to scale time to make $\tau=1$. We keep the time scale explicit. The SME is essentially Gorini–Kossakowski–Sudarshan–Lindblad (GSKL) equations \cite{gorini1976completely,lindblad1976generators} --- Lindblad equation for most --- with an additional stochastic term from measurement-related fluctuations. Averaging over the stochastic term, essentially tracing over the measurements, gives us back the Lindblad equation. These SMEs correspond to measurement efficiency $\eta=1$ where pure states remain pure under stochastic evolution. We will discuss noisy/inefficient measurements later in Subsection \ref{subsec:low-efficiency-calc}.

\begin{table}[h]
\centering
\resizebox{\textwidth}{!}{
\begin{tabular}{l l l}
\hline
Symbol & Definition & Role \\
\hline
$x$ & dimensionless measurement strength & per-step information gain \\
$\tau$ & measurement timescale (units of time) & sets the continuum-time clock \\
$\Delta t$ & time per measurement & $\Delta t/\tau = x^2/12$ (Model I), $x^2/4$ (Model II) \\
$\omega$ & precession frequency (Model II) & from $H=\omega X/2$ \\
$\phi$ & precession angle per step (Model II) & $\phi := \omega\Delta t$ \\
$a$ & scaling-invariant ratio (Model II) & $a = \phi/x^2 = \omega\tau/4$ \\
$\alpha$ & quality factor (Model II; \S\,3.3 onward) & $\alpha := \omega\tau/2 = 2a$; critical damping at $\alpha=1/2$ \\
$\eta$ & measurement efficiency & $\eta=1$: pure trajectories; $\eta<1$: noisy \\
\hline
\end{tabular}
}
\caption{\label{tab:params}Dimensionless and dimensional parameters appearing in Models~I and~II and in their continuum-time SMEs.}
\end{table}

Now that we defined the discrete time models and their continuous time versions in the weak measurement limit,  $|x|\ll 1$ , we are ready to explore the relation between measurement records and the initial state. For that we need to set up a probabilistic relation between initial state and measurements in a Bayesian manner and introduce some Information Theoretic notions.

\subsection{A word on notation}

Following the standard statistics convention, capital letters denote random variables and lowercase letters their particular realizations: $P_0$ (with realization $\rho_0$) is the random variable for the initial state, and $A_{1:T}$ (with realization $a_{1:T}$) is the random variable for the measurement record. To keep formulas uncluttered we sometimes write $\Pr[\rho_0|a_{1:T}]$, $\Pr[a_{1:T}|\rho_0]$, and $\Pr[\rho_0]$ in place of the full $\Pr[P_0=\rho_0|A_{1:T}=a_{1:T}]$, etc.

\subsection{The prior and the posterior}

A statistical approach to learning the initial state given the measurement record requires the posterior, in our case the conditional probability of the initial state given the measurement record, $\Pr[\rho_0 | a_{1:T}]$ and  a prior over initial states $\Pr[\rho_0]$.  Our posterior is
\begin{align}
    \nonumber
    \Pr\left[ \rho_{0}|a_{1:T}  \right] 
    &=
    \frac{\Pr\left[ a_{1:T} | \rho_{0} \right] \Pr[\rho_0]}{\sum_{\rho\in \mathcal{D}}\Pr\left[ a_{1:T} | \rho \right] \Pr[\rho]}
    \\
     &=
    \frac{    \Tr\left[  
     M_{a_{1:T}}
     \rho_0
    \right] \Pr[\rho_0]}
      { \sum_{\rho \in \mathcal{D}}
     \Tr\left[  
     M_{a_{1:T}}
     \rho
     \right] \Pr[\rho]}.
\end{align}
In this study, we will work with very informative priors with support on a discrete set $\mathcal{D}$ of choices of the initial density matrix. Since we know a lot about the initial states for the qubit readout problem:  very often our initial state prior is supported on the spin up $|\up\rangle\langle\up|$ and spin down $|\down\rangle\langle\down|$ states.

\subsection{Mutual information} 

Once we have the posterior $\Pr\left[ \rho_{0}|a_{1:T}  \right]$, we could have directly moved on to the task of predicting the initial state based on some criterion like accuracy. We will discuss this approach in Section \ref{sec:state_recovery}. However, we first want to deal with fundamental physics and limitations of information recovery rather than those of any particular procedure. 
To this end, we study the information physics of quantum trajectories based on weak measurements and consider what amount of information about the initial state is contained in the measurement record to begin with.
One formal characterization of this information is via the mutual information:
\begin{align}
    \label{eq:mutual-info}
    I(P_0,A_{1:T}) 
    = 
    H(P_0)+H(A_{1:T})-H(P_0,A_{1:T}),
\end{align}
where $P_0$ is the random variable taking values over $\rho_0\in \mathcal{D}$ and $A_{1:T}$ is the random variable taking values over the measurement record $a_{1:T} \in O^T$, where $O$ is a discrete set of measurement outcomes. $H(X)$ is the entropy of the random variable $X$, computed in bits:
\begin{equation}
    H(X) = \sum\limits_{x}\Pr[X=x]\log_2\frac{1}{\Pr[X=x]}
\end{equation} 
for a discrete distribution. Note that $I(P_0,A_{1:T})\le I(P_0,A_{1:T+1})$ because of the data processing inequality. Therefore, mutual information is monotonically non-decreasing in $T$. This property is not always shared by some performance measures, like the accuracy of prediction.

As $T$ becomes large, the direct approach to computing mutual information by using Eq.~\ref{eq:mutual-info} becomes untenable. Owing to the number of possible measurement records growing exponentially with $T$, quantities like $H(A_{1:T})$ and $H(P_0,A_{1:T})$ become intractable to compute. However, rewriting the mutual information as the difference between the entropy of the initial state and the conditional entropy of the initial state, given the measurement record, namely,
\begin{align}
\label{eq:mutual-info-alt}
    I(P_0,A_{1:T})
    &=
    H(P_0)-H(P_0|A_{1:T}),
\end{align}
accurate evaluation of the mutual information becomes tractable by leveraging sample estimates with concentration bounds on the (bounded) conditional entropy.
For further discussion of numerically accurate estimation of the mutual information, see section~\ref{sec:mutual-info-estimation} in the Appendix.

\begin{figure}[t!]
    \centering  
    \begin{overpic}[width=0.24\textwidth]{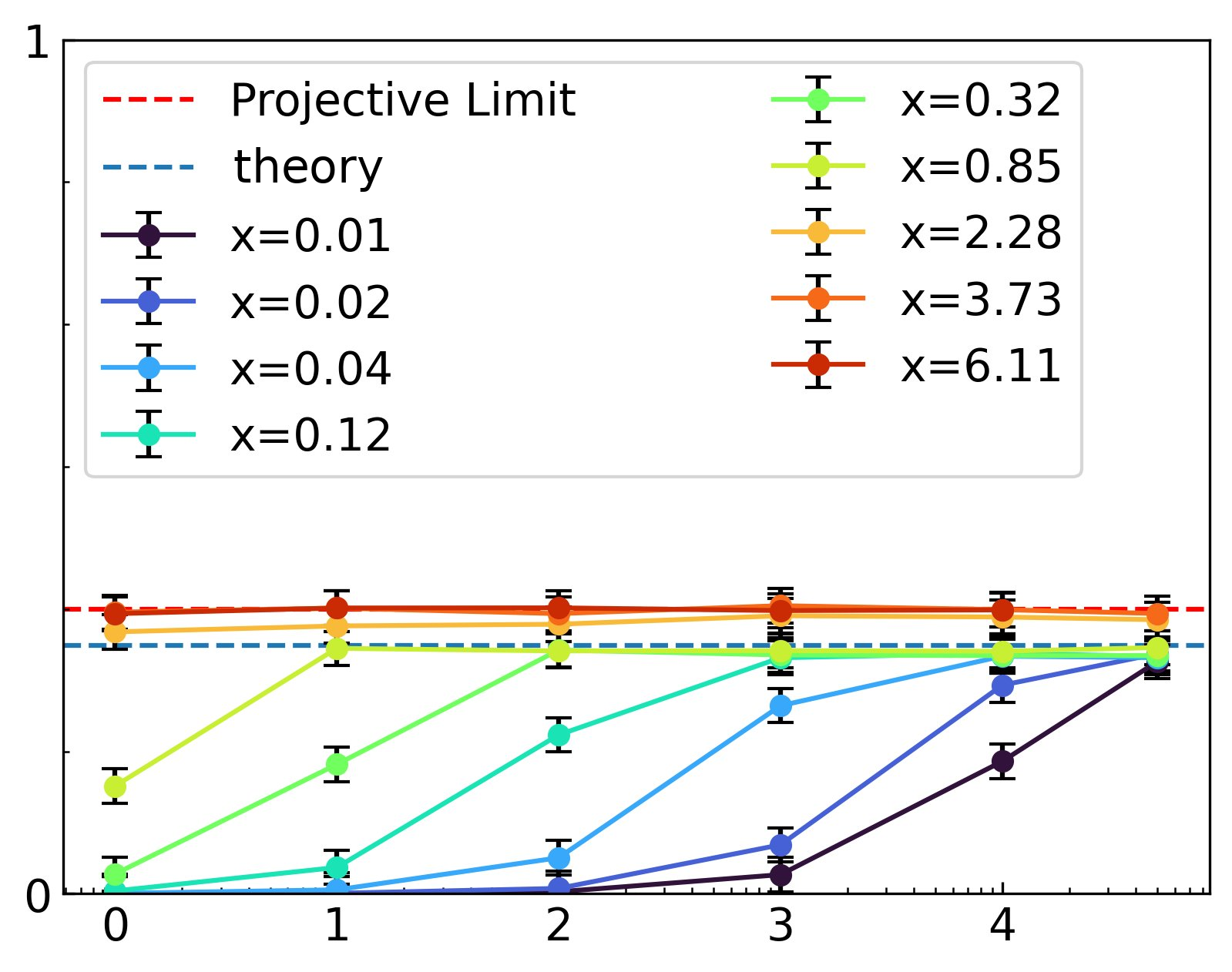}
        \put(40,45){\textbf{(a)}}
        \put(84,8){$\ell T$}
    \end{overpic}
    \begin{overpic}[width=0.24\textwidth]{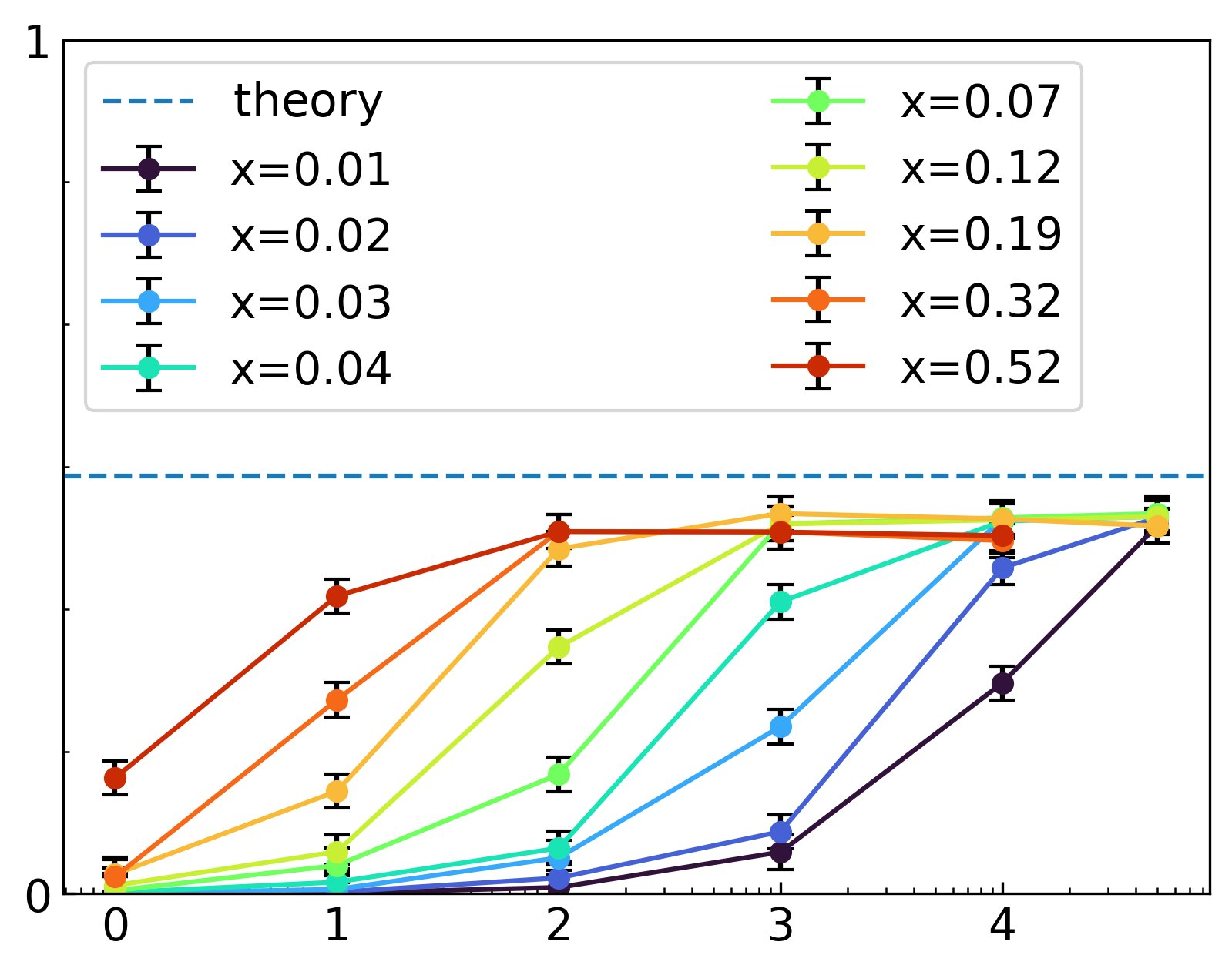}
        \put(40,55){\textbf{(b)}}
        \put(84,8){$\ell T$}
    \end{overpic}
    \begin{overpic}[width=0.24\textwidth]{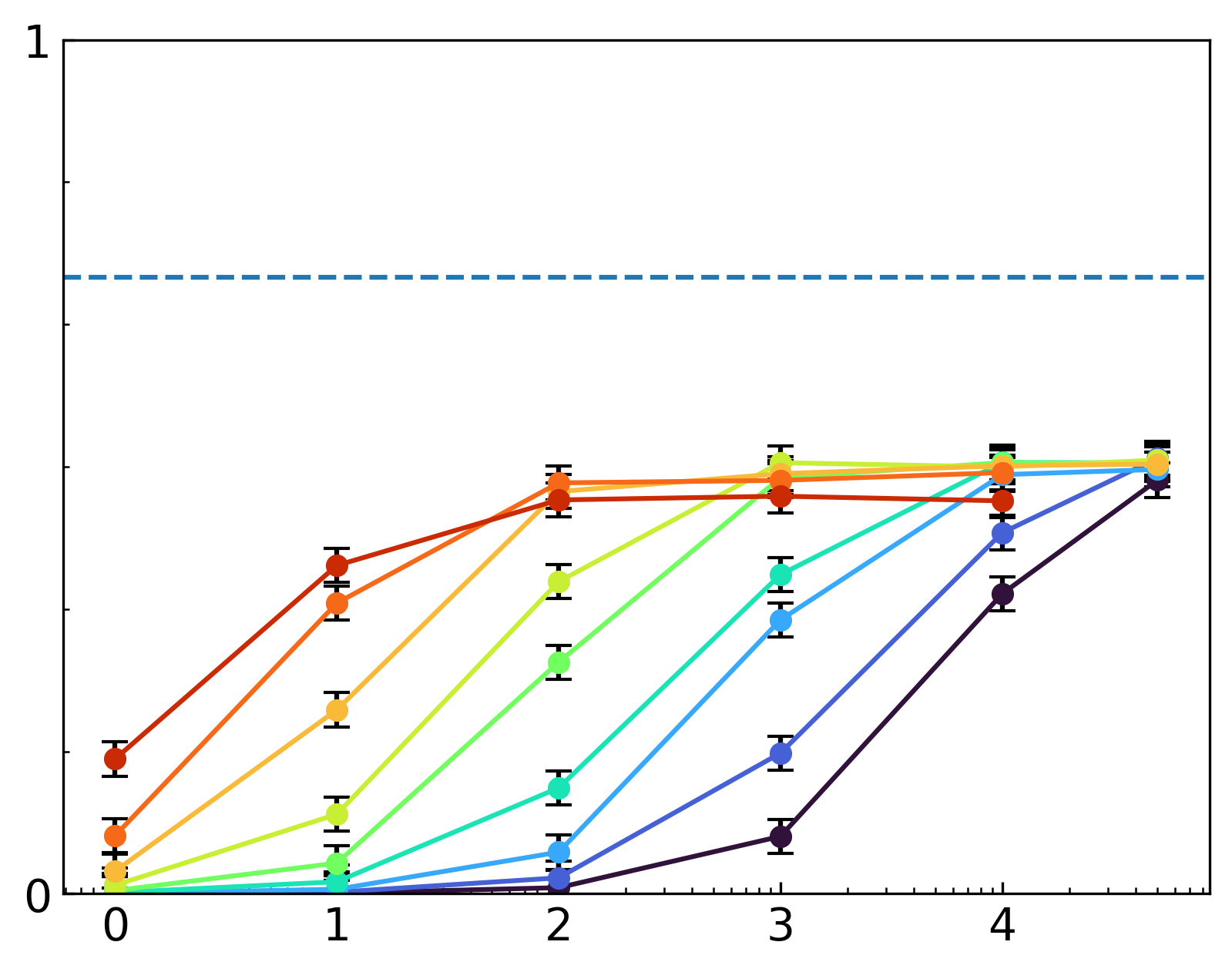}
        \put(5,65){\textbf{(c)}}
        \put(84,8){$\ell T$}
    \end{overpic}
    \begin{overpic}[width=0.24\textwidth]{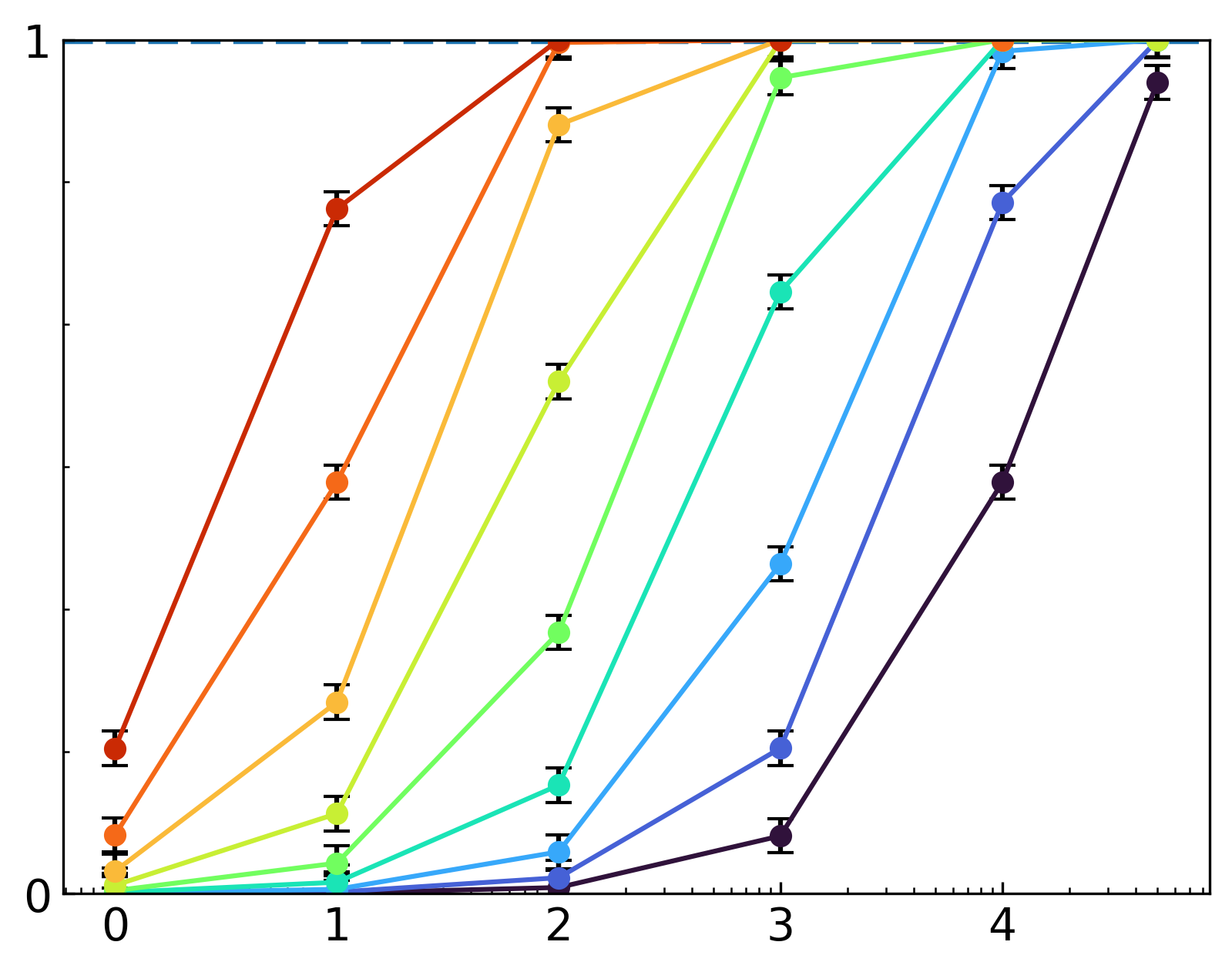}
        \put(5,65){\textbf{(d)}}
        \put(84,8){$\ell T$}
    \end{overpic}
   \\
    \begin{overpic}[width=0.24\textwidth]{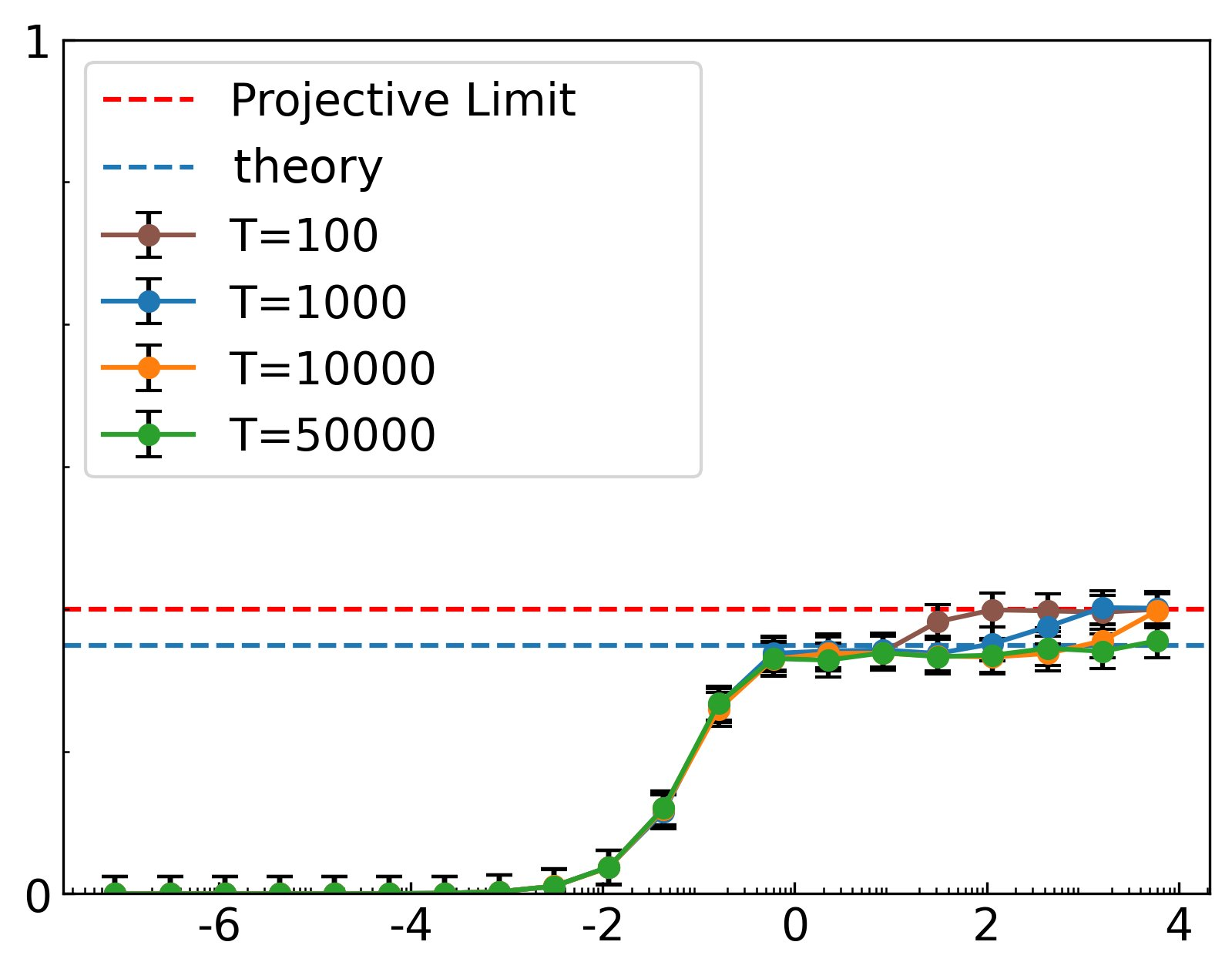}
        \put(78,65){\textbf{(e)}}
        \put(82,10){$\ell T_x$}
    \end{overpic}
    \begin{overpic}[width=0.24\textwidth]{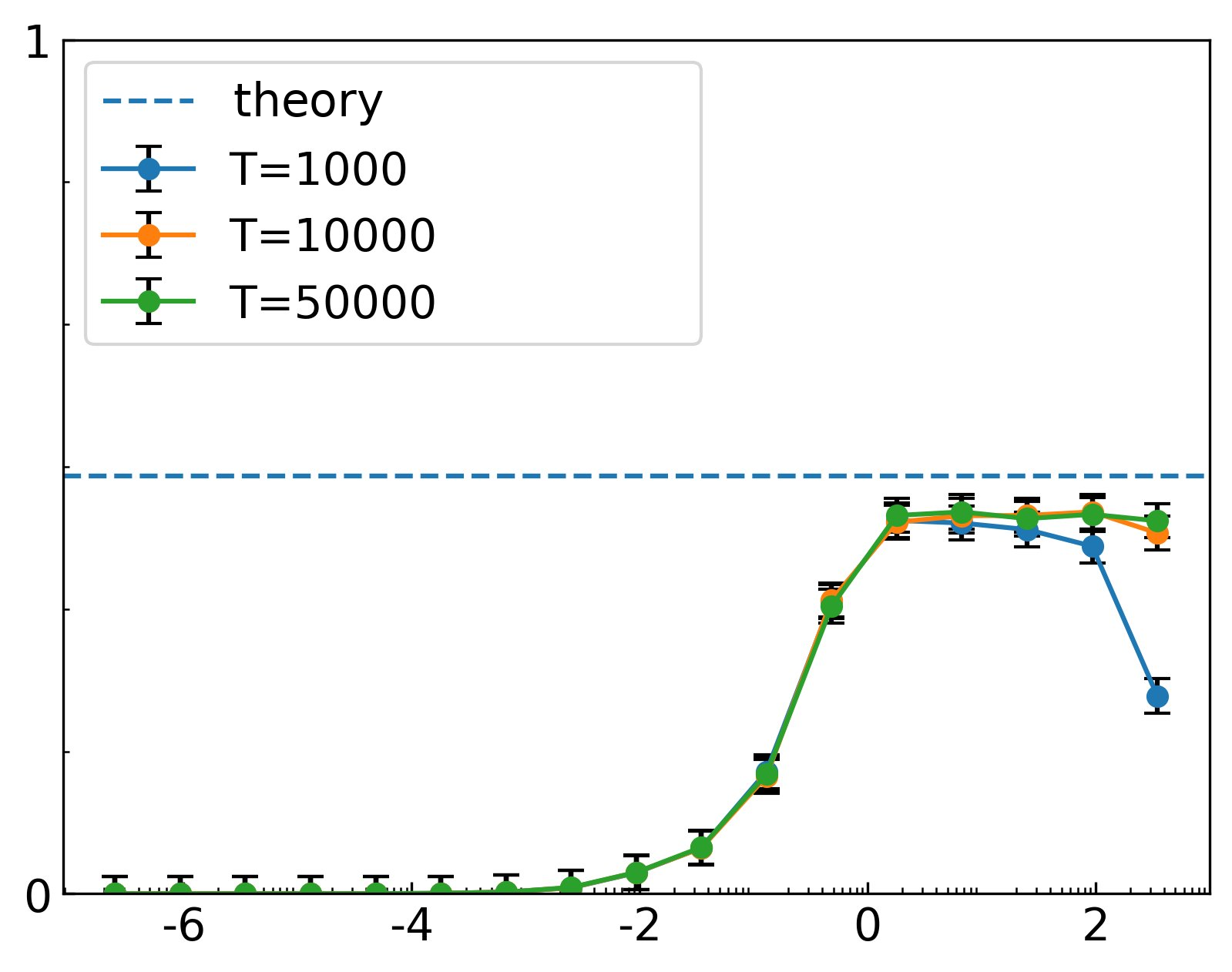}
        \put(78,65){\textbf{(f)}}
        \put(82,10){$\ell T_x$}
    \end{overpic}
    \begin{overpic}[width=0.24\textwidth]{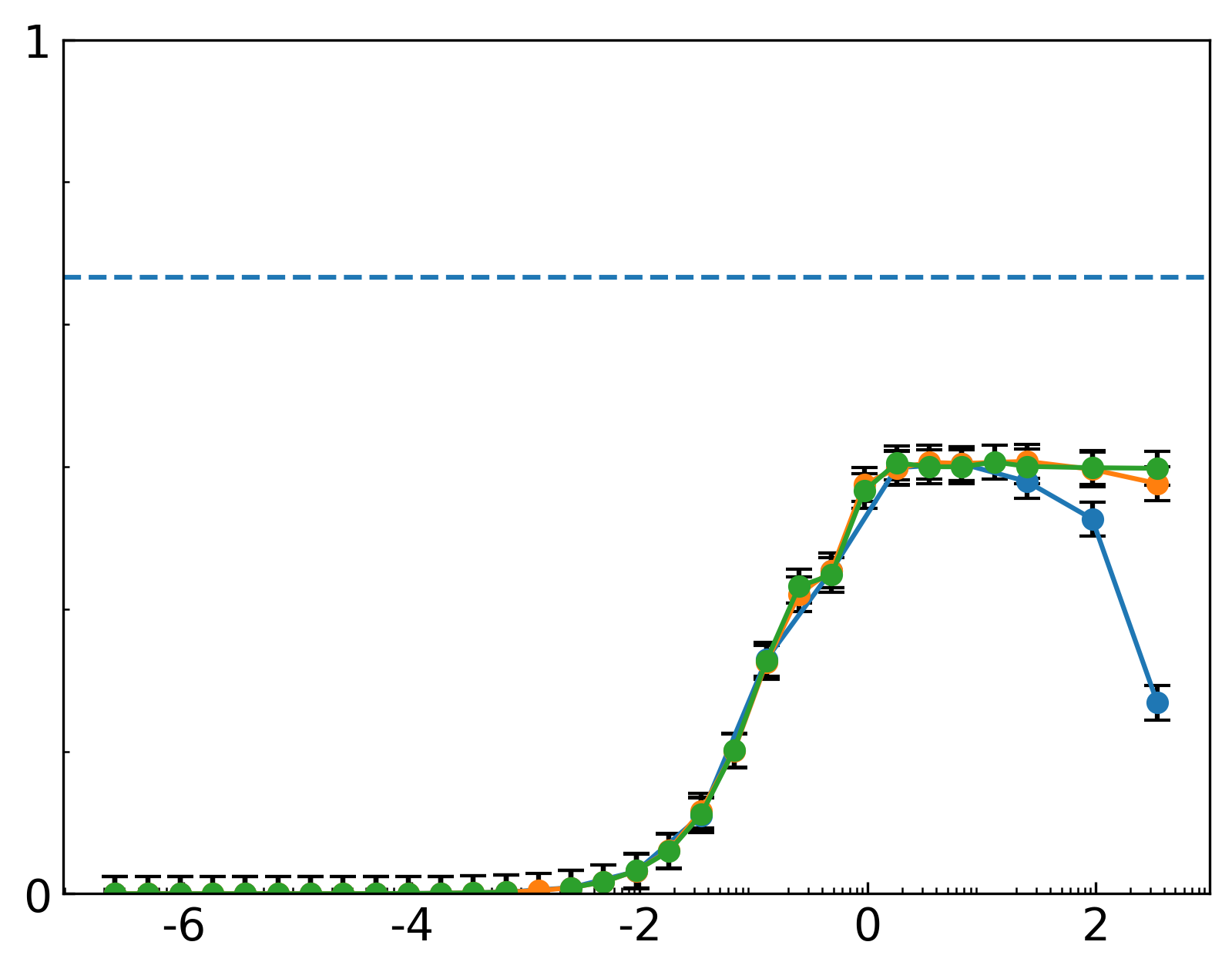}
        \put(78,65){\textbf{(g)}}
        \put(82,10){$\ell T_x$}
    \end{overpic}
    \begin{overpic}[width=0.24\textwidth]{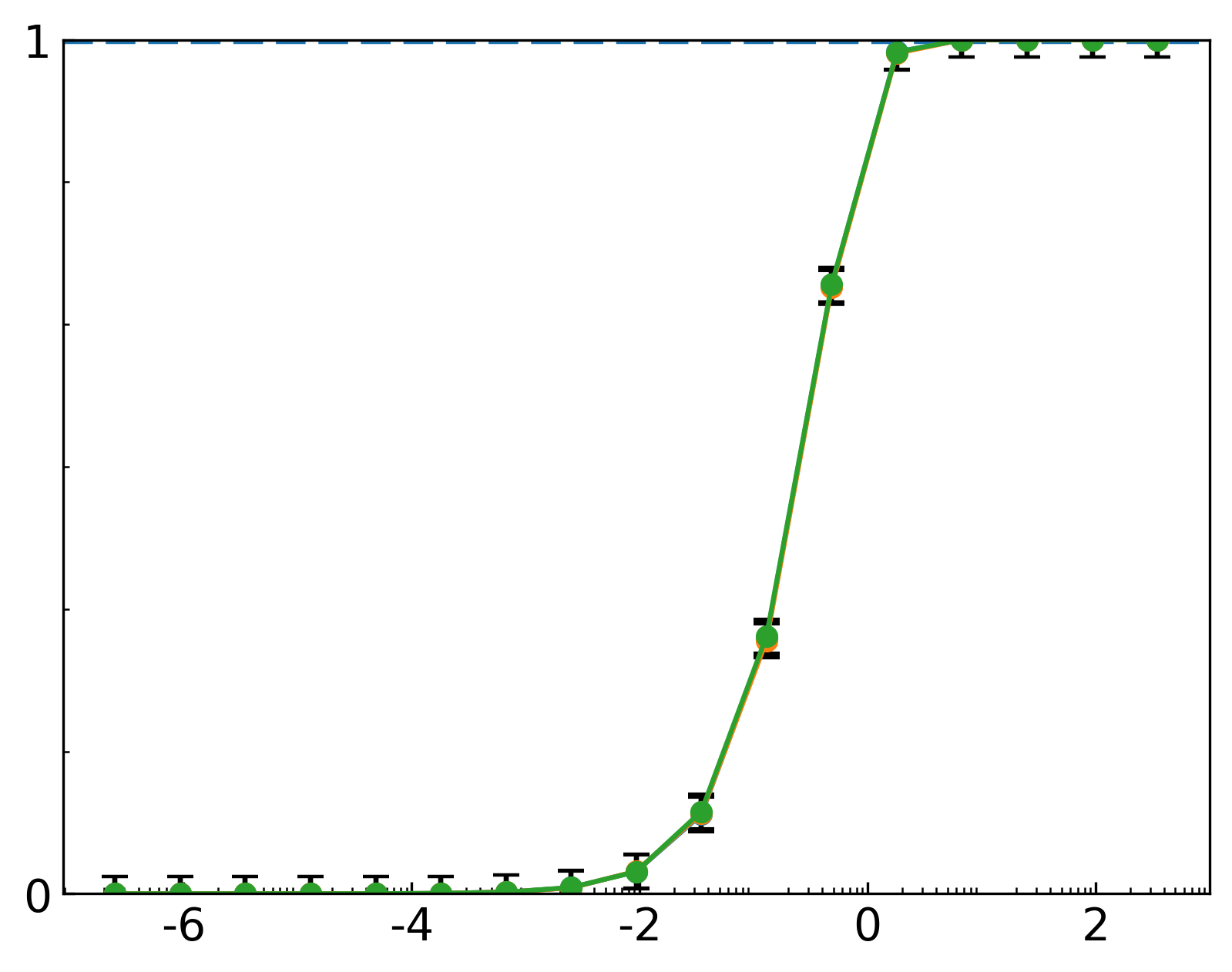}
        \put(78,65){\textbf{(h)}}
         \put(82,10){$\ell T_x$}
    \end{overpic}
    \caption{
Numerical estimates of mutual information $I(S; A_{1:T})$ from discrete sequential measurements for various measurement strengths $x$ and record lengths $T$ for both Models (note: we use $\ell T$ notation to denote $\log_{10} T$ and similarly for $T_x\equiv x^2 T$)
\\Panels $(a, e)$ are for model I, while 
 panels $(b,c,d,f,g,h)$ are for model II
    with varied strength of scaled magnetic field values $\phi/x^2=10, 1, 0$ in $(b,f),(c,g),(d,h)$, respectively
    (Eq.~\ref{eq:adefn}). We use bare (discrete) time $T$ in the top row, to contrast against scaling collapse (of the same data) in the bottom row, where $x^2T$ is used instead. Panels $(b,c,d)$ follow the same labeling convention as detailed in panel $(b)$. Similarly, panels $(f,g,h)$ follow the same labeling convention as detailed in panel $(f)$. 
    \\
    Dashed blue traces are  obtained by extrapolating analytic low efficiency calculation to $T\to \infty, \eta=1$ -- see  Secs. \ref{subsec:low-efficiency-calc}, \ref{sec:inefficient_mearurement}, \ref{sec:MI_in_weak_noise_limit} and Fig. \ref{fig:smalleffextrap}.}
    \label{fig:MI_numerical_simulations}
\end{figure}

\section{\label{sec:SME_simulation} Information Content of the Measurement Record  - Simulation}

In this section, we address fundamental aspects of information loss in our measurement schemes and formally quantify it using mutual information.
Here, we look at the mutual information $I(P_0,A_{1:T})$, as we vary the record length $T$ and strength $x$ of our measurement schemes on random initial states. 
We focus on two key effects:
\begin{enumerate}
\item The saturation of mutual information with increasing $T$,
\item The emergence of scaling with small $x$ and large $T$, consistent with the continuum limit.
\end{enumerate}

We will restrict our analysis to the qubit setting. 
For initial states, we use the (pure) up and down states (the Pauli-$Z$ basis states), thus referring to $P_0$ as $S$.
These states represent the optimal setting for information extraction: two orthogonal states, for which the Holevo bound guarantees a maximum of $1$-bit of information from measurements.
We also numerically estimate mutual information and apply the Hoeffding concentration inequality \cite{hoeffding1994probability}, which guarantees with $99\%$ probability (or failure probability $1\%$) that our mutual information estimates are accurate up to $2\%$ error, corresponding to estimates with (at least) 6623 samples.  (see Appendix section \ref{sec:mutual-info-estimation} for further details).
For reproducibility, all simulations use a fixed random-number-generator seed; the seed values, together with the full set of simulation parameters and the source code, are reported in the Data Availability statement. To avoid the exponential underflow of the unnormalized Kraus-matrix product at large $T$, we propagate \emph{normalized} conditional states, rescaling the trace to unity at every step; this keeps the matrix norm $O(1)$ and the Born-rule weight of each branch accurately tracked.

\subsection{Mutual information and measurement record length}

In Fig.~\ref{fig:MI_numerical_simulations}, we plot $I(S,A_{1:T})$ over the length of the record $T$ with curves at various measurement strengths $x$, for both model I and model II. In model II, the informationally incomplete with unitary dynamics measurement scheme, we also vary $\phi$, the field-induced rotation angle per step according to Eq.~\ref{eq:adefn}. 

For both models, in (a-d) in Fig.~\ref{fig:MI_numerical_simulations}, the numerical simulations show mutual information monotonically increases with $T$, as expected. 
Furthermore, the mutual information is generally less then $1$-bit, which is the optimal value over all measurement schemes in our setting. 
In Model I, even in the projective limit, $|x| \to \infty$, it is $\tfrac{1}{3}$ of $1$-bit. Thus, some information is inevitably lost in our measurement schemes.

It is worth noting that in our measurement schemes, there is also dependence on initial states for optimal information extraction using the weak vs. projective limits.
For initial states, polarized in the $z$ direction, increasing $x$ typically gets us more information. 
In model II, however, for certain initial states (say, polarized in the Pauli-$x$ direction), and for a fixed $\phi \neq 0$, one can have nonmonotonicity of mutual information information as a function of $x$. Later, we also see this phenomena in the noisy/inefficient measurements setting.
In those cases, we can readout the initial state configurations with higher fidelity in a weak measurement protocol than in its projective measurement limit (see Fig.~\ref{fig:nonmon} and (d) in Fig.~\ref{fig:Noisy_MI_plots}).

\subsection{Emergence of scaling functions in the very weak measurement limit with $T$ large}
Next, we move on to the mutual information scaling functions.
In subsection~\ref{subsec:sme-eta-one}, we saw that the continuum limit description emerges as we scale 
\begin{gather}
    T\sim x^{-2}, \, \phi\sim x^2.
\end{gather}
It is not guaranteed that the final outputs (like mutual information) have the same scaling. However, this SME being a one-dimensional stochastic problem, it is reasonable to assume that the naive scaling holds.

In (e-h) in Fig.~\ref{fig:MI_numerical_simulations}, we exploit the existence of the mutual information scaling function with $x^2 T$ as the scaling variable to see data collapse in the weak measurement limit.
For model II, we need to keep $a = \frac{\phi}{x^2}$ fixed (see Eq.~\ref{eq:adefn}), to get different scaling curves.
Notably, these scaling functions reveal the existence of previously unknown mutual information plateaus in the weak measurement limit.
These plateaus are below the projective limit, and correspond to plateaus as physical time $t$ ($\propto x^2T$) goes to $\infty$ in the continuum limit. Outside the weak measurement limit, the curved in (e-h) peal off the plateau. The collapse of scaled data in Fig.~\ref{fig:MI_numerical_simulations} suggests that, in the scaling limit, $I(S,A_{1:T})$ for parameters $x,\phi$, which we denote as $I(S,A_{1:T}|x,\phi)$ is well approximated by a scaling function of the form $f(x^2T,\frac{\phi}{x^2})$.
If $\lim_{b\to\infty} f(b,a) < \log{2}$, the continuous-time dynamics loses initial-state information irretrievably.

Our numerical simulations show that, in the sequential measurement setting, generically, conditioned on the sequential measurement outcomes, some information of the initial state is lost.  
In the next section, we want to provide some analytic insights into this phenomenon, of the maximum mutual information being less than $1$-bit, in our measurement schemes.

\subsection{Beyond scaling -- multiple plateaux in Model II}

\begin{figure}
\centering    \begin{overpic}[width=0.45\textwidth]{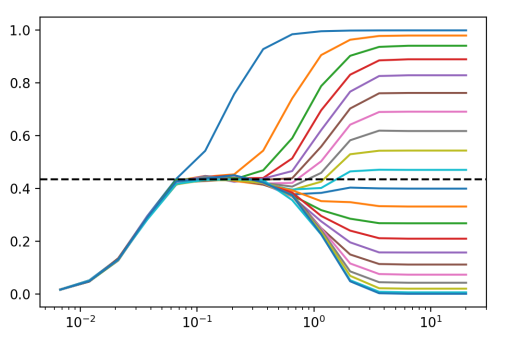}
        \put(10,57){\textbf{(a)}}
        \put(62,8){$x$}
    \end{overpic}
    \begin{overpic}[width=0.45\textwidth]{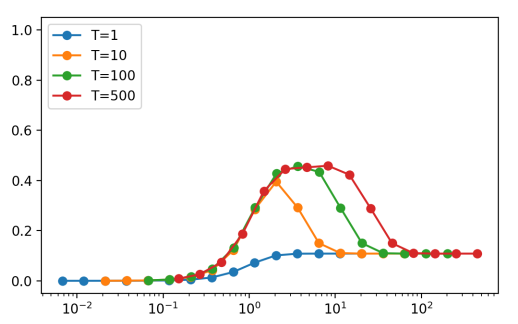}
        \put(10,36){\textbf{(b)}}
        \put(84,7){$x\sqrt{T}$}
    \end{overpic}
        \caption{
Nonmonotonic dependence of mutual information (MI) $I(S; A_{1:T})$ on measurement ($x$) and precession field ($\phi$) strengths outside of the scaling limit in Model II. 
        Panel (a) shows evolution of MI at $T=50$  over a range $\phi\in [0,\frac{\pi}{2})$ with steps of $\Delta\phi = 0.075$. The lowest field corresponds to the top curve, and as the field increases, the curves for a given $x$ monotonically decrease. Panel (b) shows MI for $\phi=\frac{\pi}{8}$ for different several $T$'s. All of the curves have been obtained from the exact analytical value when $T$ is small, or with $N=3750$ samples, that following Hoeffding's inequality provides an accuracy around $1\%$.  Black dashed line is a simple guide to the eye.
        }     
        \label{fig:nonmon}
\end{figure}

We now illustrate interesting and potentially useful regimes beyond the scaling limit explored above. Specifically, in model II we may consider keeping $\phi$ finite while taking the measurement rate to zero. Here, we expect the orientations of relative polarization of the initial state, measurement axis, and the direction of the external field to produce non-monotonic behavior. For example, weaker measurements can (and do) extract more information, see Fig.~\ref{fig:nonmon}. In these cases, weaker measurements may be advantageous for information extraction. Additionally, it would be interesting to explore extensions of our formulation to multi-qubit readouts.

\section{\label{sec:SME_theory} Information Content of the Measurement Record  - Analytical Explorations}

At this point, we do not have a rigorous way of establishing the conditions under which such information loss in measurement records occurs. However, we can still provide some analytic evidence and understanding of this phenomenon by looking at three special cases that add to our insight into initial state information loss. In 1) the case of nontrivial invariant subspace in the measurement scheme, we analyze the setting with no information loss, in 2) information in the $T$-th measurement, we look at exponentially decaying mutual information contributions from late time measurements, and most importantly, in 3) continuous time low efficiency limit, we develop a perturbative approach to estimate the mutual information plateau of the weak measurement limit in general settings.

\subsection{\label{subsec:spl-case}The special cases with a non-trivial invariant subspace}

Let us first start by focusing on a case where there is no asymptotic information loss.
Specifically, the case of incomplete measurements with no unitary evolution, or $\phi=0$, in model II with initial spin up and down states. 
In this case, all that matters for the output is the number $N_{+}$ or $N_{-}$ of $+$ and $-$ calls, corresponding to the two measurement outcomes in this model, respectively since weak measurement along $z$-axis for up or down states do not affect them at all. 
Then the conditional variables,
\begin{gather}
\begin{gathered}
    (N_+|S=\uparrow) \sim B(T,p(x))
    \\
    (N_+|S=\downarrow) \sim B(T,1-p(x))
\end{gathered}
\end{gather}
are both distributed binomially, allowing for an estimation of the mutual information.

Now consider the behavior of the mutual information in the $T\to\infty$ limit. Since for large $T$, the two conditional distributions have very little overlap, $Pr[S|N_+]$ tends to deterministic distributions for most $N_+$. We can, thus, reliably extract all the information about the initial state from $N_+$, which depends only on $A_{1:T}$, in the large $T$ limit. 

In this simple case, we have perfect asymptotic retrieval of information because the Kraus operators $K_a$s of this measurement scheme commute.
In general, if $K_a$s have an invariant subspace of dimension $2$ or greater and are simultaneaously diagonalizable, we can come up with a scheme of more than one initial state, which are not affected by the Kraus operators. and provided each measurement provides some information, repeated measurements can retrieve the whole information. This essentially the point of  \cite{oreshkov2005weak}, in a more general setting.

This observation suggests a more general conjecture: Perfect asymptotic recovery is possible only when the Kraus algebra is reducible and contains at least two distinguishable invariant sectors. By distinguishable, we mean that the probability laws of measurement records differ across these sectors.  

Since generic Kraus families generate the full matrix algebra, such perfect recovery is a fine-tuned, nongeneric phenomenon. We gave an example of perfect asymptotic recovery above. We do not have a rigorous/computable demonstration of information loss and have relied on numerical experiments for evidence. To provide further such evidence, we start by considering the mutual information between the initial state and the $T$-th measurement.

\subsection{Information in $T$-th measurement}

In mutual information, we see that in the generic case, our ability to infer the initial state from the sequential measurements remains limited, even when $T$ goes to infinity. 
If this mutual information was high, we have not forgotten the initial state. The converse is not quite true. The chain rule of mutual information \cite{Cover2006} gives
\begin{align}
    I(P_0,A_{1:T})&=\sum_{j=1}^TI(P_0,A_j|A_{j+1:T})\nonumber\\
    &=I(P_0,A_1|A_{2:T})+I(P_0,A_2|A_{3:T})+\cdots\nonumber\\
    &+I(P_0,A_{T-1}|A_T)+I(P_0,A_T).
\end{align}
Hence, $I(P_0,A_T)$ could be small but $I(P_0,A_{1:T})$ could be high. However, $I(P_0,A_T)$ is easy to calculate, so let us look at its behavior.
To understand when very late measurement adds to the knowledge of the initial state, we calculate the joint distribution of $P_0$ and $A_T$ here as,
\begin{align}
    \nonumber
    \Pr[\rho_0, a_T]
    &=
    \sum\limits_{a_{1:T-1}}
    \Pr{\left[ a_{1:T}|\rho_0 \right]}
    \Pr{\left[ \rho_0 \right]}
    \\
    &=
    \Tr{\left[ \E_{a_T}(\E^{T-1}(\rho_0)) \right]}
    \Pr\left[ \rho_0 \right]
\end{align}
where we define the superoperators $\{\E_a\}$ and $\E$ as follows,
\begin{equation}
    \E_a(\rho)
    :=
    K_a\rho K_a^\dagger,\, \mathrm{and},\, 
    \E(\rho)
    :=
    \sum\limits_{a}\E_a(\rho).
\end{equation}
Where $\E$ is a (probability conserving) quantum channel. Now in our qubit setting, we can represent these superoperators as $4\times4$ matrices $\{E_a\}$ and $E$ defined by $\E_a(\rho)\to E_a\p$ $\E(\rho)\to E\p$, see Appendix~\ref{sec:channel_transfer_matrix} for more details. since $\E$ is trace preserving, $E$ has at least one eigenvector $\p_*$ corresponding to the eigenvalue $1$, and the rest of the eigenvalues lie in the unit disc in the complex plane. If all the rest of the eignevalues have norm strictly less than $1$, 
we define the quantity $\xi$, which is like a boundary correlation length, by
\begin{equation}
    e^{-\frac{1}{\xi}}=\max\limits_{\lambda\in Eig(E),\lambda\neq 1} |\lambda|
\end{equation}

With this setup, in Appendix~\ref{sec:channel_transfer_matrix}, we show that, for $T$ large, the initial spin states $P_0=S$ and $A_T$ are nearly independent, up to small corrections. Thus we have $I(S,A_T)=O( te^{-\frac{2T}{\xi}})$, in general.  In some cases, 
the upper bounds would be sharper: $I(S,A_T)=O( e^{-\frac{2T}{\xi}})$. In either case, the mutual information between the initial state and the $T$-th measurement is going to zero exponentially as $T$ tends to infinity. Thus, for later observations to be of value, we must have nontrivial multiplicity of the eigenvalue $1$, which we could think of as $\xi=\infty$. This is indeed the case for model II, with $\phi=0$. The general condition mentioned in Subsection~\ref{subsec:spl-case}, also leads to $\xi=\infty$. In Section~\ref{sec:channel_transfer_matrix}, we explicitly compute $\xi$ for models I and II and show that, with the exception of the case model II with $\phi=0$, for small $x$, $\xi\sim x^{-2}$ which validate our scaling functions.

As we mentioned, $I(S,A_T)=O(\exp(-cT))$ for large $T$, in general, does not imply that $I(S,A_{1:T})-I(S,A_{1:T-1})$ also goes to zero in a similar manner. That said, we conjecture that, in our problem, $\xi$ being finite implies imperfect recovery of information about $S$ from arbitrarily long sequence of measurements. 

Note that, in continuous time version of the problem, the averaging of all the measurement related noise for the intermediate times lead to the Lindblad equation. Our computation of $\xi$ is thus related to figuring out the convergence time of the Lindblad equation. In general, the SME time scale may not quite be the Lindblad time scale. We may glean the accuracy of the simple continuum approximation $\xi\propto x^{-2}$ in actual discrete time simulations in Fig. \ref{fig:MI_numerical_simulations}, where apparent data collapse (upper vs. lower rows) is excellent, and also seems to agree with explicit Lindblad (low efficiency) results for $\xi(x)$ in \ref{sec:channel_transfer_matrix}.

\begin{figure*}[t!]
    \centering
    \begin{overpic}[width=0.24\textwidth]{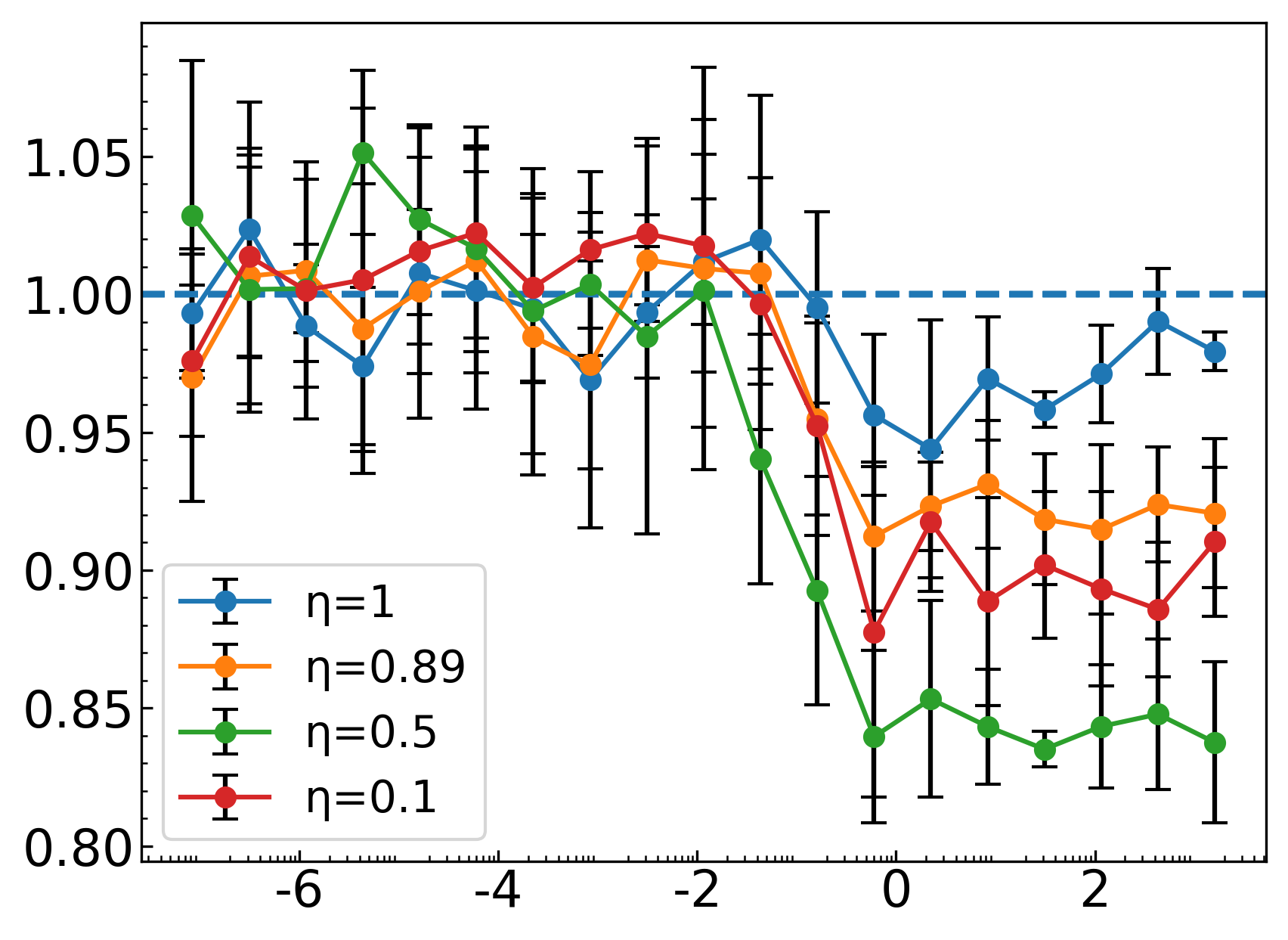}
        \put(78,62){\textbf{(a)}}
       \put(48,10){$\ell T_x$}
    \end{overpic}
    \begin{overpic}[width=0.24\textwidth]{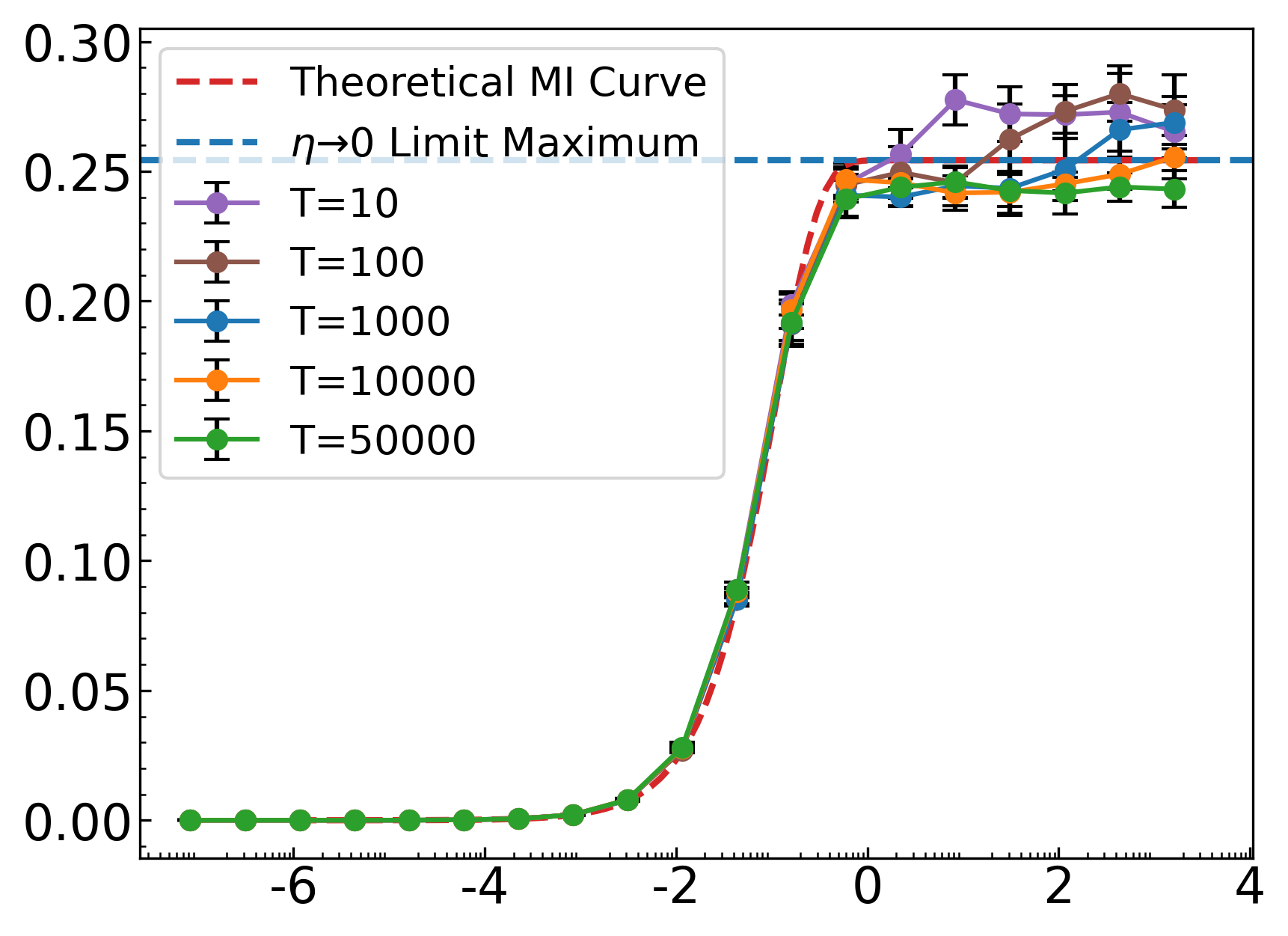}
        \put(78,27){\textbf{(b)}}
        \put(10,20){$\eta=0.89$}
        \put(74,10){$\ell T_x$}
    \end{overpic}
    \begin{overpic}[width=0.24\textwidth]{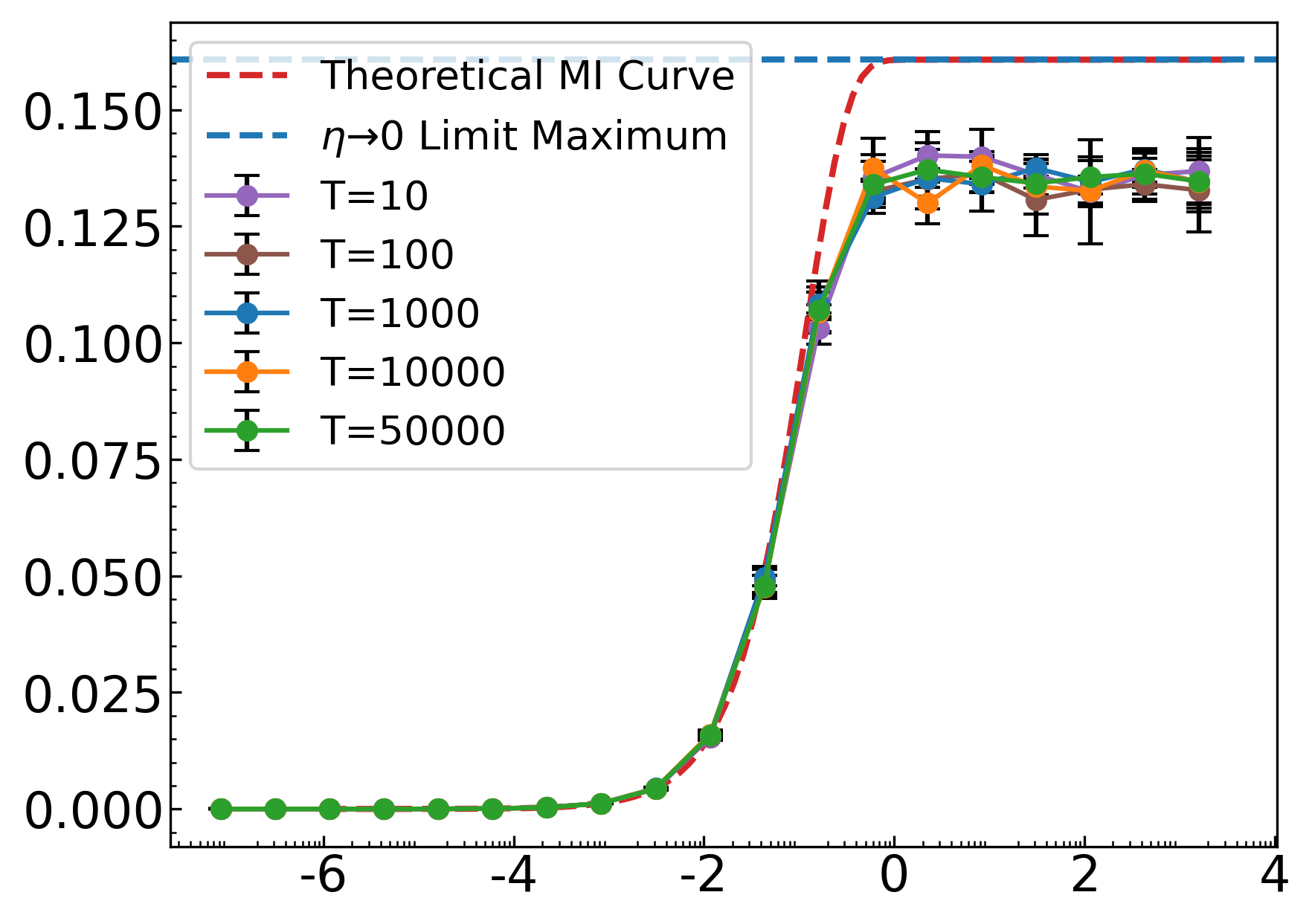}
        \put(78,27){\textbf{(c)}}
        \put(10,20){$\eta=0.50$}
        \put(74,10){$\ell T_x$}
    \end{overpic}
    \begin{overpic}[width=0.24\textwidth]{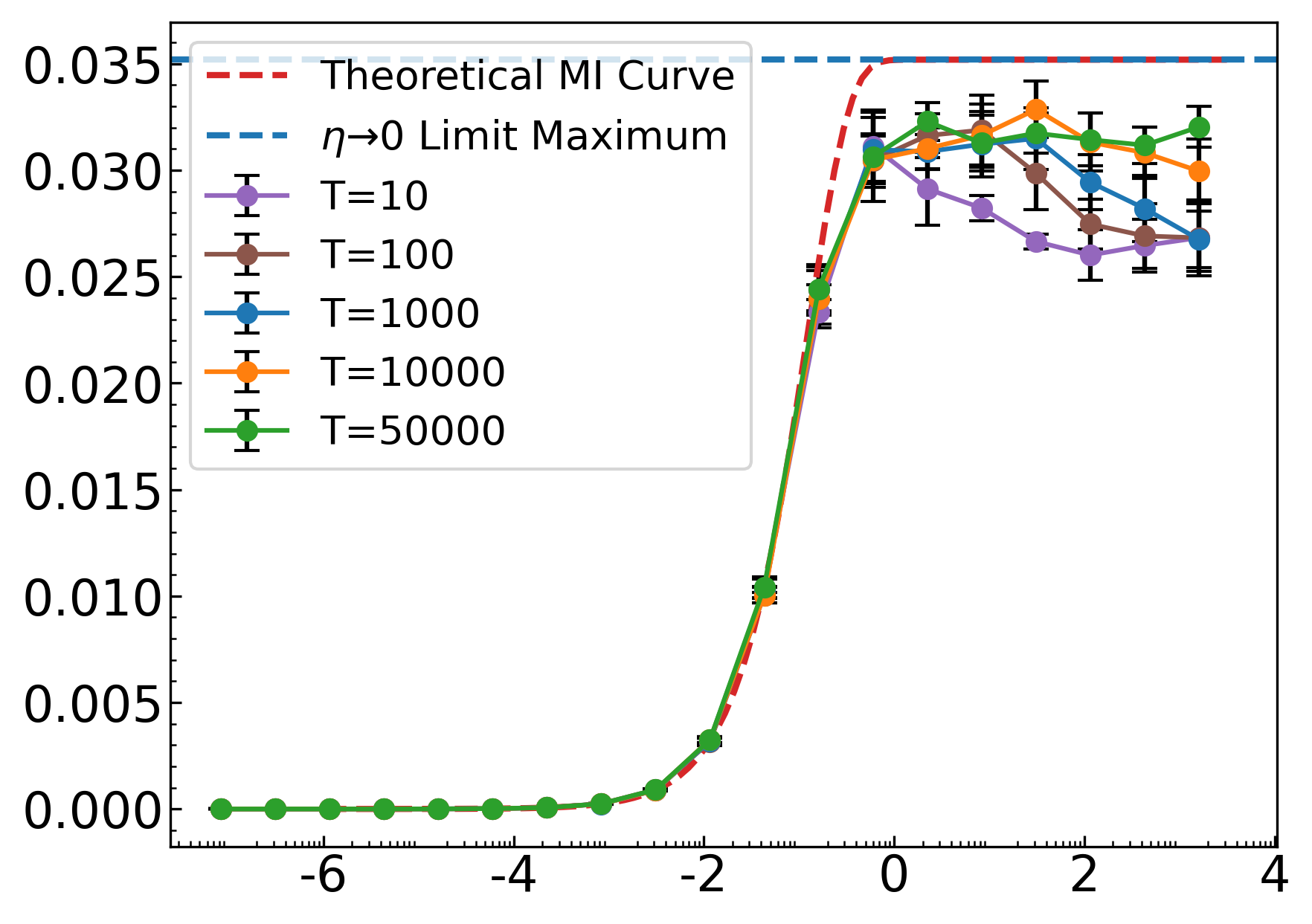}
        \put(78,27){\textbf{(d)}}
        \put(10,20){$\eta=0.10$}
        \put(74,10){$\ell T_x$}
    \end{overpic}
   \\
    \begin{overpic}[width=0.24\textwidth]{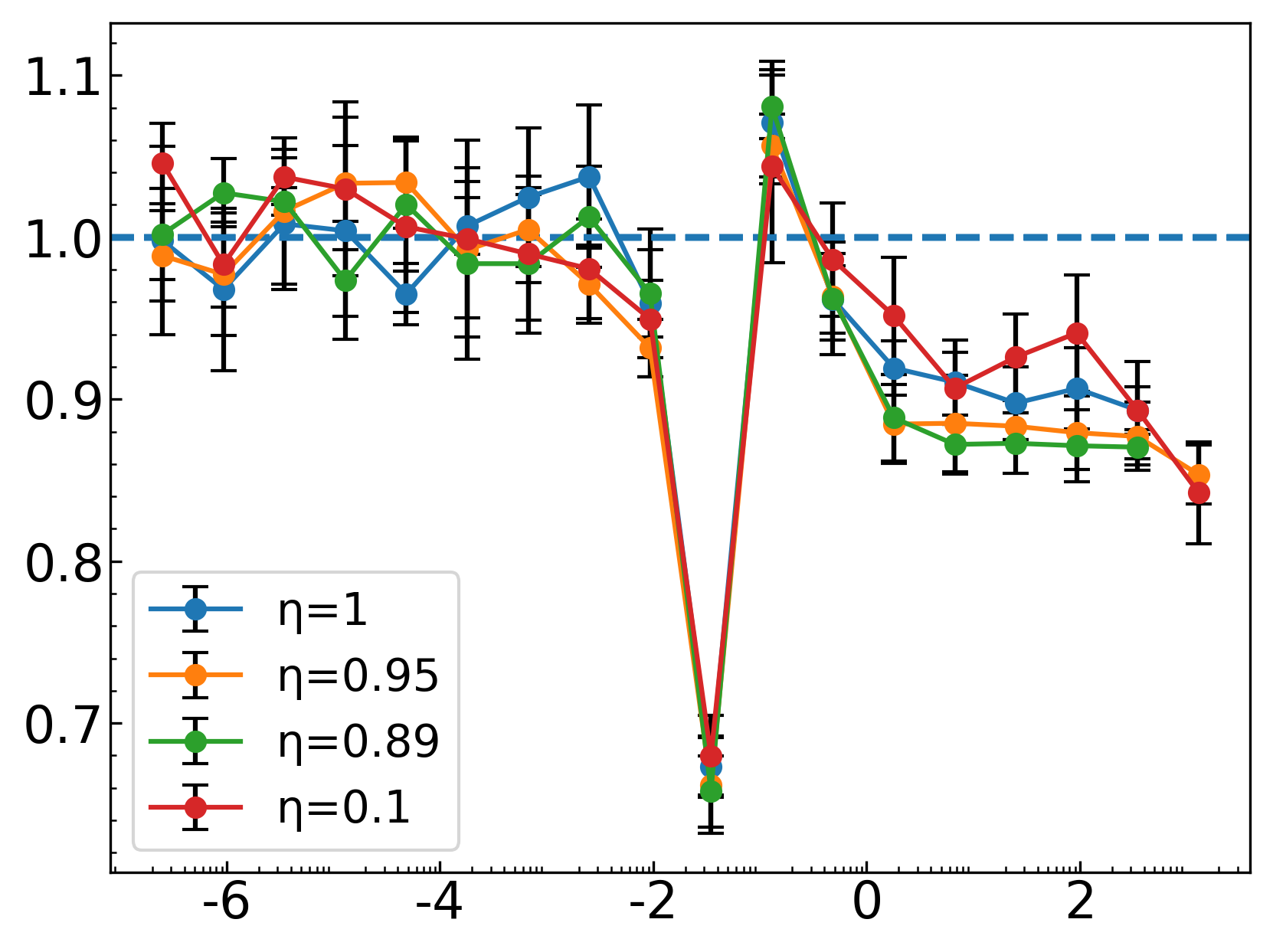}
        \put(64,22){\textbf{(e)}}
        \put(64,10){$\ell T_x$}
    \end{overpic}
    \begin{overpic}[width=0.24\textwidth]{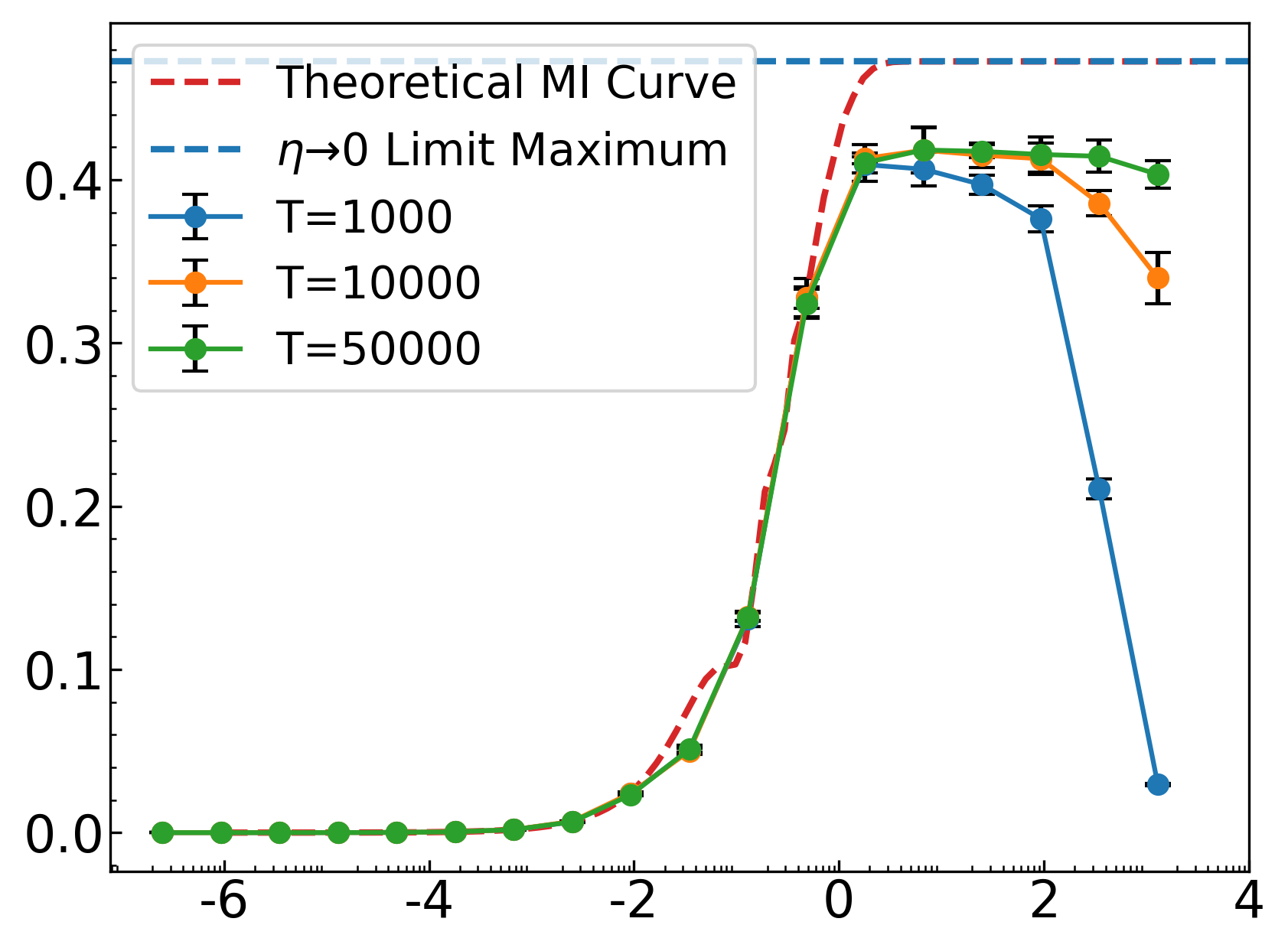}
        \put(64,22){\textbf{(f)}}
        \put(10,20){$\eta=0.95$}
        \put(64,10){$\ell T_x$}
    \end{overpic}
    \begin{overpic}[width=0.24\textwidth]{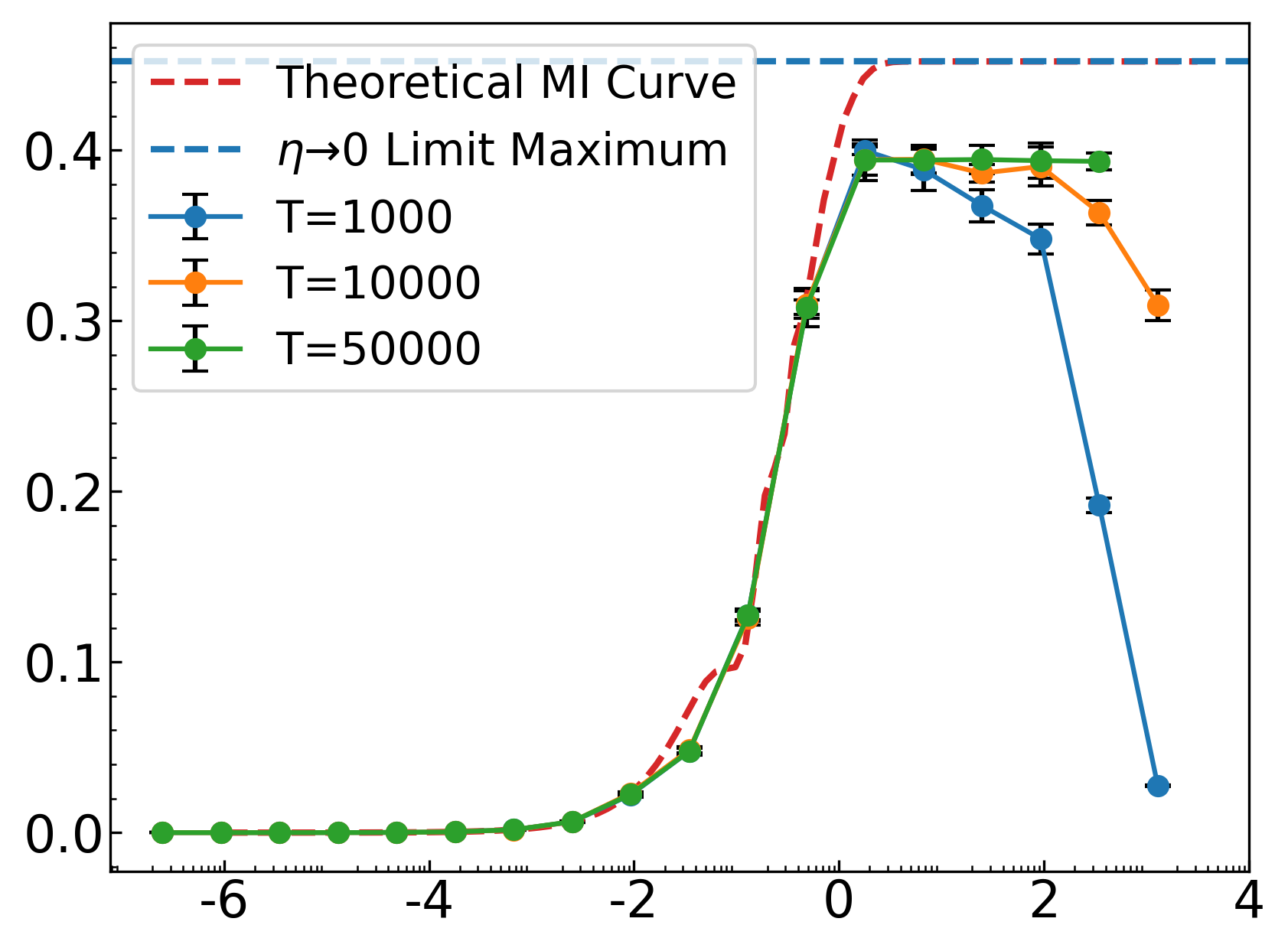}
        \put(64,22){\textbf{(g)}}
        \put(10,20){$\eta=0.89$}
        \put(64,10){$\ell T_x$}
    \end{overpic}
    \begin{overpic}[width=0.24\textwidth]{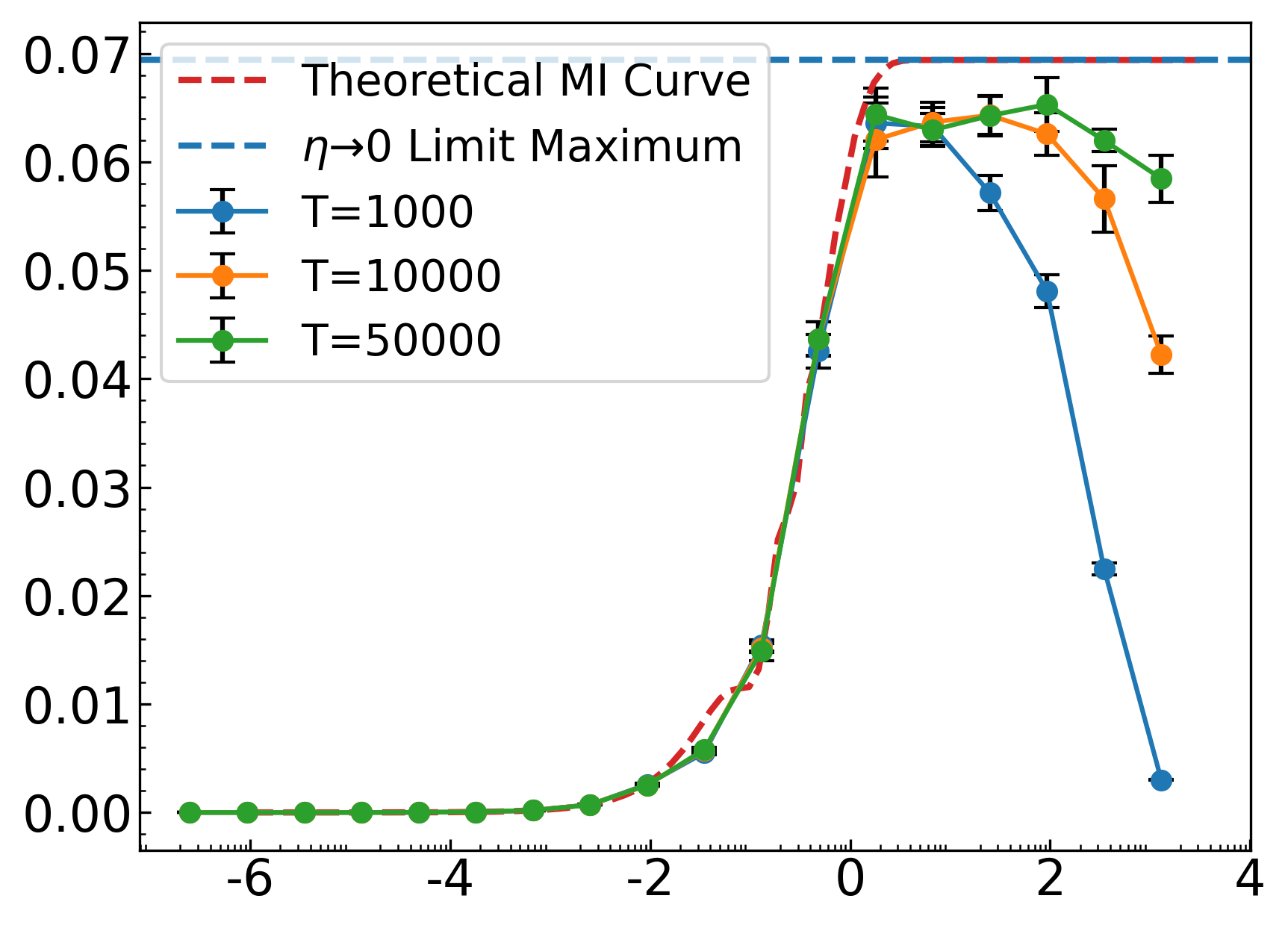}
        \put(64,22){\textbf{(h)}}
        \put(10,20){$\eta=0.10$}
        \put(64,10){$\ell T_x$}
    \end{overpic}
    \caption{
(note: $\ell T_x\equiv \log (x^2 T)$)\\Comparisons of numerical simulations and analytic (Sec. \ref{subsec:low-efficiency-calc}) calculations of mutual information MI $I(S; A_{1:T})$ for low efficiency measurements $\eta<1$ for model I (top row) and model II with $a=10$ (bottom row).
 Panels   $(b,c,d, f,g, h)$ show comparisons for indicated values of $\eta$.
    Panels $(a, e)$ are ratios of the numerical over theoretical mutual information estimates. }
    \label{fig:Noisy_MI_plots}
\end{figure*}

\subsection{\label{subsec:low-efficiency-calc} Mutual information in the continuous time low efficiency limit}

In this subsection, we present a perturbative approach to estimate the mutual information plateau of the weak measurement limit.
We consider a generalization of our continuous time measurement models, with an additional control parameter: the measurement efficiency $\eta$, where we have so far considered $\eta=1$. For small $\eta$, the SME, in some sense, is a small perturbation of the Lindblad equation. We are going to take advantage of this control parameter and explicitly calculate the leading contribution to the mutual information (see \ref{sec:inefficient_mearurement} and \ref{sec:MI_in_weak_noise_limit} for additional details of derivations and plots are presented, along with discussion of limits of applicability).

To derive the SME with nontrivial efficiency $\eta$, one concrete way to think about such systems is to start with a discrete time process, where, $p$ fraction of times the measurement output is the true measurement outcome, and $(1-p)$ fraction of times, the output is independent of the true measurement. 
In principle, this is equivalent to ignoring the measurement $(1-p)$ fraction of times, leading to a Lindblad process. 
In the continuum limit, this leads to a SME parameterized by the efficiency parameter $\eta:=p^2$.
The explicit SME derivations for our models with the efficiency parameter $\eta$ are  given in Appendix ~\ref{sec:inefficient_mearurement}. Here we present the final equations.

For model I, the SME with measurement efficiency $\eta$ is,
\begin{align} \nonumber \label{eq:model_1_noisy_sme_main_text}
    d\rho_t
    &=
    \frac{1}{\tau}
    \sum_{i}
    \left( 
    \sigma_i  \rho_t \sigma_i
    -
    \rho_t
    \right)
    \, dt
    \\
    &\quad
    +
    \sqrt{\frac{\eta}{\tau}} \sum_{i}
    \left(
    \left[ \sigma_i \rho_t + \rho_t \sigma_i  \right]
    -
    2
    \text{Tr}\left[ \sigma_i \rho_t \right] \rho_t
    \right)
    \, dW_{t i},
    \\
    dy_{t i} 
    &=
    2 \frac{\sqrt{\eta}}{\tau}
    \text{Tr}\left[ \sigma_i \rho_t \right]
    \, dt 
    +
    \frac{1}{\sqrt{\tau}}
    \, dW_{t i}.
\end{align}
and for model II, the SME with measurement efficiency $\eta$ is,
\begin{align} \nonumber \label{eq:model_2_noisy_sme_main_text}
    d \rho_t
    &= 
    -i
    \frac{\omega}{2}
    [X,\rho_t]
    \, dt
    +
    \frac{1}{\tau}
    \left(Z \rho_t Z  - \rho_t \right) 
    \, dt
    \\
    &\quad 
    +
    \sqrt{\frac{\eta}{\tau}}
    \left( 
    \left[ Z \rho_t + \rho_t Z \right]
    - 
    2 \text{Tr}\left[ Z \rho_t \right] \rho_t \right)
    \, dW_t,
    \\
    dy_t
    &
    = 2 \frac{\sqrt{\eta}}{\tau}
    \text{Tr}\left[ Z \rho_t \right] 
    \, dt 
    +
    \frac{1}{\sqrt{\tau}}
    \, dW_t .
\end{align}
where $dt$ is deterministic infinitesimal time and $dW$ is the stochastic Wiener differential.

We can now do a perturbation in $\eta$ (see Appendix~\ref{sec:MI_in_weak_noise_limit} for more details). Consider the state $\rho_t\left( \sqrt{\eta} \right)$ as a power series in $\sqrt{\eta}$,
\begin{gather}
    \rho_t\left( \sqrt{\eta} \right)
    =
    \rho_t^{(0)}
    +
    \left( \sqrt{\eta} \right)
    \rho_t^{(1)}
    +
    O(
    \left( \sqrt{\eta} \right)^2
    )
\end{gather}
where the superscript $(n)$ indicates the $n$-th order approximation.
Plugging the expansion into the SME in its general form Eq.~\ref{eq:SME} with an explicit timescale $\tau$, in the zeroth order terms, we just have the standard Lindblad evolution of $\rho_t$: 
\begin{align} \nonumber
    d\rho^{(0)}_t  
    &=
    - i 
    \left[H_0,\rho^{(0)}_t \right]
    \, dt  
    \\ \label{eq:main_text_Lindblad}
    &\quad
    +
    \frac{1}{\tau}
    \left( 
    \sum_\nu 
    L_\nu \rho^{(0)}_t L_\nu^\dag 
    -
    \tfrac{1}{2}( L_\nu^\dag L_\nu \rho^{(0)}_t + \rho^{(0)}_t  L_\nu^\dag L_\nu)
    \right)
    \, dt.
\end{align}
If the Lindblad dynamics is solvable, we can approximate, to first order, the measurement output Eq.~\ref{eq:dy} with $\rho^{(0)}_t$, from Eq.~\ref{eq:main_text_Lindblad}, as:
\begin{equation} \label{eq:dy0}
  dy^{(1)}_{t \nu}
  =
  \frac{ \sqrt{\eta_\nu} }{\tau}
  \text{Tr}(L_\nu\rho^{(0)}_t+\rho^{(0)}_t L_\nu^\dag) 
  \, dt
  +
  \frac{1}{\sqrt{\tau}}
  \, dW_{t \nu }.
\end{equation}

In the low efficiency limit, $\eta \ll 1$, these approximate dynamics allow us to estimate the mutual information between the initial state distribution and the measurement output by using the binary input Additive White Gaussian Noise (bi-AWGN) channel (see Appendix sections \ref{sec:small_efficiency_bi-AWGN} and \ref{sec:bi-AWGN_MI}). The mutual information approximation from the bi-AWGN channel (Eq.~\ref{eq:small_efficiency_AGWN_mutual_information}) is,
\begin{align} \label{eq:main_text_small_efficiency_AGWN_mutual_information}
    I(S,Y)(t)
    =&
    \int_{-\infty}^\infty\frac{e^{-\frac{z^2}{2}}}{\sqrt{2\pi}}\Big[\log 2 \nonumber \\
    -&\log\big(1+e^{-2\gamma(t)-2\sqrt{\gamma(t)}z}\big)\Big]dz,
\end{align}
with a signal-to-noise ratio SNR (Eq.~\ref{eq:gammat}),
\begin{equation}
    \gamma(t)
    =
    \sum_\nu 
    \frac{\eta_\nu}{4\tau}
    \int_0^t
    [\text{Tr}\Big((L_\nu+L_\nu^\dag)(\rho^{0\uparrow}_{s}-\rho^{0\downarrow}_{s} )\big )\Big]^2ds.
\end{equation}

Now, we use these to estimate the mutual information plateaus in our measurement models. Derivations are given in ~\ref{sec:MI_in_weak_noise_limit}. 

For Model I, $    \gamma(t)
    = 
    \frac{\eta}{2}
    \left[ 1- e^{-\frac{8}{\tau}t} \right].
$
For Model II, with the ratio $\frac{\phi}{x^2}$ (see Eq.~\ref{eq:adefn}) held fixed, we get coupled differential equations with the parameter $\alpha \coloneqq \frac{\omega \tau}{2}$, playing the role of the quality factor, for the dynamics.
The SNR $\gamma(t)$ when underdamped $\alpha > \frac{1}{2}$, or physically $\omega > \frac{1}{\tau}$, is,
{\small
\begin{align}
    \nonumber
    \gamma(t)
    &=
    \eta
    \bigg( 
    \frac{1+\alpha^2}{\alpha^2}
    +
    \frac{e^{- \frac{2}{\tau} t}}{ \alpha^2 \mu^2 }
    \bigg[ 
    - 4\alpha^4
    +
    \left( -3\alpha^2+1 \right)
    \cos{ \left( \frac{2 \mu }{\tau} t \right)}
    \\ 
    &\quad
    +
    \left( \alpha^2-1 \right)
    \mu
    \sin{ \left( \frac{2 \mu }{\tau} t \right)}
    \bigg]
    \bigg),
\end{align}
}
overdamped $\alpha < \frac{1}{2}$, or physically $\omega < \frac{1}{\tau}$, is,
{\small
\begin{align}
    \nonumber
    \gamma(t)
    &= \eta 
    \bigg( 
    \frac{1+\alpha^2}{\alpha^2}
    +
    \frac{e^{- \frac{2}{\tau} t}}{\alpha^2\mu'^2}
    \bigg[
    4\alpha^4 
    +
    \left( 3\alpha^2-1 \right) \cosh{\left( \frac{2 \mu'}{\tau} t\right)}
    \\ 
    &\quad
    +
    \left( \alpha^2-1 \right) \mu'
    \sinh{\left( \frac{2 \mu'}{\tau} t\right)}
    \bigg]
    \bigg),
\end{align}
}
and critically damped $\alpha = \frac{1}{2}$, or physically $\omega = \frac{1}{\tau}$, is,
\begin{align}
    \gamma(t)
    &= 
    \eta
    \left(
    5
    -
    e^{-\frac{2}{\tau} t}
    \left( 2\left( \frac{t}{\tau} \right)^2 + 6\frac{t}{\tau} + 5 \right)
    \right)
\end{align}
where $\mu \coloneqq \sqrt{4\alpha^2-1}$ and $\mu' \coloneqq \sqrt{1-4\alpha^2}$.
The mutual information $I(S,Y)(t)$ is computed with Eq.~\ref{eq:main_text_small_efficiency_AGWN_mutual_information} using these $\gamma(t)$.

Theses theoretical values of the mutual information in the low efficiency limit, $\eta \ll 1$ are plotted in Fig. \ref{fig:smalleffextrap} and used as guides to the eye in Figs. \ref{fig:MI_numerical_simulations} and \ref{fig:Noisy_MI_plots}. Mutual information for discrete noisy simulations and decreasing measurement efficiency $\eta$ for various $T$ are also plotted. The ratio of the discrete noisy simulation mutual information estimates over the small efficiency theoretical values in Fig.~\ref{fig:Noisy_MI_plots} are close to $1$.  This shows that the the mutual information plateau of the noisy simulation agrees with our theoretically derived values and justifies our perturbative estimate of the mutual information plateau in the low efficiency limit.

\section{\label{sec:state_recovery} Initial state recovery from sequential measurements}

Mutual information reveals the fundamental correlations and information loss between initial states and measurement records. This naturally leads us to the question: How can we extract all available information from measurements to optimally predict the initial quantum state? In what follows, we address this initial state recovery/estimation problem and present a protocol based on the Bayes-optimal predictor.
Furthermore, our protocol leverages a key insight from studying the mutual information in our sequential weak measurement schemes: more measurements are not necessarily better. Instead, there exists a specific timescale over which measurements are useful, allowing us to rule out ``physics"-agnostic learning frameworks for initial state estimation.

\subsection{The Bayes optimal classifier}
If we have access to the posterior  $\Pr\left[ \rho_{0}|a_{1:T}  \right]$, the Bayes optimal predictor/classifier $f^{\mathrm{BO}}:O^T\to \mathcal D$ (mapping $A_{1:T}\mapsto P_0$, i.e. from measurement records to initial states) is given by \cite{murphy2022probabilistic}
\begin{equation}
 f^{\mathrm{BO}}\left(a_{1:T}  \right)=\arg\max_{\rho_0\in \mathcal{D}}\Pr\left[ \rho_0|a_{1:T}  \right].   
\end{equation}
In the case of a tie, we tiebreak according to some prefixed order.

The Bayes optimal predictor is the optimal statistical approach to readout by construction. Furthermore, the accuracy of any predictor/classifier $f:O^T\to \mathcal{D}$  is given by 
\begin{equation}
 \A(f)
 :=
 \sum_{a_{1:T}\in A_T , \rho_0 \in \mathcal{D}} I\left[f(a_{1:T})=\rho_0\right]
 \Pr\left[ a_{1:T} |\rho_0 \right]
 \Pr[\rho_0].  
\end{equation}
We have $\A(f^{BO})\ge \A(f)$ for any $f$ and the Bayes optimal predictor upper-bounds the fidelity of any statistical readout scheme. 

The mutual information $I(P_0,A_{1:T})$, between the measurement record and the original density matrix, can provide an upper bound on an accuracy via Fano's inequality \cite{ash2012information, Cover2006}. This inequality implies that
\begin{align}
 &\mathcal H(1-\A(f))+(1-\A(f))\log_2(|\mathcal D|-1)\nonumber\\
 \ge &H(P_0|A_{1:T})
 = H(P_0)-I(P_0,A_{1:T})
 \label{Fano}
\end{align}
for any $f$, with $\mathcal H(p):=p\log_2 \tfrac {1}{p}+(1-p)\log_2\tfrac {1}{1-p}$. So, if the mutual information $I(P_0,A_{1:T})$ is less that entropy of the input $H(P_0)$, the accuracy $\A(f)$ has to be strictly less than 100\%. 
Further relations (or lack thereof) between accuracy and mutual information in evenly balanced supervised learning problems are discussed in \cite{meyen2016relation}. Beyond limitations of accuracy, the existence of a time-scale of information loss has serious consequences for general-purpose supervised learning methods \cite{murphy2022probabilistic}, as we will see in the next subsection.

\subsection{Information loss, inference of initial state and overfitting}
\label{subseq:overfit}

\begin{figure}
    \centering
    \includegraphics[width=0.92\linewidth]{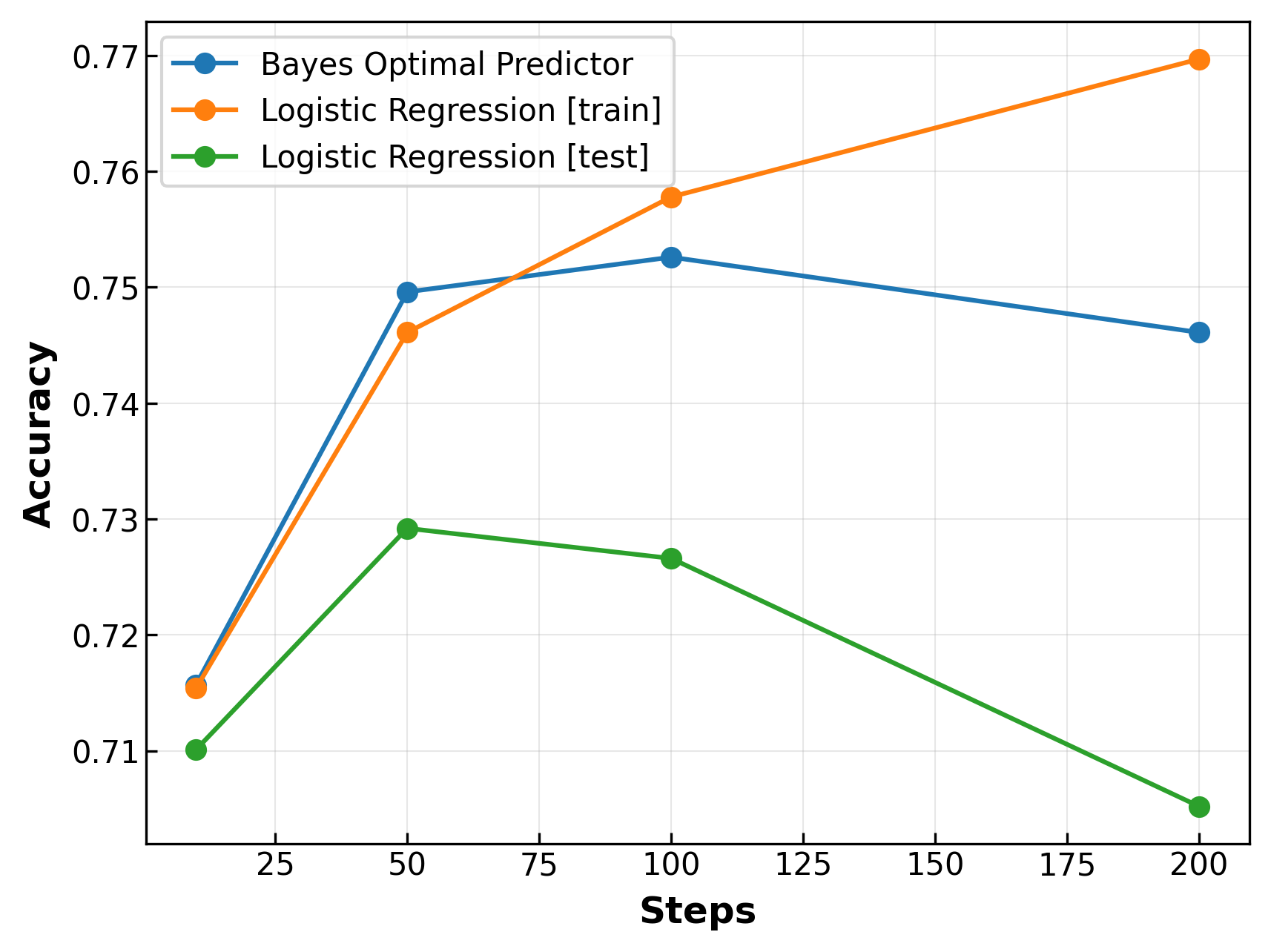}
    \caption{
    Prediction accuracy of Bayes optimal predictor and logistic regression. Measurement strength $x=0.4$ and $n=10000$ samples were used. The logistic regression models used a concatenation of one hot vectors for each measurement outcome to represent the measurement trajectory}
    \label{fig:overfit}
\end{figure}

To recover the initial state from the measurement record $A_{1:T}$, we might choose a flexible supervised learning approach. The Bayes optimal method, which has access to all information about the quantum dynamics and measurement schemes, and is aware that the measurements coming much later that $\xi$ does not have much to say about the initial state. A ``physics"-agnostic method, with limited sample size (sub-exponential in $T$) could possibly fit the initial state to the noisy late measurements during the training process and not generalize when it comes to testing. This the phenomenon we see in Fig. \ref{fig:overfit}.

We represent the measurement record as $\mathbf{X}:=[\mathbf{x}_1 \cdots\mathbf{x}_T]$, where $x_t\in \mathbb R^{|O|}$ is a one-hot column vector that encodes the measurement result at $t$. In other words, $(x_t)_a=\delta_{aa_t}$. The predictor $Y\in\{1,-1\}$ for initial spin $S$ is given by 
\begin{equation}
    Y=\text{sgn}(\Tr(\mathbf{W}^T\mathbf{X})-\theta),
\end{equation}
where $\mathbf{W}\in\mathbb R^{|O|\times T}$ and $\theta\in\mathbb R$. When $T>>\xi$, $x_t$ for larger $t$ are not informative about the initial spin $S$ but it is easy to find $\mathbf{W}$ that predict $S$ in training data well, when the sample size is much less than $|O|^T$. This is a classic case of overfitting, and as a result the accuracy on the test datasets drops significantly compared to the training data. As expected, in Fig. \ref{fig:overfit}, the train results are better than the Bayesian optimal classifier's but the test results are much worse.

\section{Conclusion}

Our study indicates that, under generic circumstances, the information about the initial state effectively saturates, past a certain number of observations, and does not allow perfect recovery. As we saw, this phenomenon leads to some  physics-agnostic statistical methods for initial state inference trying to extract signals from late measurements which are effectively independent of the initial state. This could lead to overfitting. We show that a Physics-aware Bayesian classifier avoids this problem.

We believe that this phenomenon of limited readability of the initial sate is the other side of the coin of dynamic state purification by measurements \cite{gullans2020dynamical}. Such state purification and learnability transition has recently seen a surge of interest \cite{ippoliti2023dynamical, ippoliti2024learnability}. In our measurement scheme, knowing an initial pure state and the measurement record tells us the precise pure state trajectory, up till that time. In the case of long measurement records, where later measurements have very little information about the initial state, is also the case where the final state could be predicted with great accuracy, given only the long measurement record. In future we hope to present our ongoing work on this relationship. 

We are optimistic that this line of work will eventually inform and impact actual experiments. With that in mind, Table~\ref{tab:experimental_platforms} surveys representative ranges of the per-step measurement strength $x$, the quantum efficiency $\eta$, and the coherent-drive timescale across four prominent qubit-readout architectures. Existing platforms collectively access most of the parameter regimes studied in this work --- from informationally complete strong measurement to weak, low-efficiency, dynamically driven monitoring --- so the saturation and plateau phenomena we analyze are not merely formal limits but should be accessible in present-day setups. The mapping from platform observables (dispersive shift, photon-collection efficiency, drive amplitude, etc.) to a per-step $x$ is necessarily approximate; the entries should be read as orders of magnitude rather than precise calibrations.

\begin{table}[h]
\centering
\footnotesize
\resizebox{\textwidth}{!}{
\begin{tabular}{p{2.7cm} p{2.8cm} p{2.7cm} p{2.9cm} p{2.7cm}}
\hline
Platform & Physical interpretation & Meas.\ strength ($x$) & Efficiency ($\eta$) & Coherent drive \\
\hline
Circuit QED (cQED) \cite{murch2013quantum,hatridge2013quantum} & Cavity-enhanced dispersive readout; continuous microwave driving & Variable via pulse duration; $x\sim 0.05$--$0.5$ & $\eta\sim 0.5$--$0.7$ & Continuous drive; $\omega\Delta t\sim 0.01$--$0.2$ \\
Color centers ($\mathrm{SnV}^{-}$ in diamond) \cite{TVQrosenthal2024single} & Optical spin-dependent scattering; microwave-driven Rabi rotation & Tunable via laser power; $x\sim 0.1$--$0.9$ & Low raw det.\ eff.; $\eta_{\text{raw}}\sim 10^{-3}$, $\sim 0.985$ conditioned & Stroboscopic/pulsed; $\Omega_{\text{Rabi}}/2\pi\sim 6$\,MHz \\
Neutral atoms (spin QND ensembles) \cite{smith2006efficient,hammerer2010quantum} & Dispersive Faraday rotation; collective atomic spin precession & Scaled via photon number; $x\sim 0.01$--$0.3$ & Moderate tracking; $\eta\sim 0.1$--$0.4$ & Larmor precession; $\omega\Delta t\sim 10^{-3}$--$10^{-1}$ \\
Quantum dots (semiconductor spin) \cite{elzerman2004single,delbecq2011coupling} & Charge sensing via QPC or SET; spin-to-charge conversion & Tunable via tunnel barriers; $x\sim 0.1$--$0.6$ & Historically low to mod.; $\eta\sim 0.05$--$0.2$ & Electrically driven spin; $\omega\Delta t\sim 0.01$--$0.1$ \\
\hline
\end{tabular}
}
\caption{\label{tab:experimental_platforms}Typical parameter landscapes, physical mappings, and representative literature across prominent experimental qubit-readout architectures. Parameters follow the definitions of Table~\ref{tab:params}: measurement strength per step $x$, quantum efficiency $\eta$, and a coherent-drive timescale (dimensionless $\omega\Delta t$ where the drive runs throughout the measurement; bare Rabi rate $\Omega_{\text{Rabi}}/2\pi$ for stroboscopic platforms where drive and readout are interleaved).}
\end{table}

\section{Acknowledgments}
We wish to acknowledge useful exchanges with Gerry Angelatos, Jed Pixley, Dries Sels, Bert Kappen, and Hakan T\"ureci. The Flatiron Institute is
a division of the Simons Foundation. CL acknowledges support from the National Science Foundation Graduate Research Fellowship Program under Grant No. DGE-2146755. Any opinions, findings, and conclusions or recommendations expressed in this material are those of the author and do not necessarily reflect the views of the National Science Foundation.
ABC is supported by Grant MMT24IFF-01. The funding for this grant and contract comes from the European Union’s Recovery and Resilience Facility-Next Generation, in the framework of the General Invitation of the Spanish Government’s public business entity Red.es to participate in talent attraction and retention programs within Investment 4 of Component 19 of the Recovery, Transformation, and Resilience Plan.

\section{Data availability statement}

The code and data that support the findings of this study are openly available at the following github repository URL: \href{https://github.com/vadimoganesyan/2512.14583}{https://github.com/vadimoganesyan/2512.14583}.

\section{References}

\bibliographystyle{unsrt}
\bibliography{apssamp}

\appendix

\appendix\section{Estimating Mutual Information and Initial State Estimation Accuracy}

In our analysis, we used mutual information as the central measure of the intrinsic information about initial states in sequential measurement records. However, even in our discrete setting, it is challenging to compute because it scales exponentially, with the number of measurements, in the terms required to evaluate it. 
Similarly, we also consider accuracy in initial state estimation and run into the same problem.
However, we present a method to accurately estimate these quantities from $M$ samples using Hoeffding's inequality. This yields a probabilistic error bound on our estimates: with probability at least $1-\delta$, the estimation error is at most $\varepsilon$, provided the number of samples is at least $M=\frac{1 }{2 \varepsilon^2 } \ln{\frac{2}{\delta}}$. The accuracy of our estimates scale as  $O\left( \frac{1}{\sqrt{M}} \right)$ in the number of samples.

We will work in the same qubit setting as the main text, where we restrict the set of possible initial states to the spin up $|\up\rangle\langle\up|$ and spin down $|\down\rangle\langle\down|$ states, which we will refer to by a random variable $S \in \{\up,\down\}$.

\subsection{\label{sec:mutual-info-estimation}Mutual information estimation}

We can accurately estimate mutual information $I(S,A_{1:T})$ between the initial bit $S$ and the measurement record $A_{1:T}$ using Hoeffding's inequality. First, consider the mutual information as,
\begin{align}
    \nonumber
    I(S,A_{1:T})
    &=
    H(S)
    -
    H\left( S|A_{1:T} \right)
    \\ \nonumber
    &=
    H(S)
    -
    \sum_{a_{1:T}}
    \mathrm{Pr}\left[ A_{1:T}=a_{1:T} \right]
    H\left( S|A_{1:T}=a_{1:T} \right)
    \\
    &=
    H(S)
    -
    \mathbb{E}_{ a_{1:T}  \sim \mathrm{Pr}\left[ A_{1:T} \right] }
    \left[
    H\left( S|A_{1:T}=a_{1:T} \right)
    \right]
\end{align}
where $H$ represents entropy. The first term is easy to calculate. The last term includes a sum over an exponential, in $T$, number of terms, making it hard to evaluate as $T$ becomes large. However, the last term is an expectation value so we can use an efficient sample average to estimate $H(S|A_{1:T})$ and, ultimately, the mutual information.

Next, we bound the error in our estimates. If we measure information in bits, namely 
$H(X)=-\sum_x \mathrm{Pr}[X=x]\log_2 \mathrm{Pr}[X=x]$, then 
$H(S|A_{1:T}=a_{1:T})$ 
is bounded 
$0\le H(S|A_{1:T}=a_{1:T})\le 1$.
We thus treat this quantity like a bounded random variable and apply Hoeffding's inequality \cite{hoeffding1994probability} to the $M$-sample average, estimate of $I$, $\hat I_M$, to bound the error as,
\begin{equation}
    \mathrm{Pr}
    \left[ \left| \hat I_M-I \right|\ge \varepsilon \right]
    \le 
    2 \exp{\left( -2M\varepsilon^2 \right)}. \label{eq:hoeff}
\end{equation}
Furthermore, in order to guarantee our bound holds with at least probability $1-\delta$,
$\mathrm{Pr}[|\hat I_n-I|\ge\varepsilon]\le \delta$, we need $M \ge \frac{1}{2\varepsilon^2}\ln{\frac{2}{\delta}} $ samples.

Finally, we derive the computational complexity of our estimates. This bound is independent of $T$, so the only slightly nontrivial computation is $\mathrm{Pr}[S=\sigma|A_{1:T}=a_{1:T}]$, which involves computing the measurement probability $\mathrm{Pr}[A_{1:T}=a_{1:T}|S=\sigma] = \text{Tr} \left[ M_{a_1, \cdots, a_T} | \sigma \rangle \langle \sigma |  \right]  = \langle \sigma|  M_{a_1, \cdots, a_T} |\sigma\rangle$.
Since $M_{a_1, \cdots, a_T}$ is a product of $2T$ $2\times2$ matrices, that operation takes $O(T)$ time. Combining these results, we get a computational complexity of $O(\frac{T}{\varepsilon^2}\ln(\frac{2}{\delta}))$.  This analysis may be extended to $N$-qbit readout which will modify Eq.~\ref{eq:hoeff} s.t. 
$\exp{\left( -2M\varepsilon^2 \right)}\to \exp{\left( -2M\varepsilon^2/N \right)}$

\subsection{Estimating qubit initial state estimation accuracy}

Now, we look at estimating the accuracy of our Bayes optimal predictor in single qubit initial state recovery/estimation. In initial state estimation, 
our objective is to classify the initial state correctly given a measurement record $a_{1:T}$.
Using the Bayes optimal predictor, the probability with which we can correctly classify the initial state is well-defined. However, similar to calculating mutual information, the only difficulty is that the number of measurement records scales exponentially in the number measurements $T$. Luckily, we can, again, accurately estimate the accuracy using Hoeffding's inequality.

First, the accuracy of a classifier is defined as the probability of being correct and, formally, as,
\begin{equation}
    \A 
    =
    \pi_\up \Pr[S = \up]
    +
    \pi_\down \Pr[S = \down]
\end{equation}
where $\pi_\up$ ($\pi_\down$) is the probability of $S=\up$ ($S=\down$) being predicted correctly. Marginalizing over all measurement records of length $T$,
\begin{equation}
    \pi_s
    =
    \sum\limits_{a_{1:T}} \chi_s (a_{1:T})\Pr[A_{1:T} = a_{1:T}|S = s].
\end{equation}
where $\chi_s (a_{1:T})$ is the indicator that the classifier would choose $s$, given the measurement record $a_{1:T}$. For the Bayes optimal classifier, these indicators are given by
\begin{equation}
   \chi_\up^{BO}(a_{1:T}) =\left\{ \begin{array}{ccc}
        1 & \mathrm{if} & \Pr[A_{1:T} = a_{1:T}|S = \up]>\Pr[A_{1:T} = a_{1:T}|S = \down] 
        \\
        1 & \mathrm{if} & \Pr[A_{1:T} = a_{1:T}|S = \up]=\Pr[A_{1:T} = a_{1:T}|S = \down]
        \\
        & & \&\,\mathrm{tiebreak}\,\to \sigma
        \\
        0 &\mathrm{otherwise} & 
    \end{array} \right. 
\end{equation}
where $\sigma \in \{ 0, 1\}$ is a predetermined state value and $\chi_\down^{BO}(a_{1:T}) =1-\chi_\up^{BO}(a_{1:T})$. 
With that, accuracy is reduced to
\begin{align}
    \nonumber
    \A^{BO} 
    &=
    \Pr[S = \up]\sum\limits_{a_{1:T}}\chi_\up^{BO}(a_{1:T})
    \Pr[A_{1:T} = a_{1:T}|S = \up]
    \\ \nonumber
    &\quad 
    +
    \Pr[S = \down]\sum\limits_{a_{1:T}}(1-\chi_\up^{BO}(a_{1:T}))
    \Pr[A_{1:T} = a_{1:T}|S = \down]
    \\
    &=
    \frac{1}{2}\sum\limits_{a_{1:T}}\max\{\Pr[A_{1:T} = a_{1:T}|S = \up],\Pr[A_{1:T} = a_{1:T}|S = \down]\}
\end{align}
where, in the last line, we let $\Pr[S = \up]=\Pr[S = \down]=\frac{1}{2}$.
When $T$ is too large to evaluate this sum, one could estimate the accuracy by sampling,
\begin{align}
    \A^{BO} 
    =&
    \sum_{a_{1:T}}\mathrm{Pr}[A_{1:T}=a_{1:T},S=s]\chi_s^{BO}(a_{1:T})
    \\\nonumber
    =&
    \mathbb{E}_{ a_{1:T}  \sim \mathrm{Pr}\left[ A_{1:T} \right] }
    \left[
    \chi_S^{BO}(A_{1:T})
    \right].
\end{align}
Since $\chi_S^{BO}(A_{1:T})$ take values in range $\{0,1\}$, it is a bounded random variable. Then we can accurately estimate accuracy with an $M$-sample empirical mean $\hat \A_n^{BO}$, that would be close to the true $\A^{BO}$, 
using Hoeffding's inequality with the same probabilistic estimation error bounds and scaling as in the previous section.

\section{\label{sec:channel_transfer_matrix} Mean Dynamics and Late Time Mutual Information}

Here, we look at mutual information in the $T$-th measurement of an arbitrary measurement record $a_{1:T}$. This motivates us to define the quantity $\xi$, which is like a correlation length of the mean measurement dynamics. Ultimately, $\xi$ gives us insight into the mutual information of late $T$ measurement. Finally, we calculate $\xi$ for both of the weak measurement schemes in our work. For both cases, $\xi \sim \tfrac{1}{x^2}$, suggesting things should depend on $\tfrac{T}{\xi}\propto Tx^2$. 

\subsection{Initial state and $T$-th measurement joint probability}

In the both accuracy and in mutual information, we see that in the generic case, our ability to infer the initial state from the sequential measurements remains limited, even when $T$ goes to infinity. To understand when very late measurement adds to the knowledge of the initial state, here, we calculate the joint distribution $\Pr[S =\rho_0,A_T=a_T]$ of initial state $S$ and $T$-th measurement $A_T$. Consider,
\begin{align}
    \nonumber
    \Pr[S=\rho_0,A_T=a_T]
    &=
    \sum\limits_{a_{1:T-1}}
    \Pr{\left[ a_{1:T}|\rho_0 \right]}
    \Pr{\left[ \rho_0 \right]}
    \\
    \nonumber
    &=
    \sum\limits_{a_{1:T-1}} 
    \Tr{\left[
    \left( K_{a_T} K_{a_{T-1}} \cdots K_{a_1} \right)
    \rho_0
    \left( K_{a_T} K_{a_{T-1}} \cdots K_{a_1} \right)^\dag
    \right]}
    \Pr\left[ \rho_0 \right]
    \\
    \nonumber
    &=
    \Tr{\left[
    K_{a_T}
    \left( \sum\limits_{a_{1:T-1}} 
    \left( K_{a_{T-1}} \cdots K_{a_1} \right)
    \rho_0
    \left( K_{a_{T-1}} \cdots K_{a_1} \right)^\dag
    \right)
    K_{a_T}^\dag 
    \right]}
    \Pr\left[ \rho_0 \right]
    \\
    &=
    \Tr{\left[ \E_{a_T} \left( \E^{T-1}\left( \rho_0 \right) \right) \right]}
    \Pr\left[ \rho_0 \right]
\end{align}
where we define the superoperators $\{\E_a : a \in \mathcal{O} \}$, over the set of measurement outputs $\mathcal{O}$, and $\E$ as follows,
\begin{gather}
    \E_a(\rho):=K_a\rho K_a^\dagger, 
    \, \mathrm{and},\,
    \E(\rho):=\sum\limits_{a}\E_a(\rho).
\end{gather}
with $\E$ a (probability conserving) quantum channel.

For our single qubit (spin-$\tfrac{1}{2}$) example, we can parametrize our measured state in the Pauli basis 
$
\rho
=
\tfrac{1}{2}[p_0 I+\vec p\cdot\vec\sigma]
$. and equivalently represent the state with the vector of Pauli coefficients,
\begin{align}
    \mathbf p(\rho) 
    :=
    \begin{bmatrix}
        p_0 \\ p_x \\ p_y \\ p_z
    \end{bmatrix}
\end{align}
Next, consider the action of $\{\E_a\}$ and $\E$ on the coefficients vector $\p(\rho)$. We define the corresponding $4\times4$ matrices $\{E_a\}$ and $E$ by $\E_a(\rho)\to E_a\p$ and $\E(\rho)\to E\p$. In this representation, the measurement probability are,
\begin{equation}
    \Tr\left[ \E_{a_T}\left( \E^{T-1}\left(\rho_0 \right) \right) \right]
    =
    \e_0'E_{a_T}E^{T-1}\p(\rho_0).
\end{equation}
where $\e_0 \coloneqq [1, 0, 0, 0]^{T}$.

Since $\E$ is trace preserving, $E$ has at least one eigenvector $\p_*$ corresponding to the eigenvalue $1$, and the rest of the eigenvalues lie in the unit disc in the complex plane. If all the rest of the eignevalues have norm strictly less than $1$, there is a vector $\q_*$ that is orthogonal to all the other subspaces associated with the eigenvalues distinct from $1$, but has $\q_*'\p_*=1$. Finally, the measurement probability is,
\begin{equation}
    \Pr[S=\rho_0,A_T=a_T]=\Tr(\E_{a_T}(\E^{T-1}(\rho_0)))=\e_0'E_{a_T}E^{T-1}\p(\rho_0)=(\e_0'E_{a_T}E^{T-1}\p_*)(\q_*'\p(\rho_0))+O(e^{-\frac{T}{\xi}})
\end{equation}
where, for the mean transfer superoperator $E$, we define
\begin{equation}
    e^{-\frac{1}{\xi}}
    =
    \max\limits_{\lambda\in Eig(E),\lambda\neq 1} |\lambda|
    .
\end{equation}
which is like a boundary correlation length in this problem.

For $T$ large, $S$ and $A_T$ are independent, up to small corrections. Thus we have $I(S,A_T)=O( Te^{-\frac{2T}{\xi}})$, in general.   If $(\e_0'E_{a}E^{T-1}\p_*)(\q_*'\p(\rho_0))\neq 0$ for all measurement outcomes $a$ and all $\rho_0\in \mathcal{D}$, the upper bounds would be sharper: $I(S,A_T)=O( e^{-\frac{2T}{\xi}})$. In either case, the mutual information between the initial state and the $T$-th measurement is going to zero exponentially as $T$ tends to infinity. Thus for later observations to be of value, we must have nontrivial multiplicity of the eigenvalue $1$, which we could think of as $\xi=\infty$.

$I(S,A_T)=O(\exp(-cT))$ for large $T$, in general, does not imply that $I(S,A_{1:T})-I(S,A_{1:T-1})$ also goes to zero in a similar manner. That said, we conjecture that, in our problem, $\xi$ being finite implies imperfect recovery of information about $S$ from arbitrarily long sequence of measurements.

\subsection{Informationally complete measurements}

Now we calculate $\xi$ for our first measurement scheme. First we can calculate the individual measurement matrices as, 
\begin{equation}
\begin{aligned}
    E_{X,y} 
    =
    \frac{1}{6}
    \begin{pmatrix}
        1 & y\tanh x  &  0 & 0   \\
        y\tanh x  & 1 & 0 &  0 \\
        0 & 0 & \sech x & 0 \\
        0 & 0 & 0 & \sech x
    \end{pmatrix},
    \\
    E_{Y,y} 
    =
    \frac{1}{6}
    \begin{pmatrix}
        1 & 0 & y\tanh x  &  0   \\
        0 & \sech x  & 0 &  0 \\
        y\tanh x & 0 & 1 & 0 \\
        0 & 0 & 0 & \sech x
    \end{pmatrix},
    \\
    E_{Z,y} 
    =
    \frac{1}{6}
    \begin{pmatrix}
        1 & 0 &  0 &
        y\tanh x   \\
        0 & \sech x & 0 &  0 \\
        0 & 0 & \sech x & 0 \\
        y\tanh x & 0 & 0 & 1
    \end{pmatrix},
\end{aligned}
\end{equation}
and, after averaging over all outcome, we get the mean as,
\begin{equation}
    E
    = 
    \begin{pmatrix}
        1 & 0  &  0 & 0   \\
        0 & \frac{1+2\sech x}{3} & 0 &  0 \\
        0 & 0 & \frac{1+2\sech x}{3} & 0 \\
        0 & 0 & 0 & \frac{1+2\sech x}{3}
    \end{pmatrix}.
\end{equation}
In this case, $\xi^{-1}=\ln \tfrac{3}{1+2\sech x}.$

\subsection{Informationally incomplete measurements with unitary evolution}

For our next measurement model, as usual, define $\phi:=\omega\Delta t$.
We calculate the individual measurement matricies as,
\begin{equation}
    E_y
    =
    \begin{pmatrix}
        1 & y\tanh x\sin\phi &  0 &
        y\tanh x\cos\phi   \\
        0 & \sech x\cos\phi & 0 &  -\sech x\sin\phi \\
        0 & 0 & \sech x & 0 \\
        y\tanh x & \sin \phi & 0 & \cos\phi
    \end{pmatrix}
\end{equation}
and, after averaging over all outcome, we get the mean as,
\begin{equation}
    E
    =
    \begin{pmatrix}
        1 & 0  &  0 & 0   \\
        0 & \sech x\cos\phi & 0 & -\sech x\sin\phi \\
        0 & 0 & \sech x & 0 \\
        0 & \sin \phi & 0 & \cos\phi
    \end{pmatrix}
\end{equation}
The only way for this system to have more than one unit modulus eigenvalue is either to have $\cos\phi=\pm1$, namely field rotations are multiples of $180^\circ$, or $\sech x\to 1$, which corresponds to projective measurements. For each $x$, there is a range of $\phi$ away from $0^\circ$ where this matrix has two complex eigenvalues conjugate to each other. In that regime $\xi^{-1}=2\ln \cosh x.$

In general, for any regime the two complex eigenvalues are
\begin{equation}
    \lambda_\pm = \frac{1}{2} \left( \left( 1 + \sech(x) \right)\cos\phi \pm \sqrt{\cos^2\phi\left( 1 + \sech(x)\right)^2 - 4\sech(x)} \right)
\end{equation}

Notice, the case of incomplete measurements with $\phi=0$ illustrates our earlier point that $\xi = \infty $ and later observations are of value in the $T\to\infty$ limit. This also corresponds the the case where our measurement scheme convergences to a projective measurement.

\section{\label{sec:WML} SME Dynamics of the Weak Measurement Limit}

In this section, we derive the continuous-time Stochastic Master Equations (SME) that describe the dynamics of our sequential weak measurement schemes in the weak measurement limit, measurement strength $x \ll 1$.
We consider the three weak measurement models discussed in the main text: 1) informationally incomplete, 2) informationally complete, and 3) informationally incomplete with unitary dynamics.

\subsection{Stochastic master equation in the weak measurement limit}

In general, the continuous time dynamics of our weak measurement schemes are Stochastic Master Equations in the diffusive limit \cite{barchielli2009quantum,ROUCHON2022252}, given by,
\begin{align} \label{eq:SME}
    \nonumber
    d\rho_t  
    &= 
    - i \Big[H_0,\rho_t\Big] \, dt  
    +
    \frac{1}{\tau}
    \left(
    \sum_\nu L_\nu \rho_t L_\nu^\dag 
    -
    \tfrac{1}{2}
    \left( 
    L_\nu^\dag L_\nu \rho_t + \rho_t  L_\nu^\dag L_\nu 
    \right)
    \right)
    \, dt 
    \\ \nonumber
    &\quad
    + 
    \sum_\nu \sqrt{\frac{\eta_\nu}{\tau} }
    \Big( L_\nu \rho_t + \rho_t L_\nu^\dag  
    -
    \text{Tr}\left[
    L_\nu\rho_t+\rho_t L_\nu^\dag \right] \rho_t \Big)
    \, dW_{\nu t}
    \\
    &= 
    -i [H_0, \rho_t ] \, dt
    +
    \sum_\nu
    \left(
    \frac{1}{\tau}
    \mathcal{D}[L_\nu]\rho_t 
    \, dt
    +
    \sqrt{\frac{\eta_\nu}{\tau} }
    \mathcal{H}[L_\nu] \rho_t 
    \, dW_{\nu t}
    \right)
\end{align}
where $H_0$ stands for the Hamiltonian, $L_\nu$'s are related to the measurement Kraus operators, $\eta_\nu$'s are the efficiencies, $\tau$ is a measurement timescale, $dt$ is infinitesimal time, and $W_{\nu t}$ are independent Wiener processes.
Furthermore, for notational convenience, we use the Lindblad and measurement superoperators, defined as,
\begin{align}
    \mathcal{D}[L_\nu]\rho_t
    =
    L_\nu \rho_t L_\nu^\dag - \tfrac{1}{2}( L_\nu^\dag L_\nu \rho_t + \rho_t  L_\nu^\dag L_\nu)
    \\
    \mathcal{H}[L_\nu]\rho_t
    =
    L_\nu \rho_t + \rho_t L_\nu^\dag  - \text{Tr}\left[ L_\nu\rho_t+\rho_t L_\nu^\dag \right] \rho_t
\end{align}
The corresponding measurement output is,
\begin{equation} \label{eq:dy}
  dy_{\nu t}
  =
  \frac{\sqrt{\eta_\nu}}{\tau} \,
  \text{Tr}\left[ L_\nu\rho_t+\rho_t L_\nu^\dag \right] 
  \, dt 
  +
  \frac{1}{\sqrt{\tau}}
  dW_{\nu t}.
\end{equation}
where the same Wiener processes $\{W_{\nu t}\}$  are shared by the  state dynamics  and the classical measurement output.

\subsection{Sequential weak measurements}

Our sequential weak measurement schemes are derived from universal weak measurements \cite{oreshkov2005weak} of observables $\mathcal{O}$ with eigenvalues in the set $ \{1, -1\}$. We specifically consider measurements of the single-qubit Pauli operators $\sigma \in \{ X, Y, Z\}$. In the sequential measurement process, the measurements at each step are identical generalized measurements decribed by the set of measurement Kraus operators $\{ K^{\sigma}_{y} (x) \}$ where,
\begin{equation}
    K^{\sigma}_{y} (x)
    =
    \sqrt{
    \frac{1}{2}
    \left(
    \mathbb{I}
    + 
    y \tanh{(x)} 
    \sigma
    \right)
    }
\end{equation}
with binary measurement outputs $y \in \{\pm 1\}$ and measurement strength $x \in \mathbb{R}$, which is held fixed throughout the measurement process. Note, these are equivalent to the Kraus operators used in the main text, Eq.~\ref{eq:Kraus-O}, but in a more convenient form for our derivations. The statistics of the measurement output $y$ are,
\begin{align}
    p( y | \rho)
    &= 
    \frac{1}{2}
    \left(
    1
    + 
    y \tanh{(x)}
    \text{Tr}\left[
    \sigma \rho
    \right]
    \right)
    \\
    \mathbb{E}\left[ y | \rho \right]
    &=
    \tanh{(x)} 
    \text{Tr}\left[ \sigma \rho \right]
    \\
    \mathbb{E}\left[ y^2 | \rho \right]
    &=
    1
\end{align}
and the (single step) post measurement state is,
\begin{align}
    \rho^\prime
    &= 
    \frac{ K^{\sigma}_{y} \rho \left(K^{\sigma}_{y}\right)^\dag }
    {\text{Tr}\left[ K^{\sigma}_{y} \rho \left(K^{\sigma}_{y}\right)^\dag \right]}
\end{align}
For small measurement strength, these Kraus operators slightly bias the measurement outcomes to one of the eigenstate of $\sigma$ with small measurement feedback on the measured state. 
In the long run under multiple identical sequential measurements, they will converge to a projective measurement with the same statistics as a projective measurement of the  Pauli operator $\sigma$ (see \cite{oreshkov2005weak} for more details).

\subsection{Informationally incomplete weak measurements} 
\label{WML_1}

We first derive the Stochastic Master Equation (SME) in the context of an informationally incomplete weak measurement scheme, equivalent to universal weak measurements. Our construction is the sequence of weak Pauli measurements, described in the last section, and, without loss of generality, consider weak measurements of just the qubit Pauli-$Z$ operator, described by the set of Kraus operators $\{ K^{Z}_{y}(x) \}$. We start by expanding the discrete measurement evolution in the weak measurement limit. We define $\epsilon \coloneqq \frac{1}{2}\tanh(x) \ll 1$, then the discrete measurement evolution of a state $\rho$ up to $O(\epsilon^2)$ is, 
\begin{align}
    \nonumber
    \rho^\prime
    &= 
    \frac{ K^{Z}_{y} \rho \left(K^{Z}_{y}\right)^\dag }
    {\text{Tr}\left[ K^{Z}_{y} \rho \left(K^{Z}_{y}\right)^\dag \right]}
 =
    \frac{
    \left[ 
    \sqrt{
    \frac{1}{2}
    \left(
    \mathbb{I}
    + 2\epsilon y Z
    \right)
    }
    \right]
    \rho
    \left[ 
    \sqrt{
    \frac{1}{2}
    \left(
    \mathbb{I}
    + 2\epsilon y Z
    \right)
    }
    \right]^\dag
    }
    {
    \frac{1}{2}
    \left(
    1
    + 2\epsilon y
    \text{Tr}\left[ Z \rho \right]
    \right)
    }
    \\ \nonumber
    &=
    \left[
    \left( \mathbb{I} + \frac{1}{2}(2\epsilon y Z) - \frac{1}{8}(2\epsilon y Z)^2 + O\left(\epsilon^3\right) \right)
    \rho
    \left( \mathbb{I} + \frac{1}{2}(2\epsilon y Z) - \frac{1}{8}(2\epsilon y Z)^2 + O\left(\epsilon^3\right) \right)
    \right]
    \\ \nonumber
    &\quad\cdot 
    \left[
    1 - \left(2\epsilon y \text{Tr}\left[ Z \rho \right] \right) + \left(2\epsilon y \text{Tr}\left[ Z \rho \right] \right)^2
    + O\left(\epsilon^3\right)
    \right]
    \\ \nonumber
    &\approx 
    \left[
    \rho
    + \left( \epsilon y \right) \left( Z \rho + \rho Z \right)
    + \left( \epsilon y \right)^2 
    \left( Z \rho Z - \rho \right)
    \right]
    \cdot
    \left[
    1 - \left(2\epsilon y \text{Tr}\left[ Z \rho \right] \right) + \left(2\epsilon y \text{Tr}\left[ Z \rho \right] \right)^2
    \right]
    \\
    &\approx
    \rho
    +
    \left( \epsilon y \right)^2 
    \left( Z \rho Z - \rho \right) 
    +
    \left( 
    [ Z\rho + \rho Z ] 
    -
    2 \text{Tr}\left[ Z \rho \right] \rho 
    \right)
    \left(
    \left( \epsilon y \right)
    -
    \left( \epsilon y \right)^2 2 \text{Tr}\left[ Z \rho \right]
    \right)
\end{align}

\noindent 
where the evolved state $\rho^\prime$  depends on the scaled measurement output $\epsilon y$ with mean and variance,
\begin{align}
    \mathbb{E}\left[ \epsilon y | \rho \right]
    &= 
    2\epsilon^2
     \text{Tr}\left[ Z \rho \right]
    \\
    \mathbb{E}\left[ (\epsilon y)^2 | \rho \right]
    &= 
    \epsilon^2
    \\
    \text{Var}\left[ \epsilon y | \rho \right]
    &=
    \epsilon^2
    +
    O\left( \epsilon^4\right)
\end{align} 
Therefore, up to $O\left( \epsilon^2\right)$, we can make a Gaussian approximation of the (single step) scaled measurement output as
\begin{align}
    \epsilon y = 2\epsilon^2  \text{Tr}\left[ Z \rho \right] + \eta
\end{align}
where $\eta \sim \mathcal{N}(0, \epsilon^2)$ is random Gaussian noise. 
Furthermore, for $N$ sequential measurements ($\{ y_k : k \in [N] \}$ with corresponding i.i.d. noise $\{\eta_k\}$) in a time window of size $T$, the final discrete evolution on the state $\rho$ up to $O\left( \epsilon^2\right)$ only depends on the sum of the scaled measurement outputs $\sum_{i=1}^N \epsilon y_k $ with the moments,
\begin{align}
    \mathbb{E}\left[\left. \sum_{k=1}^N \epsilon y_k \right| \rho \right]
    &= 
    2N
    \epsilon^2
     \text{Tr}\left[ \sigma \rho \right]
    \\
    \mathbb{E}\left[\left. \left(\sum_{k=1}^N \epsilon y_k \right)^2 \right| \rho \right]
    &=
    \mathbb{E}\left[ \sum_{k=1}^N \left(\epsilon y_k \right)^2 \bigg| \rho \right]  
    +
    O(\epsilon^3)
    =
    N \epsilon^2 
    +
    O(\epsilon^3)
\end{align}
We can break up a physical time window $T$ into $N$ equal steps of $\Delta t$. Introducing a time scale $\tau$, we identify at $O\left( \epsilon^2\right)$,  $N \epsilon^2=\tfrac{T}{\tau}$. Hence $\epsilon^2 = \tfrac{\Delta t}{\tau} $ and the individual measurement noise $\eta_k = \tfrac{\Delta W_k}{\sqrt \tau},$ where $\Delta W_k \sim \mathcal{N}(0, \Delta t)$  are i.i.d Wiener increments. In this time window, then define an effective measurement output $\Delta y = \epsilon y$ and we can model the discrete measurement evolution of $\rho$ up to $O(\epsilon^2)$ by,
\begin{align}
    \Delta \rho
    =
    \rho^\prime - \rho
    &= \frac{1}{\tau}
    \left( Z \rho Z  - \rho \right) 
    \Delta t
    +
    \frac{1}{\sqrt\tau}
    \left( 
    \left[ Z\rho + \rho Z \right]
    - 
    2 \text{Tr}\left[ Z \rho \right] \rho
    \right)
    \Delta W
    \\
    \Delta y 
    &=
    \frac{2}{\tau}
    \text{Tr}\left[ Z \rho \right] 
    \Delta t
    +
    \frac{1}{\sqrt\tau} 
    \Delta W
\end{align}

\noindent
In the continuous time limit, $\Delta t \to 0 $, we get the Stochastic Master Equation for the density matrix $\rho$,
\begin{align}
    d \rho
    &= \frac{1}{\tau}
    \left(Z \rho Z  - \rho \right) dt
    +\frac{1}{\sqrt\tau}
    \left( 
    \left[ Z\rho + \rho Z \right]
    - 2 \text{Tr}\left[ Z \rho \right] \rho
    \right)
    dW
    \\
    dy &= \frac{2}{\tau} \text{Tr}\left[ Z \rho \right] dt + \frac{1}{\sqrt\tau}dW
\end{align}
where $dt$ is deterministic infinitesimal time, $dW$ is the stochastic Wiener differential, and with an explicit timescale $\tau$.

\subsection{ Informationally Complete Measurements}

Next we derive the SME in the informationally complete measurement scheme where we consider simultaneous weak measurements of the qubit Pauli $X$, $Y$, and $Z$ operators.
In this case, we model the measurement outcomes as vectors $\vec{y} \in \mathbb{R}^3$, where each component $y_\sigma\in \{ +1, 0, -1\}$ corresponds to a measurement outcome along the $\sigma$ axis. This measurement scheme is described by six Kraus operators 
\begin{align}
    K^{\sigma}_{\vec{y}} (x)
    =
    \sqrt{
    \frac{1}{6}
    \left(
    \mathbb{I}
    + \tanh{(x)} \vec{y} \cdot \vec{\sigma}
    \right)
    }
\end{align}
for all $\sigma \in \{X, Y, Z\}$ and the corresponding measurement outputs are $\vec{y} \in \{ \pm \hat{e}_x, \pm \hat{e}_y, \pm \hat{e}_z \}$ where  $\hat{e}_\sigma$ is the unit vector in the $+\sigma$ direction.

Similar to the previous case, in the weak measurement limit, let $\epsilon = \frac{1}{2}\tanh(x) \ll 1$. Then the discrete measurement evolution of a state $\rho$ up to $O(\epsilon^2)$ is, 
\begin{align}
    \nonumber
    \rho^\prime 
    &=
    \frac{ K^{\sigma}_{\vec{y}} \rho \left(K^{\sigma}_{\vec{y}}\right)^\dag }
    {\text{Tr}\left[ K^{\sigma}_{\vec{y}} \rho \left(K^{\sigma}_{\vec{y}}\right)^\dag \right]}\\ \nonumber
    &=
    \frac{
    \left[ 
    \sqrt{
    \frac{1}{6}
    \left(
    \mathbb{I}
    + \tanh{(x)} \vec{y} \cdot \vec{\sigma}
    \right)
    }
    \right]
    \rho
    \left[ 
    \sqrt{
    \frac{1}{6}
    \left(
    \mathbb{I}
    + \tanh{(x)} \vec{y} \cdot \vec{\sigma}
    \right)
    }
    \right]^\dag
    }
    {
    \frac{1}{6}
    \left(
    1
    + 
    \tanh{(x)} 
    \text{Tr}\left[ \vec{y} \cdot \vec{\sigma} \rho \right]
    \right)
    }
    =
    \frac{
    \left(
    \mathbb{I}
    + 
    2\epsilon \vec{y} \cdot \vec{\sigma}
    \right)^\frac{1}{2}
    \rho
    \left(
    \mathbb{I}
    + 2\epsilon \vec{y} \cdot \vec{\sigma}
    \right)^\frac{1}{2}
    }
    {
    \left(
    1
    + 2\epsilon 
    \text{Tr}\left[ \vec{y} \cdot \vec{\sigma} \rho \right]
    \right)
    }
    \\ \nonumber
    &=
    \left[
    \left( 
    \mathbb{I} 
    +
    \frac{1}{2}
    \left(2\epsilon \vec{y} \cdot \vec{\sigma}\right) 
    - 
    \frac{1}{8}
    \left( 2\epsilon \vec{y} \cdot \vec{\sigma} \right)^2
    +
    O\left( \epsilon^3 \right) 
    \right)
    \rho
    \left( 
    \mathbb{I} 
    + 
    \frac{1}{2}
    \left(2\epsilon \vec{y} \cdot \vec{\sigma} \right) 
    -
    \frac{1}{8}
    \left( 2\epsilon \vec{y} \cdot \vec{\sigma} \right)^2
    +
    O\left( \epsilon^3 \right) 
    \right)
    \right]
    \\ \nonumber
    &\quad\cdot 
    \left[
    1 
    - 
    \left(2\epsilon \text{Tr}\left[ \vec{y} \cdot \vec{\sigma} \rho \right] \right) 
    + 
    \left(2\epsilon \text{Tr}\left[ \vec{y} \cdot \vec{\sigma} \rho \right] \right)^2
    +
    O\left( \epsilon^3 \right) 
    \right]
    \\ \nonumber
    &=
    \left[
    \left( 
    \mathbb{I} 
    +
    \left( \epsilon \vec{y} \cdot \vec{\sigma} \right) 
    - 
    \frac{1}{2}(\epsilon)^2
    +
    O\left( \epsilon^3 \right) 
    \right)
    \rho
    \left( 
    \mathbb{I} 
    + 
    \left(\epsilon \vec{y} \cdot \vec{\sigma} \right) 
    -
    \frac{1}{2}(\epsilon)^2
    +
    O\left( \epsilon^3 \right) 
    \right)
    \right]
    \\ \nonumber
    &\quad\cdot 
    \left[
    1 
    - 
    \left(2\epsilon \text{Tr}\left[ \vec{y} \cdot \vec{\sigma} \rho \right] \right) 
    + 
    \left(2\epsilon \text{Tr}\left[ \vec{y} \cdot \vec{\sigma} \rho \right] \right)^2
    +
    O\left( \epsilon^3 \right) 
    \right]
    \\ \nonumber
    &\approx 
    \left[
    \rho
    + 
    \left( \epsilon \right) 
    \left(
    \left(\vec{y} \cdot \vec{\sigma} \right)  \rho + 
    \rho \left(\vec{y} \cdot \vec{\sigma} \right)  \right)
    + 
    \left( \epsilon \right)^2 
    \left(
    \left( \vec{y} \cdot \vec{\sigma} \right)  \rho \left( \vec{y} \cdot \vec{\sigma} \right) 
    -
    \rho
    \right)
    \right]
    \cdot
    \left[
    1 - \left(2\epsilon \text{Tr}\left[ \vec{y} \cdot \vec{\sigma} \rho \right] \right) + \left(2\epsilon \text{Tr}\left[ \vec{y} \cdot \vec{\sigma} \rho \right] \right)^2
    \right]
    \\ \nonumber
    &\approx 
    \rho
    +
    \left( \epsilon \right)^2
    \left( 
    \left( \vec{y} \cdot \vec{\sigma} \right)  \rho \left( \vec{y} \cdot \vec{\sigma} \right) 
    -
    \rho
    \right)
    \\ \nonumber
    &\quad
    +
    \left( \epsilon \right)
    \left(
    \left(\vec{y} \cdot \vec{\sigma} \right)  \rho + \rho \left(\vec{y} \cdot \vec{\sigma} \right)  
    - 2
    \text{Tr}\left[ \vec{y} \cdot \vec{\sigma} \rho \right]
    \rho
    \right)
    -
    \left( \epsilon \right)^2
    2 \text{Tr}\left[ \vec{y} \cdot \vec{\sigma} \rho \right]
    \left(
    \left(
    \vec{y} \cdot \vec{\sigma} \right)  \rho + 
    \rho \left(\vec{y} \cdot \vec{\sigma}
    \right)
    -
    2 \text{Tr}\left[ \vec{y} \cdot \vec{\sigma} \rho \right]
    \rho
    \right)
    \\
    &=
    \rho
    +
    (\epsilon)^2
    \left( 
    \left( \vec{y} \cdot \vec{\sigma} \right)  \rho \left( \vec{y} \cdot \vec{\sigma} \right) 
    -
    \rho
    \right)
    +
    \left(
    \left[ \left(\vec{y} \cdot \vec{\sigma} \right)  \rho + \rho \left(\vec{y} \cdot \vec{\sigma} \right)  \right]
    -
    2
    \text{Tr}\left[ \vec{y} \cdot \vec{\sigma} \rho \right] \rho
    \right)
    \left(
    (\epsilon) - (\epsilon)^2 2 \text{Tr}\left[ \vec{y} \cdot \vec{\sigma} \rho \right]
    \right)
\end{align}

\noindent
We know that $\mathbb{E}[y_i y_j|\rho ]= \frac{1}{3} \delta_{ij}+O(\epsilon^2)$ since a measurement output vector $\vec{y}$ only has one nonzero value, then
\begin{align}
    \nonumber
    \rho^\prime 
    &=
    \rho
    +
    (\epsilon)^2
    \left( 
    \sum_{ij} y_i y_j
    \sigma_i  \rho \sigma_j
    -
    \rho
    \right)
    +
    \sum_{ij}
    \left(
    \left[ \sigma_i \rho + \rho \sigma_i  \right]
    -
    2
    \text{Tr}\left[ \sigma_i \rho \right] \rho
    \right)
    \left(
    (\epsilon) y_i - 2 (\epsilon)^2 y_i y_j \text{Tr}\left[ \sigma_j \rho \right]
    \right)
    \\
    &\approx
    \rho
    +
    \frac{\epsilon^2}{3}
    \sum_{i}
    \left( 
    \sigma_i  \rho \sigma_i
    -
    \rho
    \right)
    +
    \sum_{i}
    \left(
    \left[ \sigma_i \rho + \rho \sigma_i  \right]
    -
    2
    \text{Tr}\left[ \sigma_i \rho \right] \rho
    \right)
    \left(
    \epsilon y_i 
    -
    \frac{\epsilon^2}{3}  2 \text{Tr}\left[ \sigma_i \rho \right]
    \right)
\end{align}

\noindent
Following similar definitions as in the informationally incomplete measurements case in the previous Appendix section \ref{WML_1}, let $\frac{\epsilon^2}{3} = \tfrac{\Delta t}{\tau}$ and $ \Delta y_i = \epsilon y_i=
    \tfrac{\epsilon^2}{3}  2\text{Tr}\left[ \sigma_i \rho \right]+\eta_i
    $ where $\vec{\eta}\sim \mathcal{N}(0,\tfrac{\epsilon^2}{3}I)$. Introducing  $\Delta W_i \sim \mathcal{N}\left( 0 , \Delta t\right)$, the Wiener increment, with the components $i=1,2,3$ being independent, we can model the discrete measurement evolution of $\rho$ up to $O(\epsilon^2)$ by,
\begin{align}
    \Delta \rho
    &=\frac{1}{\tau}
    \sum_{i}
    \left( 
    \sigma_i  \rho \sigma_i
    -
    \rho
    \right)
    \Delta t
    +\frac{1}{\sqrt\tau}
    \sum_{i}
    \left(
    \left[ \sigma_i \rho + \rho \sigma_i  \right]
    -2
    \text{Tr}\left[ \sigma_i \rho \right] \rho
    \right)
    \Delta W_i
    \\
    \Delta y_i 
    &=
    \frac{2}{\tau} \text{Tr}\left[ \sigma_i \rho \right] \Delta t
    +
    \frac{1}{\sqrt\tau}\Delta W_i 
\end{align}
In the continuous time limit $\Delta t \to 0 $, we get the Stochastic Master Equation SME for the density matrix $\rho$,
\begin{align}
    d\rho
    &=\frac{1}{\tau}
    \sum_{i}
    \left( 
    \sigma_i  \rho \sigma_i
    -
    \rho
    \right)
    dt
    +\frac{1}{\sqrt\tau}
    \sum_{i}
    \left(
    \left[ \sigma_i \rho + \rho \sigma_i  \right]
    -
    2
    \text{Tr}\left[ \sigma_i \rho \right] \rho
    \right)
    dW_i
    \\
    dy_i 
    &=
     \frac{2}{\tau} \text{Tr}\left[ \sigma_i \rho \right]dt + \frac{1}{\sqrt\tau}dW_i
\end{align}
where $dt$ is deterministic infinitesimal time and $dW_i$ are independent stochastic Wiener differentials.

\subsection{ Informationally Incomplete Weak Measurements with Unitary Dynamics }

Finally, we derive the SME with unitary dynamics. Again, consider weak measurements of the qubit Pauli-$Z$ operator, the same as the informationally incomplete measurements in Appendix section \ref{WML_1}, but now under a transverse magnetic field. The weak measurements are described by the set of Kraus operators $\{ K^{Z}_{y}(x) \}$ and the transverse magnetic fields is described by the unitary time evolution operator $U(\phi) = \exp{\left(- i \Delta t \left[ \frac{\omega}{2} X \right] \right)} = \exp{\left(- i \frac{\phi}{2} X \right)}$ where we define the precession angle, the angle of rotation about the $x$-axis on the Bloch sphere, over time $\Delta t$ to be $\phi := \omega\Delta t$. This mixed dynamics can be effectively described by new Kraus operators  $\tilde{K}^{Z}_{y}(x, \phi) = K^{Z}_{y}(x) U(\phi)$.

In the weak measurement limit, let $\epsilon = \frac{1}{2}\tanh(x) \ll 1$ and $\phi \ll 1$, then we look at the discrete measurement evolution of a state $\rho$ up to $O(\epsilon^2)$ and  $O(\phi)$.
We can consider the measurement and unitary dynamics sequentially since,
\begin{align}
    \rho^\prime
    =
    \frac{ 
    \tilde{K}^{Z}_{y}(x, \phi ) 
    \rho
    \tilde{K}^{Z}_{y}(x, \phi )^\dag }
    {
    \text{Tr}\left[ 
    \tilde{K}^{Z}_{y}(x, \phi ) 
    \rho
    \tilde{K}^{Z}_{y}(x, \phi )^\dag \right]
    }
    =
    \frac{ 
    K^{Z}_{y}(x) 
    \left[ U(\phi) \rho U(\phi)^\dag \right]
    K^{Z}_{y}(x)^\dag
    }
    {
    \text{Tr}\left[ 
    K^{Z}_{y}(x) 
    \left[ U(\phi) \rho U(\phi)^\dag \right]
    K^{Z}_{y}(x)^\dag
    \right]
    }
    =
    \frac{ 
    K^{Z}_{y}(x) 
    \rho(\phi)
    K^{Z}_{y}(x)^\dag
    }
    {
    \text{Tr}\left[ 
    K^{Z}_{y}(x) 
    \rho(\phi)
    K^{Z}_{y}(x)^\dag
    \right]
    }
\end{align}
Therefore, up to $O(\epsilon^2)$,
\begin{align}
    \rho^\prime
    &=
    \rho(\phi)
    +
    \left( y \epsilon \right)^2 
    \left( Z \rho(\phi) Z - \rho(\phi) \right) 
    +
    \left( 
    [ Z\rho(\phi) + \rho(\phi) Z ] 
    -
    2 \text{Tr}\left[ Z \rho(\phi) \right] \rho(\phi)
    \right)
    \left(
    \left( y \epsilon \right)
    -
    \left( y \epsilon \right)^2 2 \text{Tr}\left[ Z \rho(\phi) \right]
    \right)
\end{align}
and up to $O(\phi)$,
\begin{align}
    \rho^\prime
    &=
    \left( \rho + \left( - \frac{1}{2} i \phi  \right)
    \left[ X,\rho \right] \right)
    +
    \left( y \epsilon \right)^2 
    \left( Z \rho Z - \rho \right) 
    +
    \left( 
    [ Z\rho(\tau) + \rho Z ] 
    -
    2 \text{Tr}\left[ Z \rho \right] \rho
    \right)
    \left(
    \left( y \epsilon \right)
    -
    \left( y \epsilon \right)^2 2 \text{Tr}\left[ Z \rho \right]
    \right)
\end{align}
where we assumed $\epsilon$ and $\phi$ are related so their products are higher order terms and this will be justified shortly.

Using the results in Appendix section \ref{WML_1}, 
with $\frac{\phi}{x^2} \coloneqq a = \frac{\omega\tau}{4}$ kept fixed as we take the continuous time limit,
then we can model the discrete measurement evolution of $\rho$ up to $O(\epsilon^2)$ and $O(\phi)$ by,
\begin{align}
    \Delta \rho
    &= 
    - i \frac{\omega}{2} [X,\rho] \Delta t
    +
    \frac{1}{\tau}
    \left(Z \rho Z  - \rho \right) \Delta t
    +
    \frac{1}{\sqrt{\tau}}
    \left( 
    \left[ Z\rho + \rho Z \right]
    - 2 \text{Tr}\left[ Z \rho \right] \rho
    \right)
    \Delta W
    \\
    \Delta y 
    &=
    \frac{2}{\tau}
    \text{Tr}\left[ Z \rho \right] \Delta t
    +
    \frac{1}{\sqrt{\tau}} \Delta W
\end{align}
In the continuous time limit $\Delta t \to 0 $, we get the Stochastic Master Equation for the density matrix $\rho$,
\begin{align}
    d \rho
    &= 
    -i\frac{\omega}{2}[X,\rho] dt
    +
    \frac{1}{\tau}
    \left(Z \rho Z  - \rho \right) dt
    +
    \frac{1}{\sqrt{\tau}}
    \left( 
    \left[ Z\rho + \rho Z \right]
    - 2 \text{Tr}\left[ Z \rho \right] \rho
    \right)
    dW
    \\
    dy 
    &= 
    \frac{2}{\tau}
    \text{Tr}\left[ Z \rho \right] dt 
    +
    \frac{1}{\sqrt{\tau}}
    dW
\end{align}
where $dt$ is deterministic infinitesimal time, $dW$ is the stochastic Wiener differential, and $\tau$ is an explicit timescale.

\section{\label{sec:inefficient_mearurement} SME Dynamics with Inefficient Measurements}

Here we extend our weak measurement limit SMEs to the setting where we have inefficient or noisy measurements. 
Following \cite{ROUCHON2022252}, we define our inefficient measurements to be characterized by a error kernel $\Pr[y|b]$, the probability of showing the measurement outcome to be $y$, when the `true' measurement outcome is $b$.
Ultimately, the SMEs will include an `efficiency' parameter $\eta$ which characterizes the amount of inefficiency or noise in the measurement output.

Consider a weak generalized measurement with $n$ measurement outputs and with corresponding Kraus operators $\{ K_b \}$. 
Then, to derive our error kernel, we use a symmetric noise model for the noisy measurement outputs, where with probability $p_{success}$ we get the correct output and, otherwise, with (equal) probability $\frac{1}{n-1}(1-p_{success})$ we get any of the other $n-1$ possible outputs. Furthermore we can parameterize $p_{success}$ in terms of an efficiency parameter $\sqrt{\eta}$ such that
$
p_{success}
=
\frac{1 + (n-1) \sqrt{\eta} }{n}
$.
This defines our error kernel or noise matrix as,
\begin{align}
    \Pr[y|b]
    =
    \beta_{yb} 
    =
    \frac{1}{n} (1- \sqrt{\eta} ) + \sqrt{\eta} \delta_{yb}
\end{align}

Next, with this noise model for our measurement outputs we want to expand the discrete time noisy dynamics in the weak measurement limit. By Bayes rule, the noisy measurements lead to the post measurement state evolution defined by,
\begin{gather}
    \rho' 
    =
    \frac{ \mathcal{E}_y\left( \rho  \right) }
    { \text{Tr}\left[ \mathcal{E}_y\left( \rho  \right) \right]}
    \\
    y 
    \sim 
    \Pr[y| \rho ]
    =
    \text{Tr}\left[ \mathcal{E}_y\left( \rho  \right) \right]
\end{gather}
where we define the superoperators,
$
\mathcal{E}_y\left( \rho  \right)
=
\sum_{b}
\beta_{yb} K_b \rho K^\dag_b
$ 
and 
$
\mathcal{E}\left( \rho  \right)
=
\sum_{y}\mathcal{E}_y\left( \rho \right)
$.

We now specialize to the case in $y$ taking values in the set $\V:=\{\alpha_b\}_{b=1}^n\subset \mathbb{R}^D$, and $K_b:=\sqrt{\tfrac{1}{n}\big(\mathbb I+\tanh(x)\alpha_b\cdot\sigma\big)}=\sqrt{\tfrac{1}{n}\big(\mathbb I+2\epsilon\alpha_b\cdot\sigma\big)}$. We choose  $||\alpha_b||=1, \forall b$,$\sum_b\alpha_b=0$, and $\tfrac{1}{n}\sum_b\alpha_b\alpha_b^\dag=\tfrac{1}{D}\mathbb I_D$, where $\mathbb I_D$ is the $D\times D$ identity matrix, in contrast to $\mathbb I:=\mathbb I_2$. $D=3, n=6$ corresponds to model I, with $\sigma=(X, Y, Z)$ and $D=1, n=2$ corresponds to model II with $\sigma=(Z)$.

Before expanding, we can save some time by noticing that,
\begin{align}
    \nonumber
    \Pr[ y| \rho ]
    =
    \text{Tr}\left[ \mathcal{E}_y\left( \rho_{}  \right) \right]
    &=
    \sum_{b}
    \beta_{yb} \text{Tr}\left[ K_b \rho K^\dag_b \right]
    \\
    \nonumber
    &=
    \sum_{b}
    \beta_{yb}
    \left(
    \frac{1}{n}\left( 1 + 2\epsilon\alpha_b\cdot\text{Tr}\left[ \sigma \rho \right] \right)
    \right)
    \\
    &=
    \frac{1}{n}\left( 1 + 2\epsilon
    \bar{\alpha}(y)
     \text{Tr}\left[ \sigma \rho \right] \right)
     \label{eq:inefficient-measurement-prob}
\end{align}
where we define the (pseudo-)mean measurement output $\bar{\alpha}(y) \coloneqq \sum_b \beta_{yb} \alpha_b=\mathbb{E}_{b \sim \beta_{y\cdot}}\left[ \alpha_b \right]$. Since $\beta$ is a symmetric kernel, we can pretend that for each $y$, there is a distribution over $b$. We can also define $Q(y):=\mathbb{E}_{b \sim \beta_{y\cdot}}\left[ \alpha_b \alpha_b^\dag\right]$. This probability in Eq. \ref{eq:inefficient-measurement-prob} has a similar form to those in our original noiseless SME derivations.

Now, without loss of generality, we follow the treatment for informationally incomplete weak measurements, without unitary dynamics, setting in Appendix section \ref{WML_1}. There, we derived the noiseless post measurement evolution, up to order $O\left( \epsilon^2 \right)$, as
$
K^{\sigma}_{y} \rho \left(K^{\sigma}_{y}\right)^\dag
\approx 
\rho
+ \left( \epsilon y \right) \left( \sigma \rho + \rho \sigma \right)
+ \left( \epsilon y \right)^2 
\left( \sigma \rho \sigma - \rho \right)
$. In our general setting, extending to inefficient  measurement outputs, we get,
\begin{align}
    \nonumber
    \mathcal{E}_y\left( \rho  \right)
    &=
    \sum_{b}
    \beta_{yb} K_b \rho K^{ \dag}_b
    \\
    \nonumber
    &\approx 
    \sum_{b}
    \beta_{yb}
    \left[ 
    \rho
    +  \epsilon  \left((\alpha_b\cdot \sigma) \rho + \rho (\alpha_b\cdot\sigma) \right)
    + \epsilon^2 
    \left((\alpha_b\cdot \sigma )\rho (\alpha_b\cdot\sigma) - \rho \right)
    \right]
    \\
    &=
    \rho
    + 
    \epsilon  \left((\bar{\alpha}(y)\cdot \sigma) \rho + \rho (\bar{\alpha}(y)\cdot\sigma) \right)
    + 
    \epsilon^2 \left(
    \sum_{ij}
     Q_{ij}(y) \sigma_i \rho \sigma_j - \rho \right)
\end{align}
Explicitly evaluating these expectation values we get,
\begin{gather}
    \bar{\alpha}(y)
    \coloneqq 
    \mathbb{E}_{b \sim \beta_{y\cdot}}\left[ \alpha_b \right]
    =
    \sqrt{\eta} y
    \\
    Q_{ij}(y)
    \coloneqq 
    \mathbb{E}_{b \sim \beta_{y\cdot}}\left[ \alpha_{bi}\alpha_{bj} \right]
    =\frac{1-\sqrt\eta}{D}\delta_{ij}+\sqrt\eta y_iy_j
\end{gather}

Putting all this together, the discrete time inefficient measurement dynamics in the weak measurement limit is,
\begin{align}
    \nonumber
     \mathcal{E}_y\left( \rho  \right)
    &=
     \sum_{b} \beta_{yb} K_b \rho K^\dag_b 
    \\
    \nonumber
    &\approx\frac{1}{n}
    \left[
    \rho
    + \epsilon \sqrt{\eta}\sum_i y_i  \left( \sigma_i  \rho + \rho \sigma_i  \right)
    + \epsilon^2 
    \left(\sum_{ij} \left( \frac{1-\sqrt\eta}{D}\delta_{ij}+\sqrt\eta y_iy_j\right)\sigma_i \rho \sigma_j  - \rho \right)
    \right]
    \\%
     &=\frac{1}{n}\left[
    \rho
    + \epsilon \sqrt{\eta}\sum_i y_i  \left( \sigma_i  \rho + \rho \sigma_i  \right)
    + \frac{\epsilon^2}{D} 
    \sum_{i} \left(\sigma_i \rho \sigma_i  - \rho \right)
    \right],
\end{align}
where, in the last step, we replaced $y_iy_j$ by its approximate expectation value $\tfrac{1}{D}\delta_{ij}$, as usual.

\begin{align}
    \nonumber
    \rho'
    &=
    \frac{ \mathcal{E}_y\left( \rho  \right) }
    { \text{Tr}\left[ \mathcal{E}_y\left( \rho  \right) \right]}
    =
    \frac{ \sum_{b} \beta_{yb} K_b \rho K^\dag_b }
    { 
    \frac{1}{n}\left( 1 +
    2\epsilon\sqrt\eta y\cdot \text{Tr}\left[ \sigma \rho \right] \right)
    }
    \\
    \nonumber
    &\approx
    \left[
    \rho
    + \epsilon \sqrt{\eta}\sum_i y_i  \left( \sigma_i  \rho + \rho \sigma_i  \right)
    + \frac{\epsilon^2}{D} 
    \sum_{i} \left(\sigma_i \rho \sigma_i  - \rho \right)
    \right]
    \\
    &\cdot
    \left[
    1 - 2\epsilon \sqrt\eta \sum_jy_j \text{Tr}\left[ \sigma_j  \rho \right] + 4\epsilon^2 \eta \left(\sum_jy_j \text{Tr}\left[ \sigma_j  \rho \right] \right)^2
    \right]
\end{align}

Continuing keeping terms up to $O(\epsilon^2)$ and replacing $y_iy_j$ by $\tfrac{1}{D}\delta_{ij}$,

\begin{align}
    \nonumber
    \rho'
    &\approx
    \left[
    \rho
    + \epsilon \sqrt{\eta}\sum_i y_i  \left( \sigma_i  \rho + \rho \sigma_i  \right)
    + \frac{\epsilon^2}{D} 
    \sum_{i} \left(\sigma_i \rho \sigma_i  - \rho \right)
    \right]
    \\
    \nonumber
    &\cdot
    \left[
    1 - 2\epsilon \sqrt\eta \sum_jy_j \text{Tr}\left[ \sigma_j  \rho \right] + \frac{4\epsilon^2 \eta }{D}\sum_j \left(\text{Tr}\left[ \sigma_j  \rho \right] \right)^2
    \right]
    \\%
    \nonumber
    &\approx\rho
    + \epsilon \sqrt{\eta}\sum_i y_i  \left( \sigma_i  \rho + \rho \sigma_i  \right)
    + \frac{\epsilon^2}{D} 
    \sum_{i} \left(\sigma_i \rho \sigma_i  - \rho \right)
    \\
    \nonumber    &-2\epsilon\sqrt\eta\sum_iy_i\text{Tr}\left[\sigma_i\rho\right]\rho+ \frac{4\epsilon^2 \eta }{D}\sum_i \left(\text{Tr}\left[ \sigma_i  \rho \right] \right)^2\rho-2\epsilon^2\eta\sum_{ij} y_i y_j \left( \sigma_i  \rho + \rho \sigma_i  \right) \text{Tr}\left[ \sigma_j  \rho \right] 
        \\%
    \nonumber
    &\approx\rho
    + \epsilon \sqrt{\eta}\sum_i y_i  \left( \sigma_i  \rho + \rho \sigma_i  \right)
    + \frac{\epsilon^2}{D} 
    \sum_{i} \left(\sigma_i \rho \sigma_i  - \rho \right)
    \\
    \nonumber  &-2\epsilon\sqrt\eta\sum_iy_i\text{Tr}\left[\sigma_i\rho\right]\rho+ \frac{4\epsilon^2 \eta }{D}\sum_i \left(\text{Tr}\left[ \sigma_i  \rho \right] \right)^2\rho-\frac{2\epsilon^2\eta}{D}\sum_i \left( \sigma_i  \rho + \rho \sigma_i  \right) \text{Tr}\left[ \sigma_i  \rho \right]
        \\%
    &=\rho
    +  \sqrt{\eta}\sum_i  \left(\left( \sigma_i  \rho + \rho \sigma_i  \right)-2\text{Tr}\left[ \sigma_i  \rho \right]\right)\left(\epsilon y_i-\frac{2\epsilon^2\sqrt\eta}{D}\text{Tr}\left[ \sigma_i  \rho \right]\right)
        + \frac{\epsilon^2}{D} 
    \sum_{i} \left(\sigma_i \rho \sigma_i  - \rho \right).
\end{align}

With the identifications $\Delta\rho:=\rho^\prime-\rho$, $\tfrac{\Delta t}{\tau}:=\tfrac{\epsilon^2}{D}$, $\tfrac{\Delta W_i}{\sqrt\tau}:=\epsilon y_i-\frac{2\epsilon^2\sqrt\eta}{D}\text{Tr}\left[ \sigma_i  \rho \right]$, we get

\begin{align}
    \Delta \rho
    &=
    \frac{1}{\tau}
    \sum_i\left(\sigma_i \rho \sigma_i - \rho \right)
    \, \Delta t
    +
    \sqrt{ \frac{\eta}{\tau} } \sum_i
    \left( 
    \left[ \sigma_i  \rho + \rho \sigma_i  \right]
    - 
    2 \text{Tr}\left[ \sigma_i  \rho \right] \rho
    \right)
    \, \Delta W_i
    \\
    \Delta y_i 
    &=
    2 \frac{\sqrt{\eta}}{\tau}
    \text{Tr}\left[ \sigma_i  \rho \right] 
    \, \Delta t 
    +
    \frac{1}{\sqrt\tau}
    \, \Delta W_i
\end{align}

Specializing to the case $D=3$, we get the SME for the inefficient measurement version of model I, and, taking $D=1$ case gives us SME for model II with inefficient measurement, without the magnetic field term (which can be added easily).

\section{\label{sec:MI_in_weak_noise_limit} Mutual Information in the Small Efficiency Limit} 

Here, we present a perturbative approach to estimate the mutual information plateau of the weak measurement limit. We use the generalization of our continuous time SME measurement models with the additional control parameter --- the measurement efficiency $\eta$ --- to explicitly calculate the leading contribution to this mutual information.

\subsection{ Expansion in the efficiency parameter }

The continuous time dynamics of a quantum state in the weak measurement limit with noisy measurements is given generally by the SME \ref{eq:SME} and measurement output \ref{eq:dy}. Without loss of generality, consider this general SME, including an explicit timescale $\tau$, for a single measurement channel,
\begin{align}
    d \rho_t
    &= 
    -i
    [H, \rho_t ] \, dt
    +
    \frac{1}{\tau}
    \mathcal{D}[L]\rho_t \, dt
    +
    \sqrt{\frac{\eta}{\tau}} \,
    \mathcal{H}[L]\rho_t \, dW_{t}
    \\
    dy
    &= 
    \frac{\sqrt{\eta}}{\tau} \, 
    \text{Tr}\left[ \rho_t \left( L + L ^\dag \right) \right] \, dt
    +
    \frac{1}{\sqrt{\tau}}
    dW_{t}
\end{align}
where $L$ is related to the measurement operator, $\eta$ is the measurement efficiency parameter, and $\mathcal{D}[L]$ and $\mathcal{H}[L]$ are the Lindblad and measurement superoperators acting on $\rho_t$ respectively.

Now consider the state $\rho_t\left( \sqrt{\eta} \right)$ as a power series in $\sqrt{\eta}$, as
$
\rho_t\left( \sqrt{\eta} \right)
=
\rho_t^{(0)}
+
\left( \sqrt{\eta} \right) \rho_t^{(1)}
+
\left( \sqrt{\eta} \right)^2 \rho_t^{(2)}
+
O\left(  \left( \sqrt{\eta} \right)^3 \right)
$. Then we can get the state dynamics in orders of efficiency $d\rho_t^{(n)}$, where $n$ denotes the $n$-th order approximation, as
\begin{align}
    \nonumber
    d \rho_t
    &=
    -i
    \left[ 
    H, 
    \left( \rho_t^{(0)} +  \sqrt{\eta}\rho_t^{(1)} + \dots \right)
    \right] \, dt
    +
    \frac{1}{\tau}
    \mathcal{D}[L]
    \left( \rho_t^{(0)} +  \sqrt{\eta}\rho_t^{(1)} + \dots \right)
    \, dt
     \\ \nonumber
    &+
    \sqrt{\frac{\eta}{\tau}} \,
    \mathcal{H}[L]
    \left( \rho_t^{(0)} +  \sqrt{\eta}\rho_t^{(1)} + \dots \right)
    \, dW_{t}
    \\
    \label{eq:d_rho_expansion}
    &=
    -i 
    \left[ H, \rho_t^{(0)} \right] 
    \, dt
    +
    \frac{1}{\tau}
    \mathcal{D}[L] \rho_t^{(0)}
    \, dt
    \nonumber
    \\ 
    &+
    \sqrt{\eta}
    \left(
    -i \left[ H, \rho_t^{(1)} \right] \, dt
    +
    \frac{1}{\tau}
    \mathcal{D}[L] \rho_t^{(1)} \, dt
    +
    \frac{1}{\sqrt{\tau}}
    \mathcal{H}[L] \rho_t^{(0)}
    \, dW_{t}
    \right)
    +
    O\left(  \left( \sqrt{\eta} \right)^2 \right)
\end{align}
and the measurement dynamics in orders of efficiency $dy^{(n)}$ as,
\begin{align}
    \nonumber
    dy
    &=
    \frac{\sqrt{\eta}}{\tau} \,
    \text{Tr}\left[ 
    \left( \rho_t^{(0)} +  \sqrt{\eta}\rho_t^{(1)} + \dots \right)
    \left( L + L^\dag \right) \right] \, dt
    +
    \frac{1}{\sqrt{\tau}}
    dW_{t}
    \\
    \label{eq:d_y_expansion}
    &=
    \frac{1}{\sqrt{\tau}}
    dW_{t}
    +
    \frac{\sqrt{\eta}}{\tau}
    \left(
    \text{Tr}\left[ 
    \rho^{(0)}
    \left( L + L^\dag \right) \right] \, dt
    \right)
    +
    O\left(  \left( \sqrt{\eta} \right)^2 \right)
\end{align}
Therefore, the evolution of the state $\rho_t$ to zeroth-order in efficiency, $\rho^{(0)}_t$, is just described by the standard Lindblad evolution of $\rho_t$, out of Eq.~\eqref{eq:d_rho_expansion}, as,
\begin{align} \label{eq:Lindblad}
    \nonumber
    d\rho^{(0)}_t  
    &=
    -i 
    \left[ H, \rho^{(0)} \right] 
    \, dt
    +
    \frac{1}{\tau}
    \mathcal{D}[L] \rho^{(0)}
    \, dt
    \\
    &=
    - 
    i 
    \Big[H_0,\rho^{(0)}_t\Big] 
    \, dt  
    + 
    \frac{1}{\tau}
    \left(
    L \rho^{(0)}_t L^\dag 
    -
    \tfrac{1}{2}
    \left(
    L^\dag L \rho^{(0)}_t + \rho^{(0)}_t  L^\dag L
    \right)
    \right)
    \, dt.
\end{align}
and, the measurement dynamics to first-order in efficiency, $dy^{(1)}_{t}$, out of Eq.~\eqref{eq:d_y_expansion}, as,
\begin{equation} \label{eq:measurement_output_first_order}
  dy^{(1)}_{t}
  =
  \frac{\sqrt{\eta}}{\tau}
  \text{Tr} \left[ L \rho^{(0)}_t+\rho^{(0)}_t L ^\dag \right] \,
  dt 
  +
  \frac{1}{\sqrt{\tau}}
  dW_{ t}
\end{equation} 

If the Lindblad dynamics $d\rho^{(0)}_t$ is solvable, then we can plug it into the first-order measurement dynamics $dy^{(1)}_{t}$
and solve for the corresponding approximate measurement output. In the next section, we will see that in the small efficiency limit, these approximate dynamics allow us to estimate the mutual information between the initial state distribution $S$ and the approximate measurement output!

\subsection{Mutual information
via bi-AWGN channel mapping \label{sec:small_efficiency_bi-AWGN}}

In the small efficiency limit, $\sqrt{\eta} \ll 0$, we can approximate the measurement SME dynamics in small orders of efficiency. Then, we can use the solutions to the approximate dynamics (Eqs.~\ref{eq:Lindblad} and \ref{eq:measurement_output_first_order}) to estimate the mutual information between the initial state distribution $S$ and the measurement record $A_{1:T}$. Furthermore, we can get an explicit expression for this estimate in terms of the solutions to the zeroth-order Lindblad dynamics   
using result for the binary input additive white Gaussian noise (bi-AWGN) channel (summarized below in \ref{sec:bi-AWGN_MI} and also Ref. \cite{choi2006adaptive}). 

First, for a fixed time window $T$ scaled by an explicit time scale $\tau$, define $N$ intervals of size $ \frac{ \Delta t }{\tau} \coloneqq \frac{ 1 }{\tau} \frac{T}{N}$. Then we can re-discretize the continuous measurement output in this window, to get a discrete measurement output vector $\vec{\Delta y} \in \mathbb{R}^N$ with the components, indexed by integers $t \in [N]$, as,
$
    \Delta y_{t} 
    =
    v_{t} \frac{ \Delta t }{\tau} + \frac{1}{\sqrt{\tau}} \Delta W_{t}
$
where 
$
v_{t} 
= 
\sqrt{\eta} \, \text{Tr}\left[ \rho_t \left( L + L ^\dag \right) \right]
$ is the mean value of the measurement operator given the state at time $t$ and $\Delta W_t \sim \mathcal{N}(0, \Delta t)$  are i.i.d Wiener increments. 
We approximate the dynamics 
$d\rho_t \approx d\rho^{(0)}_t$, and then the solution is $\rho^{(0)}_t$ of the Lindblad evolution Eq.~\ref{eq:Lindblad}. In this case, the mean dynamics is independent of the noise $\Delta W_t$ and the solution only depends on the initial states. Therefore, any specific instance of the discrete measurement vector $\vec{\Delta y}$ is just the mean vector with Isotropic gaussian noise. 

In our qubit setting, we have two mean vectors $v^{ \uparrow}_t$ and $v^{\downarrow}_t$ corresponding to initial spin up $\rho^{(0) \uparrow}_0 = |\up\rangle\langle\up|$ and spin down $\rho^{(0) \downarrow}_0 = |\down\rangle\langle\down|$ states, respectively. In this case, our mean vectors are separated by a single vector defining the axis of variation. We exploit this and transform the measurement vector ${\Delta y}_t$ onto this axis of variation. Now, the only relevant component, with every other component the same, of our discrete measurement output is,
$    \Delta y_{||}
    =
    v_{||} \frac{ \Delta t }{\tau} + \frac{1}{ \sqrt{\tau} } \Delta W_{||},
$ where, due to symmetry, $\Delta W_{||}$ is still isotropic gaussian noise.
Now, only the $\Delta y_{||}$ component has information about the initial state. This component is just a binary variable with additive gaussian nose.
This is exactly the bi-AWGN channel setting (see \ref{sec:bi-AWGN_MI}), for which we can compute the mutual information.

\begin{figure}
    \centering
 \begin{overpic}[width=0.45\textwidth]{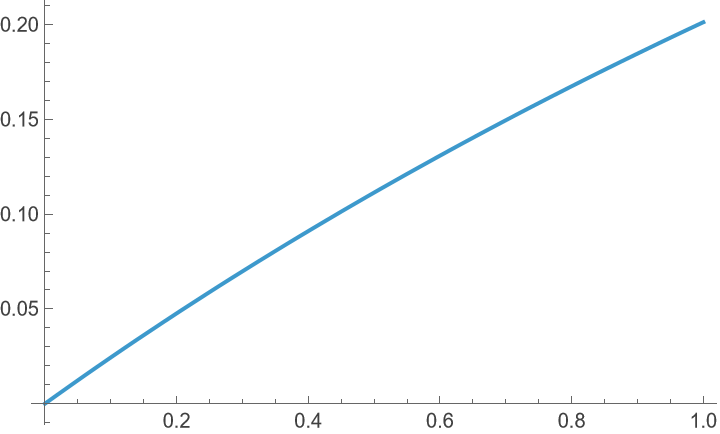}
        \put(88,42){\textbf{(a)}}
        \put(12,42){\textbf{$max I_1$}}
        \put(92,10){$\eta$}
    \end{overpic}
 \begin{overpic}[width=0.45\textwidth]{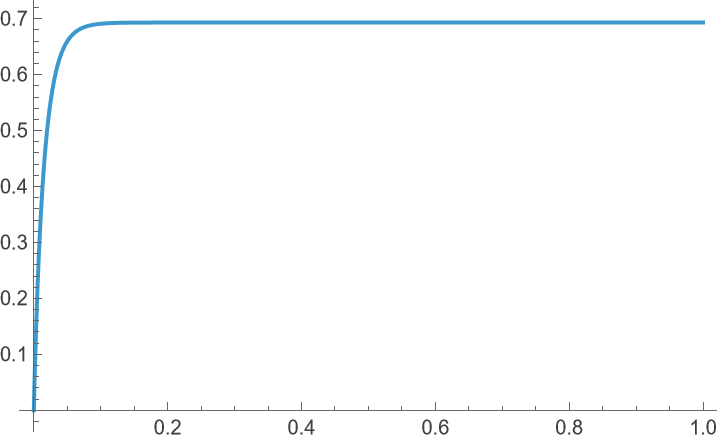}
        \put(88,42){\textbf{(b)}}
        \put(12,42){\textbf{$max I_2(\alpha=0.1)$}}
        \put(92,10){$\eta$}
    \end{overpic}
    
 \begin{overpic}[width=0.45\textwidth]{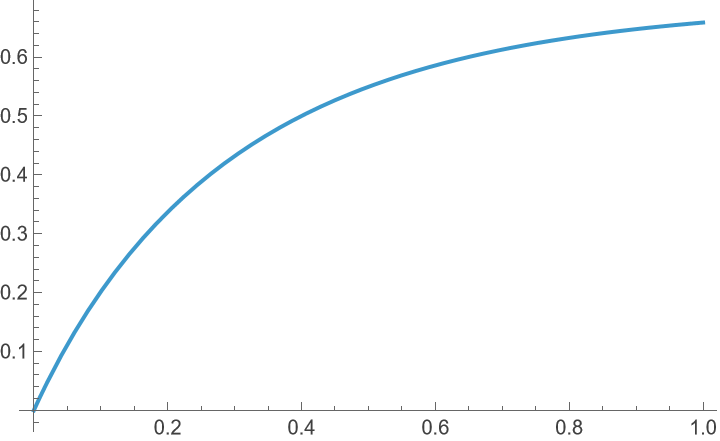}
        \put(88,42){\textbf{(c)}}
        \put(8,52){\textbf{$max I_2(\alpha=1/2)$}}
        \put(92,10){$\eta$}
    \end{overpic}
 \begin{overpic}[width=0.45\textwidth]{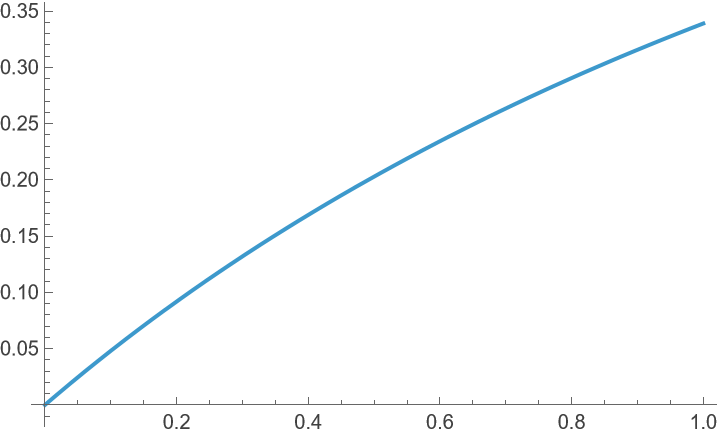}
        \put(88,42){\textbf{(d)}}
        \put(12,42){\textbf{$max I_2(\alpha=10)$}}
        \put(92,10){$\eta$}
    \end{overpic}
    \caption{We display saturation values (as $t\to \infty$) of mutual information in model 1, in panel (a) (from Eq. \ref{eq:model1formula}); and model 2 in panels (b,c,d) at (weak,medium,strong) precession, respectively, as indicated by values of $\alpha$ (see Eqs. \ref{eq:model2Aformula}, \ref{eq:model2Bformula}, \ref{eq:model2Cformula}). Recall, $\alpha=1/2$ is a transition from overdamped to underdamped dynamics in model 2. }
    \label{fig:smalleffextrap}
\end{figure}

The end result is an explicit formula 
\begin{gather}
    \label{eq:small_efficiency_AGWN_mutual_information}
    I(S,Y)(t)
    =
    \int_{-\infty}^{\infty}
    \frac{e^{-\frac{z^2}{2}}}{\sqrt{2\pi}}\Big[\log 2-\log\big(1+e^{-2\gamma(t)-2\sqrt{\gamma(t)}z}\big)\Big]
    \, dz,
    \\
    \label{eq:gammat}
    \gamma(t)
    =
    \sum_\nu 
    \frac{\eta_\nu}{4\tau}
    \int_0^t  ds
    \Bigg[ 
    \text{Tr}\left[
    (L_\nu+L_\nu^\dagger)(\rho^{(0) \uparrow}_{s}-\rho^{(0) \downarrow}_{s} ) \right]
    \Bigg]^2
\end{gather}
where $\gamma(t)$ is like a signal-to-noise ratio (SNR) in our problem and $\rho^{(0) \uparrow}_t$ and $\rho^{(0) \downarrow}_t$ are the Lindblad solutions corresponding to the initial spin up $\rho^{(0) \uparrow}_0$ and spin down $\rho^{(0) \downarrow}_0$ states, respectively.  In typical situations we expect the integrand in Eq.~\ref{eq:gammat} decays exponentially, so $\gamma(t\to \infty)$ saturates quickly and determines the maximum information content. In some cases, e.g. of commuting measurements (and dynamics), the integrand remains constant and $\gamma(t\to \infty)$ diverges, thus producing maximal mutual information of $I=\log 2$, up to small correction $\propto e^{-\gamma(t)/2}$.
While only strictly justified as a calculation of the linear $I\propto \eta$ initial growth, this approach seems produce surprisingly accurate extrapolations to $\eta=1$. We calculate (see \ref{sec:model1application} and \ref{sec:model2application}) and plot the results for both models in Fig. \ref{fig:smalleffextrap}. We compare extrapolated theoretical values against direct simulations in Figs. \ref{fig:MI_numerical_simulations} (at $\eta=1$) and Fig.~\ref{fig:Noisy_MI_plots} (several $\eta$'s), respectively. We have also plotted and compared (favorably) the full time dependence of these solutions in the latter figure.

\subsection{ Application to Model I}
\label{sec:model1application}
Here, we estimate the mutual information in our informationally complete weak measurement model, Eq.~\ref{eq:model_1_noisy_sme_main_text}, by solving for the Lindblad dynamics, in the small efficiency limit, and using the bi-AWGN channel.
The Lindblad Master equation for this measurement scheme is,
\begin{equation}
    d\rho^{(0)}_t
    =
    \frac{1}{\tau}
    \sum_{i}
    \left( 
    \sigma_i  \rho^{(0)}_t \sigma_i
    -
    \rho^{(0)}_t
    \right)
    dt
\end{equation}
We solve this in the Pauli representation defined by $\rho_t = \frac{1}{2}\left[ \rho_0(t) \mathbb{I} + \vec{\rho}(t) \cdot \vec{\sigma} \right] \equiv \left( \rho_0(t), \rho_x(t), \rho_y(t), \rho_z(t) \right)$. Then the Lindblad Master equation in this representation is,
\begin{align}
    \nonumber
    d
    \begin{bmatrix}
        \rho_0 \\
        \rho_x \\
        \rho_y \\
        \rho_z
    \end{bmatrix}
    &=
    \frac{1}{\tau}
    \left[
    \left( 
    \begin{bmatrix}
        \rho_0 \\
        \rho_x \\
        - \rho_y \\
        - \rho_z
    \end{bmatrix}
    -
    \begin{bmatrix}
        \rho_0 \\
        \rho_x \\
        \rho_y \\
        \rho_z
    \end{bmatrix}
    \right)
    +
    \left( 
    \begin{bmatrix}
        \rho_0 \\
        - \rho_x \\
        \rho_y \\
        - \rho_z
    \end{bmatrix}
    -
    \begin{bmatrix}
        \rho_0 \\
        \rho_x \\
        \rho_y \\
        \rho_z
    \end{bmatrix}
    \right)
    +
    \left( 
    \begin{bmatrix}
        \rho_0 \\
        - \rho_x \\
        - \rho_y \\
        \rho_z
    \end{bmatrix}
    -
    \begin{bmatrix}
        \rho_0 \\
        \rho_x \\
        \rho_y \\
        \rho_z
    \end{bmatrix}
    \right)
    \right]
    dt
    \\
    &=
    -\frac{2}{\tau}
    \left(
    \begin{bmatrix}
        0 \\
        0 \\
        \rho_y \\
        \rho_z
    \end{bmatrix}
    +
    \begin{bmatrix}
        0 \\
        \rho_x \\
        0 \\
        \rho_z
    \end{bmatrix}
    +
    \begin{bmatrix}
        0 \\
        \rho_x \\
        \rho_y \\
        0
    \end{bmatrix}
    \right)
    dt
    =
    -\frac{4}{\tau}
    \begin{bmatrix}
        0 \\
        \rho_x \\
        \rho_y \\
        \rho_z
    \end{bmatrix}
    dt
\end{align}
or alternatively, $d\rho_i(t) = - \frac{4}{\tau} \rho_i(t) dt$ where $i\in \{x, y, z\}$.
Therefore the components of the density matrix $\rho(t)$ are
\begin{align}
    \rho_0(t) &= \rho_0(0)
    \\
    \rho_i(t)
    &=
    \rho_i(0) e^{-\frac{4}{\tau}t}
\end{align}

\noindent
Solving for initial up and down spin ($\rho^{(0) \uparrow}(0) =  \frac{1}{2}\left[ \mathbb{I} + Z \right] $ and $\rho^{(0) \downarrow}(0) =  \frac{1}{2}\left[ \mathbb{I} - Z \right]$) states we get,
\begin{align}
    \rho^{(0) \uparrow}(t)
    &= 
    \frac{1}{2}\left[ \mathbb{I} + e^{-\frac{4}{\tau}t} Z \right]
    \\
    \rho^{(0) \downarrow}(t)
    &= 
    \frac{1}{2}\left[ \mathbb{I} - e^{-\frac{4}{\tau}t} Z \right]
\end{align}
and the difference is $\rho^{(0) \uparrow}(t) - \rho^{(0) \downarrow}(t) = e^{-\frac{4}{\tau}t} Z $. Therefore the SNR $\gamma(t)$ is
\begin{align}
    \gamma(t)
    &= 
    \sum_i
    \frac{\eta_i}{4\tau}
    \int_0^t
    \left(
    2\text{Tr}\left[
    \sigma_i 
    \left( e^{-\frac{4}{\tau}s} Z \right)
    \right]
    \right)^2ds
    =
    \frac{\eta_z}{4\tau}
    \int_0^t
    \left(
    2\text{Tr}\left[
    e^{-\frac{4}{\tau}s} \mathbb{I}
    \right]
    \right)^2ds
    \nonumber \\
    &=
    \frac{4 \eta_z}{\tau}
    \int_0^t
    e^{-\frac{8}{\tau}s} 
    ds
    =
    \frac{\eta_z}{2}
    \left[ 1- e^{-\frac{8}{\tau}t} \right]\to     \frac{\eta}{2}
\end{align}
We can get the time-dependent mutual information from  eq.~\ref{eq:small_efficiency_AGWN_mutual_information} for this $\gamma(t)$ as,
\begin{align} 
    I(S,Y)(t)
    &=
    \int_{-\infty}^\infty
    \frac{e^{-\frac{z^2}{2}}}{\sqrt{2\pi}}
    \left[
    \log 2-\log{\left(1+e^{-2\gamma(t)-2\sqrt{\gamma(t)}z}\right)}
    \right]dz
    \nonumber
    \\ 
    &
    =
    \log{2}
    -
    \int_{-\infty}^\infty
    \frac{e^{-\frac{z^2}{2}}}{\sqrt{2\pi}}\log{\left(1+e^{-2\gamma(t)-2\sqrt{\gamma(t)}z}\right)}
    dz.
    \label{eq:model1formula}
\end{align}

\excludecomment{commentA}
\begin{commentA}

\subsection{ Informationally incomplete measurements with unitary dynamics }

For our next model, informationally incomplete measurements with unitary dynamics,
the Lindblad Master Equation is,
\begin{equation}
    d\rho^{(0)}_t
    =
    (-i)
    [X, \rho_t ] \alpha dt
    +
    \left( 
    Z  \rho_t  Z
    -
    \rho_t
    \right)
    dt
\end{equation}
where for simplicity we absorb all auxiliary parameters into $\alpha$ and set the timescale scale $\tau=1$.
We again solve this in the Pauli representation, where the Lindblad Master Equation in this representation is,
\begin{align}
    \nonumber
    d
    \begin{bmatrix}
        \rho_0 \\
        \rho_x \\
        \rho_y \\
        \rho_z
    \end{bmatrix}
    &=
    \left( 
    (-i) (-2) \alpha
    \begin{bmatrix}
        0 \\
        0 \\
        i \rho_z \\
        -i \rho_y
    \end{bmatrix}
    +
    (-2)
    \begin{bmatrix}
        0 \\
        \rho_x \\
        \rho_y \\
        0
    \end{bmatrix}
    \right)
    dt
    \\
    &=
    -2
    \begin{bmatrix}
        0 \\
        \rho_x \\
        \alpha \rho_z + \rho_y \\
        -\alpha \rho_y
    \end{bmatrix}
    dt
\end{align}
Therefore the components of the density matrix $\rho^{(0)}(t)$ for general $\alpha$ are,

\begin{align}
    \rho_0(t) &= \rho_0(0)
    \\
    \rho_x(t)
    &=
    e^{-2t} \rho_x(0)
    \\
    \rho_y(t)
    &=
    \frac{e^{-t}}{ 2\omega }
    \left[ 
    \frac{1}{2} 
    \left( 
    \omega
    \left( 
    e^{\omega t}
    +
    e^{-\omega t}
    \right) 
    +
    \left( 
    e^{-\omega t}
    -
    e^{\omega t}
    \right) 
    \right)
    \rho_y(0) 
    +
    \alpha
    \left( 
    e^{-\omega t}
    -
    e^{\omega t}
    \right)
    \rho_z(0) 
    \right]
    \\
    \rho_z(t)
    &=
    \frac{e^{-t}}{ 2\omega }
    \left[ 
    \frac{1}{2} 
    \left( 
    \omega
    \left( 
    e^{\omega t}
    +
    e^{-\omega t}
    \right) 
    -
    \left( 
    e^{-\omega t}
    -
    e^{\omega t}
    \right) 
    \right)
    \rho_z(0) 
    -
    \alpha
    \left( 
    e^{-\omega t}
    -
    e^{\omega t}
    \right)
    \rho_y(0) 
    \right]
\end{align}
where $\omega \coloneqq \sqrt{4\alpha^2-1}$. 

\noindent
In particular we parameterize, 
\begin{enumerate}
    \item 
    For $\alpha > \frac{1}{2}$,
    let $\omega \coloneqq \sqrt{4\alpha^2-1}$,
    \begin{align}
        \rho_y(t)
        &=
        \frac{e^{-t}}{ \omega }
        \left[ 
        \left(
        \omega
        \cos{\left( \omega t \right)} 
        -
        \sin{\left( \omega t \right)} 
        \right)
        \rho_y(0)
        -
        2 \alpha
        \sin{\left( \omega t \right)}
        \rho_z(0)
        \right]
        \\
        \rho_z(t)
        &=
        \frac{e^{-t}}{ \omega }
        \left[ 
        \left(
        \omega 
        \cos{\left( \omega t \right)} 
        +
        \sin{\left( \omega t \right)} 
        \right)
        \rho_z(0)
        +
        2 \alpha
        \sin{\left( \omega t \right)}
        \rho_y(0)
        \right]
    \end{align}

    \item 
    For $\alpha < \frac{1}{2}$, 
    let $\omega \coloneqq \sqrt{1-4\alpha^2}$,
    \begin{align}
        \rho_y(t)
        &=
        \frac{e^{-t}}{ \omega }
        \left[ 
        \left( 
        \omega
        \cosh{\left( \omega t \right)} 
        -
        \sinh{\left( \omega t \right)} 
        \right)
        \rho_y(0)
        -
        2 \alpha
        \sinh{\left( \omega t \right)} 
        \rho_z(0)
        \right]
        \\
        \rho_z(t)
        &=
        \frac{e^{-t}}{ \omega }
        \left[ 
        \left( 
        \omega
        \cosh{\left( \omega t \right)} 
        +
        \sinh{\left( \omega t \right)} 
        \right)
        \rho_z(0)
        +
        2 \alpha
        \sinh{\left( \omega t \right)} 
        \rho_y(0)
        \right]
    \end{align}
    \item For $\alpha = \frac{1}{2}$,
    \begin{align}
        \rho_y(t)
        &=
        e^{-t}
        \left(
        \left( 
        1-t
        \right)
        \rho_y(0)
        -
        t
        \rho_z(0)
        \right)
        \\
        \rho_z(t)
        &=
        e^{-t}
        \left(
        \left( 
        t+1
        \right)
        \rho_z(0) 
        +
        t
        \rho_y(0)
        \right)
    \end{align}
\end{enumerate}

\noindent
Solving for initial up and down spin states $\left( \rho^{0\uparrow}(0) =  \frac{1}{2}\left[ \mathbb{I} + Z \right] \right. $ and $\left. \rho^{0\downarrow}(0) =  \frac{1}{2}\left[ \mathbb{I} - Z \right] \right)$ we get,
\begin{enumerate}
    \item $\alpha > \frac{1}{2}$:
    \begin{align*}
        \rho^{0\uparrow}(t)
        &= 
        \frac{1}{2}\left[ 
        \mathbb{I} 
        +
        \frac{-2 \alpha e^{-t} }{\omega}
        \sin{\left( \omega t \right)}
        Y
        +
        \frac{ e^{-t} }{\omega}
        \left(
        \omega
        \cos{\left( \omega t \right)}
        +
        \sin{\left( \omega t \right)}
        \right)
        Z
        \right]
        \\
        \rho^{0\downarrow}(t)
        &= 
        \frac{1}{2}\left[ 
        \mathbb{I} 
        +
        \frac{2 \alpha e^{-t} }{\omega}
        \sin{\left( \omega t \right)}
        Y
        +
        \frac{ -e^{-t} }{\omega}
        \left(
        \omega
        \cos{\left( \omega t \right)}
        +
        \sin{\left( \omega t \right)}
        \right)
        Z
        \right]
    \end{align*}
    and the difference is
    \begin{align}
        \rho^{0\uparrow}(t) - \rho^{0\downarrow}(t) 
        &= 
        \left[ 
        \frac{-2 \alpha e^{-t} }{\omega}
        \sin{\left( \omega t \right)}
        Y
        +
        \frac{ e^{-t} }{\omega}
        \left(
        \omega
        \cos{\left( \omega t \right)}
        +
        \sin{\left( \omega t \right)}
        \right)
        Z
        \right]
    \end{align}
    Therefore the SNR $\gamma(t)$ is,
    \begin{align}
        \gamma(t)
        &=
        \eta
        \left( 
        \frac{1+\alpha^2}{\alpha^2}
        +
        \frac{e^{-2t}}{ a^2 \left(4\alpha^2-1 \right)}
        \left( 
        - 4\alpha^4
        +
        \left( -3\alpha^2+1 \right)
        \cos{ \left( 2\sqrt{4\alpha^2-1} \ t \right)}
        +
        \left( \alpha^2-1 \right)
        \sqrt{4\alpha^2-1}
        \sin{ \left( 2\sqrt{4\alpha^2-1} \ t \right)}
        \right)
        \right)
    \end{align}

    \item $\alpha < \frac{1}{2}$:
    \begin{align}
        \rho^{0\uparrow}(t)
        &= 
        \frac{1}{2}\left[ 
        \mathbb{I} 
        +
        \frac{-2 \alpha e^{-t} }{\omega}
        \sinh{\left( \omega t\right)}
        Y
        +
        \frac{ e^{-t} }{\omega}
        \left(
        \omega
        \cosh{\left( \omega t\right)}
        +
        \sinh{\left( \omega t\right)}
        \right)
        Z
        \right]
        \\
        \rho^{0\downarrow}(t)
        &= 
        \frac{1}{2}\left[ 
        \mathbb{I} 
        +
        \frac{2 \alpha e^{-t} }{\omega}
        \sinh{\left( \omega t\right)}
        Y
        +
        \frac{ -e^{-t} }{\omega}
        \left(
        \omega
        \cosh{\left( \omega t\right)}
        +
        \sinh{\left( \omega t\right)}
        \right)
        Z
        \right]
    \end{align}
    and the difference is
    \begin{align}
        \rho^{0\uparrow}(t) - \rho^{0\downarrow}(t) 
        &= 
        \left[ 
        \frac{-2 \alpha e^{-t} }{ \omega }
        \sinh{\left( \omega t\right)}
        Y
        +
        \frac{ e^{-t} }{ \omega }
        \left(
        \omega
        \cosh{\left( \omega t\right)}
        +
        \sinh{\left( \omega t\right)}
        \right)
        Z
        \right]
    \end{align}
    Then the SNR $\gamma(t)$ is,
    \begin{align*}
        \gamma(t)
        &= \eta 
        \left( 
        \frac{1+a^2}{a^2}
        +
        \frac{e^{-2t}}{a^2(1-4a^2)}
        \left( 
        4a^4 
        +
        \left( 3a^2-1 \right) \cosh{\left( 2 \sqrt{1-4a^2} t\right)}
        +
        \left( a^2-1 \right) \sqrt{1-4a^2} \sinh{\left( 2\sqrt{1-4a^2} t\right)}
        \right) 
        \right)
    \end{align*}
    \item $\alpha = \frac{1}{2}$:
    \begin{align}
        \rho^{0\uparrow}(t)
        &= 
        \frac{1}{2}\left[ 
        \mathbb{I} 
        +
        \left( - t e^{-t} \right)
        Y
        +
        \left(  (1+t) e^{-t} \right)
        Z
        \right]
        \\
        \rho^{0\downarrow}(t)
        &= 
        \frac{1}{2}\left[ 
        \mathbb{I} 
        +
        \left( t e^{-t} \right)
        Y
        +
        \left( - (1+t) e^{-t} \right)
        Z
        \right]
    \end{align}
    and the difference is
    \begin{align}
        \rho^{0\uparrow}(t) - \rho^{0\downarrow}(t) 
        &= 
        \left[
        \left( - t e^{-t} \right)
        Y
        +
        \left(  (1+t) e^{-t} \right)
        Z
        \right]
    \end{align}
    Then the SNR $\gamma(t)$ is,
    \begin{align*}
        \gamma(t)
        &= 
        \eta
        \left(
        5
        -
        e^{-2t}
        \left( 2t^2 + 6t + 5 \right)
        \right)
    \end{align*}
\end{enumerate}
We then can get the time-dependent mutual information  from eq.~\ref{eq:AGWN_mutual_information} for these $\gamma(t)$.

\end{commentA}

\subsection{Application to Model II}
\label{sec:model2application}
For our next model, informationally incomplete measurements with unitary dynamics, Eq.~\ref{eq:model_2_noisy_sme_main_text},
the Lindblad Master equation is,
\begin{equation}
    d\rho^{(0)}_t
    =
    -i \frac{\omega}{2}
    [X, \rho^{(0)}_t ] dt
    +
    \frac{1}{\tau}
    \left( 
    Z  \rho^{(0)}_t  Z
    -
    \rho^{(0)}_t
    \right)
    dt
\end{equation}
We again solve this in the Pauli representation, where the Lindblad Master equation in this representation is,
\begin{align}
   & d
    \begin{bmatrix}
        \rho_0 \\
        \rho_x \\
        \rho_y \\
        \rho_z
    \end{bmatrix}
    =
    \left( 
    \left( -i\frac{\omega}{2} \right) (-2)
    \begin{bmatrix}
        0 \\
        0 \\
        i \rho_z \\
        -i \rho_y
    \end{bmatrix}
    +
    \left( \frac{1}{\tau} \right) 
    (-2)
    \begin{bmatrix}
        0 \\
        \rho_x \\
        \rho_y \\
        0
    \end{bmatrix}
    \right)
    dt
    =
    -\frac{2}{\tau}
    \begin{bmatrix}
        0 \\
        \rho_x \\
        \alpha \rho_z + \rho_y \\
        -\alpha \rho_y
    \end{bmatrix}
    dt
\end{align}
Where we define the parameter $ \alpha \coloneqq \frac{\omega\tau}{2}$, playing the role of the quality factor, for the dynamics. Therefore the components of the time-dependent density matrix $\rho^{(0)}(t)$ for general $\alpha$ are,

\begin{align}
    \rho_0(t) &= \rho_0(0)
    \\
    \rho_x(t)
    &=
    e^{-\frac{2}{\tau} t} \rho_x(0)
    \\
    \rho_y(t)
    &=
    \frac{e^{- \frac{1}{\tau} t}}{ \mu }
    \left[ 
    \frac{1}{2} 
    \left( 
    \mu
    \left( 
    e^{ \frac{\mu}{\tau} t}
    +
    e^{- \frac{\mu}{\tau} t}
    \right) 
    +
    \left( 
    e^{- \frac{\mu}{\tau} t}
    -
    e^{ \frac{\mu}{\tau} t}
    \right) 
    \right)
    \rho_y(0) 
    +
    \alpha
    \left( 
    e^{- \frac{\mu}{\tau} t}
    -
    e^{ \frac{\mu}{\tau} t}
    \right)
    \rho_z(0) 
    \right]
    \\
    \rho_z(t)
    &=
    \frac{e^{- \frac{1}{\tau} t}}{ \mu }
    \left[ 
    \frac{1}{2} 
    \left( 
    \mu
    \left( 
    e^{\frac{\mu}{\tau} t}
    +
    e^{-\frac{\mu}{\tau} t}
    \right) 
    -
    \left( 
    e^{-\frac{\mu}{\tau} t}
    -
    e^{\frac{\mu}{\tau} t}
    \right) 
    \right)
    \rho_z(0) 
    -
    \alpha
    \left( 
    e^{-\frac{\mu}{\tau} t}
    -
    e^{\frac{\mu}{\tau} t}
    \right)
    \rho_y(0) 
    \right]
\end{align}
where we define $\mu \coloneqq \sqrt{1 - 4\alpha^2}$. 

\noindent
In particular we use the forms, 
\begin{enumerate}
    \item 
    For $\alpha > \frac{1}{2}$,
    let $\mu \coloneqq \sqrt{4\alpha^2-1}$,
    \begin{align}
        \rho_y(t)
        &=
        \frac{e^{-\frac{1}{\tau}t}}{ \mu }
        \left[ 
        \left(
        \mu
        \cos{\left( \frac{\mu}{\tau} t \right)} 
        -
        \sin{\left( \frac{\mu}{\tau} t \right)} 
        \right)
        \rho_y(0)
        -
        2 \alpha
        \sin{\left( \frac{\mu}{\tau} t \right)}
        \rho_z(0)
        \right]
        \\
        \rho_z(t)
        &=
        \frac{e^{-\frac{1}{\tau} t}}{ \mu }
        \left[ 
        \left(
        \mu 
        \cos{\left( \frac{\mu}{\tau} t \right)} 
        +
        \sin{\left( \frac{\mu}{\tau} t \right)} 
        \right)
        \rho_z(0)
        +
        2 \alpha
        \sin{\left( \frac{\mu}{\tau} t \right)}
        \rho_y(0)
        \right]
    \end{align}

    \item 
    For $\alpha < \frac{1}{2}$, 
    let $\mu \coloneqq \sqrt{1-4\alpha^2}$,
    \begin{align}
        \rho_y(t)
        &=
        \frac{e^{- \frac{1}{\tau} t}}{ \mu }
        \left[ 
        \left( 
        \mu
        \cosh{\left( \frac{\mu}{\tau} t \right)} 
        -
        \sinh{\left( \frac{\mu}{\tau} t \right)} 
        \right)
        \rho_y(0)
        -
        2 \alpha
        \sinh{\left( \frac{\mu}{\tau} t \right)} 
        \rho_z(0)
        \right]
        \\
        \rho_z(t)
        &=
        \frac{e^{- \frac{1}{\tau} t}}{ \mu }
        \left[ 
        \left( 
        \mu
        \cosh{\left( \frac{\mu}{\tau} t \right)} 
        +
        \sinh{\left( \frac{\mu}{\tau} t \right)} 
        \right)
        \rho_z(0)
        +
        2 \alpha
        \sinh{\left( \frac{\mu}{\tau} t \right)} 
        \rho_y(0)
        \right]
    \end{align}
    \item For $\alpha = \frac{1}{2}$,
    \begin{align}
        \rho_y(t)
        &=
        e^{-\frac{1}{\tau} t}
        \left(
        \left( 
        1- \frac{t}{\tau}
        \right)
        \rho_y(0)
        -
        \frac{t}{\tau}
        \rho_z(0)
        \right)
        \\
        \rho_z(t)
        &=
        e^{- \frac{1}{\tau} t}
        \left(
        \left( 
        \frac{t}{\tau}  +1
        \right)
        \rho_z(0) 
        +
        \frac{t}{\tau}
        \rho_y(0)
        \right)
    \end{align}
\end{enumerate}

\noindent
Solving for initial up and down spin states $\left( \rho^{(0) \uparrow}(0) =  \frac{1}{2}\left[ \mathbb{I} + Z \right] \right. $ and $\left. \rho^{(0) \downarrow}(0) =  \frac{1}{2}\left[ \mathbb{I} - Z \right] \right)$ we get,
\begin{enumerate}
    \item For $\alpha > \frac{1}{2}$:
    \begin{align*}
        \rho^{(0) \uparrow}(t)
        &= 
        \frac{1}{2}\left[ 
        \mathbb{I} 
        +
        \frac{-2 \alpha e^{- \frac{1}{\tau} t} }{\mu}
        \sin{\left( \frac{\mu}{\tau}  t \right)}
        Y
        +
        \frac{ e^{- \frac{1}{\tau} t} }{\mu}
        \left(
        \mu
        \cos{\left( \frac{\mu}{\tau} t \right)}
        +
        \sin{\left( \frac{\mu}{\tau} t \right)}
        \right)
        Z
        \right]
        \\
        \rho^{(0) \downarrow}(t)
        &= 
        \frac{1}{2}\left[ 
        \mathbb{I} 
        +
        \frac{2 \alpha e^{- \frac{1}{\tau} t} }{\mu}
        \sin{\left( \frac{\mu}{\tau} t \right)}
        Y
        +
        \frac{ -e^{- \frac{1}{\tau} t} }{\mu}
        \left(
        \mu
        \cos{\left( \frac{\mu}{\tau} t \right)}
        +
        \sin{\left( \frac{\mu}{\tau} t \right)}
        \right)
        Z
        \right]
    \end{align*}
    and the difference is,
    \begin{align}
        \rho^{(0) \uparrow}(t) - \rho^{(0) \downarrow}(t) 
        &= 
        \left[ 
        \frac{-2 \alpha e^{- \frac{1}{\tau} t} }{\mu}
        \sin{\left( \frac{\mu}{\tau} t \right)}
        Y
        +
        \frac{ e^{- \frac{1}{\tau} t} }{\mu}
        \left(
        \mu
        \cos{\left( \frac{\mu}{\tau} t \right)}
        +
        \sin{\left( \frac{\mu}{\tau} t \right)}
        \right)
        Z
        \right]
    \end{align}
    Therefore the SNR $\gamma(t)$ is,
    \begin{align}
        \gamma(t)
        &=
        \eta
        \left( 
        \frac{1+\alpha^2}{\alpha^2}
        +
        \frac{e^{- \frac{2}{\tau} t}}{ a^2 \left(4\alpha^2-1 \right)}
        \left( 
        - 4\alpha^4
        +
        \left( -3\alpha^2+1 \right)
        \cos{ \left( 2 \frac{\sqrt{4\alpha^2-1}}{\tau}  t \right)}\right.\right.
        \nonumber \\ &+
\left.\left.     \left( \alpha^2-1 \right)
        \sqrt{4\alpha^2-1}
        \sin{ \left( 2 \frac{\sqrt{4\alpha^2-1}}{\tau}  t \right)}
        \right)
        \right)
                    \label{eq:model2Cformula}
    \end{align}

    \item For $\alpha < \frac{1}{2}$:
    \begin{align}
        \rho^{(0) \uparrow}(t)
        &= 
        \frac{1}{2}\left[ 
        \mathbb{I} 
        +
        \frac{-2 \alpha e^{- \frac{1}{\tau} t} }{\mu}
        \sinh{\left( \frac{\mu}{\tau} t\right)}
        Y
        +
        \frac{ e^{- \frac{1}{\tau} t} }{\mu}
        \left(
        \mu
        \cosh{\left( \frac{\mu}{\tau} t\right)}
        +
        \sinh{\left( \frac{\mu}{\tau} t\right)}
        \right)
        Z
        \right]
        \\
        \rho^{(0) \downarrow}(t)
        &= 
        \frac{1}{2}\left[ 
        \mathbb{I} 
        +
        \frac{2 \alpha e^{- \frac{1}{\tau} t} }{\mu}
        \sinh{\left( \frac{\mu}{\tau} t\right)}
        Y
        +
        \frac{ -e^{- \frac{1}{\tau} t} }{\mu}
        \left(
        \mu
        \cosh{\left( \frac{\mu}{\tau} t\right)}
        +
        \sinh{\left( \frac{\mu}{\tau} t\right)}
        \right)
        Z
        \right]
    \end{align}
    and the difference is,
    \begin{align}
        \rho^{(0) \uparrow}(t) - \rho^{(0) \downarrow}(t) 
        &= 
        \left[ 
        \frac{-2 \alpha e^{- \frac{1}{\tau} t} }{ \mu }
        \sinh{\left( \frac{\mu}{\tau} t\right)}
        Y
        +
        \frac{ e^{- \frac{1}{\tau} t} }{ \mu }
        \left(
        \mu
        \cosh{\left( \frac{\mu}{\tau} t\right)}
        +
        \sinh{\left( \frac{\mu}{\tau} t\right)}
        \right)
        Z
        \right]
    \end{align}
    Then the SNR $\gamma(t)$ is,
    \begin{align*}
        \gamma(t)
        &= \eta 
        \left( 
        \frac{1+\alpha^2}{\alpha^2}
        +
        \frac{e^{- \frac{2}{\tau} t}}{\alpha^2(1-4\alpha^2)}
        \left( 
        4\alpha^4 
        +
        \left( 3\alpha^2-1 \right) \cosh{\left( 2 \frac{\sqrt{1-4\alpha^2}}{\tau} t\right)}
\right.\right. \nonumber \\    
&+
\left.\left.    
\left( \alpha^2-1 \right) \sqrt{1-4\alpha^2} \sinh\left( 2 \frac{\sqrt{1-4\alpha^2}}{\tau} t\right)
        \right) 
        \right)
                    \label{eq:model2Aformula}
    \end{align*}
    \item For $\alpha = \frac{1}{2}$:
    \begin{align}
        \rho^{(0) \uparrow}(t)
        &= 
        \frac{1}{2}\left[ 
        \mathbb{I} 
        +
        \left( - \frac{t}{\tau} e^{- \frac{1}{\tau} t} \right)
        Y
        +
        \left(  \left( 1+\frac{t}{\tau} \right) e^{- \frac{1}{\tau} t} \right)
        Z
        \right]
        \\
        \rho^{(0) \downarrow}(t)
        &= 
        \frac{1}{2}\left[ 
        \mathbb{I} 
        +
        \left( \frac{t}{\tau} e^{- \frac{1}{\tau} t} \right)
        Y
        +
        \left( - \left( 1+\frac{t}{\tau} \right) e^{- \frac{1}{\tau} t} \right)
        Z
        \right]
    \end{align}
    and the difference is,
    \begin{align}
        \rho^{(0) \uparrow}(t) - \rho^{(0) \downarrow}(t) 
        &= 
        \left[
        \left( - \frac{t}{\tau} e^{- \frac{1}{\tau} t} \right)
        Y
        +
        \left(  \left( 1+ \frac{t}{\tau} \right) e^{- \frac{1}{\tau} t} \right)
        Z
        \right]
    \end{align}
    Then the SNR $\gamma(t)$ is,
    \begin{align}  
        \gamma(t)
        &= 
        \eta
        \left(
        5
        -
        e^{- \frac{2}{\tau} t}
        \left( 2\left(\frac{t}{\tau}\right)^2 + 6\frac{t}{\tau} + 5 \right)
        \right)
            \label{eq:model2Bformula}
  \end{align}  
\end{enumerate}
We then can get the time-dependent mutual information  from Eq.~\ref{eq:small_efficiency_AGWN_mutual_information} for these $\gamma(t)$. We also numerically solve these and plot the results in the main text in Fig.~\ref{fig:Noisy_MI_plots}.

\section{\label{sec:bi-AWGN_MI}  Brief review of bi-AWGN channel}

The binary Additive Gaussian White Noise (bi-AWGN) is a 
classical communication channel of noisy observations of a clean signal. It is defined with the (noisy observations) output $Y_t=\sqrt{\rho}X_t+Z_t$ with (clean signal) input $X_t \in \{\pm 1\}$ and (noise) $Z_t\sim\mathcal{N}(0,1)$ with $Z_t$ being independent for each $t$. For a single time point with $X$ taking values $\pm 1$ with equal probability, while the output is given by $Y=\sqrt{\rho}X+Z$, where $Z\sim\mathcal{N}(0,1)$ with $Z$ being independent of $X$, we can derive the expression for the mutual information as,
\begin{equation}
\label{bi-AWGN}
    I(X,Y)
    =
    \int_{-\infty}^\infty\frac{e^{-\frac{z^2}{2}}}{\sqrt{2\pi}}\Big[\log 2-\log\big(1+e^{-2\rho-2\sqrt{\gamma}z}\big)\Big]dz.
\end{equation}

If $X,Z,Y$ take values in $\mathbb R^n$, with $X$ taking two ``values" $\mu_\pm\in \mathbb R^n$ with equal probability, while the output is given by $Y=\sqrt{\gamma}X+Z$, $Z\sim\mathcal{N}(0,\sigma^2 I_n)$ with $Z$ being independent of $X$, the mutual information would be given by the same formula as Eq. \ref{bi-AWGN} with,
$   \gamma
    =
    \bigg(\frac{||\mu_+-\mu_-||}{2\sigma}\bigg)^2.
$

\end{document}